\documentclass[aps,prl,graphicx,singlecolumn,a4paper]{article}

\usepackage{graphicx}
\usepackage{lineno,hyperref}
\usepackage{upgreek} 
\usepackage{mathcomp} 
\usepackage{textcomp} 
\usepackage{longtable}
\usepackage{tabularx}
\usepackage{color}

\newcommand{\sFF}{Figure\,\ref}

\begin{document}


\begin{center}
\vspace{1cm}
\LARGE{\textbf{Positrons in Surface Physics}} \\
\vspace{1cm}
\large Christoph Hugenschmidt \\
\vspace{1cm}
\normalsize FRM\,II and Physik-Department E21, Technische Universit\"at M\"unchen, Lichtenbergstra{\ss}e 1, 85748, Germany \\
\vspace{1cm}
Email: christoph.hugenschmidt$@$frm2.tum.de \\
\vspace{2cm}

\end{center}




Within the last decade powerful methods have been developed to study surfaces using bright low-energy positron beams.
These novel analysis tools exploit the unique properties of positron interaction with surfaces, which comprise the absence of exchange interaction, repulsive crystal potential and positron trapping in delocalized surface states at low energies. 
By applying reflection high-energy positron diffraction (RHEPD) one can benefit from the phenomenon of total reflection below a critical angle that is not present in electron surface diffraction. 
Therefore, RHEPD allows the determination of the atom positions of (reconstructed) surfaces with outstanding accuracy. 
The main advantages of positron annihilation induced Auger-electron spectroscopy (PAES) are the missing secondary electron background in the energy region of Auger-transitions and its topmost layer sensitivity for elemental analysis. 
In order to enable the investigation of the electron polarization at surfaces low-energy spin-polarized positrons are used to probe the outermost electrons of the surface.
Furthermore, in fundamental research the preparation of well defined surfaces tailored for the production of bound leptonic systems  plays an outstanding role.
In this report,  it is envisaged to cover both, the fundamental aspects of positron surface interaction and the present status of surface studies using modern positron beam techniques. 
\vspace{2cm}


\vspace{1cm}


 \newpage
 \setcounter{secnumdepth}{3}
 \setcounter{tocdepth}{3}
 \pdfbookmark[1]{Table of contents}{toc}
 \tableofcontents
 \newpage      
\section{Introduction}
\label{Intro}
In  surface science the structure and elemental composition ideally of the topmost atomic layer of a solid and all kinds of phenomena related to the surface are subject of most research activities.
For this a zoo of well-known standard analysis tools based on spectroscopic, microscopic, scattering and diffraction methods are commonly applied.

In the 1980's, positron diffraction experiments \cite{Ros80, Wei83} as well as measurements of positron annihilation induced Auger electrons \cite{Wei88} have been carried out for the first time. 
Several techniques using low-energy positrons were shown to exhibit, among other features, outstanding surface sensitivity.
In contrast, complementary methods, such as electron diffraction, classical Auger-electron and photo electron spectroscopy suffer from additional signals originating from layers below.
For this reason, open questions with regard to the elemental analysis, atom positions and electron polarization at the surface have been successfully addressed by positron beam experiments. 
The application of the positron as a surface probe, however, has become established only slowly mainly due to the comparable low intensities provided by positron beam systems built in several  laboratories around the world. 
Therefore, great efforts have been made to improve both, the beam intensity of low-energy positrons and the efficient detection of the messenger particles, i.e.\,positrons, electrons or photons.

In the last decades remarkable advances have been made in the production of evermore intense positron beams.
A great step forward has been the development of  strong positron sources at large-scale research facilities yielding intensities in the order of $10^8-10^9$ moderated positrons per second. 
Currently, two positron beam systems operated at an electron linac (KEK, Japan) \cite{Hyo11, Wad12, Wad13} and at a research reactor (NEPOMUC, Germany) \cite{Hug08b, Hug13b} are dedicated to surface experiments. 
In the light of their great potential novel positron beam techniques have been developed successfully such as (total) reflection high-energy positron diffraction ((T)RHEPD) \cite{Kaw98} and time dependent positron annihilation induced Auger-electron spectroscopy (PAES) \cite{May10c}.
On the other hand the measurement times for surface studies could be reduced considerably by improving detection efficiencies, e.g.\,for recording electron spectra,  and by development of new techniques such as the combination of the time-of-flight method and adiabatic magnetic beam guiding systems \cite{Lei89, Ohd99}.

In solid state physics and material science positron annihilation spectroscopy (PAS) has become an established non-destructive technique for defect spectroscopy of bulk materials since lattice defects such as vacancies, dislocations or precipitates determine the mechanical, optical and electronic properties of materials to a large extent. 
The most widespread methods are positron annihilation lifetime spectroscopy (PALS) and Doppler-broadening spectroscopy (DBS) of the positron electron annihilation line where one profits from the positrons high affinity to open-volume defects.
Coincident DBS (CDBS) provides additional information on the chemical surrounding  of open volume defects or the presence of precipitates.
Since the positron typically diffuses over hundreds of lattice constants prior to annihilation, the positron can be regarded as an extremely sensitive nano-probe to detect e.g.\,single vacancy concentrations as low as 10$^{-7}$ vacancies per atom.
On the other hand, in defect-free crystals the electronic structure can be revealed by the measurement of the angular correlation of annihilation radiation (ACAR).
For PAS in solid state physics refer to standard references such as \cite{hauto1979} and \cite{Dupasquier1995}.

With the advent of monoenergetic positron beams, the near surface region of samples became object of research. 
By variation of the positron energy up to 30\,keV the mean positron implantation depth can be chosen  between a few nm  and several $\mathrm{\tcmu}$m.
The according probed volume is hence defined by the positron implantation profile blurred by the positron diffusive motion, which is in the order of 100\,nm for defect-free crystals, and by the positron beam diameter which typically ranges between 10\,$\mathrm{\tcmu}$m and a few mm.
Thus, the investigation of the type and the concentration of open volume defects as a function of positron implantation depth, and, provided that a sufficient small beam diameter is achieved, spatially resolved PAS have become feasible.
Therefore, since many years numerous positron beam experiments have been performed addressing a wide field of scientific issues with regard to the structure and defects in the  near surface region, thin films, multi-layers, and interfaces. 
Such depth dependent PAS studies comprise: 
(i) element selective analysis of metallic clusters and layers buried in Al with CDBS \cite{Hug08a, Pik11}, and, using ACAR, the detection of semi-coherently Li clusters embedded in single crystalline MgO, which have been formed after Li ion implantation \cite{Fal02}, 
(ii) in oxide systems, the characterization of different types of metallic vacancies present in pulsed laser deposited thin film perovskites (see e.g.\,\cite{Kee10}), and imaging of the spatial variation of the oxygen deficiency, and hence the local variation of the transition temperature for superconductivity,  in single crystalline YBa$_2$Cu$_3$O$_{7-\delta}$ thin films \cite{Rei15a},
(iii) in polymer films, the determination of the free volume \cite{Egg08} and the open volume in bioadhesive \cite{Rae10},
(iv) thin film annealing and alloying at a Au/Cu interface \cite{Rei14a} as well as interdiffusion in epitaxial single-crystalline Au/Ag thin films \cite{Noa16},
(v) irradiation induced defects e.g.\,in Mg \cite{Sta07b} and in silica \cite{Osh09a, Bru10} as well as defect annealing after He implantation in InN and GaN \cite{Reu10} and H induced defects in Pd films \cite{Ciz09},
(vi) defect imaging of the fatigue damage in Cu \cite{Gre97},  around a scratch on a GaAs sample \cite{Dav01}, and in Al samples after plastic deformation \cite{Hug09b}.
For further details of near-surface studies with positron refer to appropriate reviews (see e.g.\,\cite{Sch88, Pus94} and \cite{Coleman2000}).


Within the scope of this review  it is intended to focus on positron studies of the topmost atomic layers, i.e.\,positron surface investigations \textit{par excellence}. 
Thus, low-energy  positron beams with a kinetic energy perpendicular to the surface of typically a few eV have to be applied in order to examine the features of reconstructed surfaces, adsorbates, single adsorbed layers and their spacing to the substrate as well as cover layers with a nominal thickness in the (sub-)monolayer (ML) range. 
The studies presented here comprise investigations of the elemental composition, the atom positions and the electron polarization in the outermost atomic layer. 
 
The knowledge of slight displacements of the atomic positions in the topmost surface layers is crucial for the correct description of the electronic structure and magnetic properties of crystal surfaces.
For this reason, (total) reflection high-energy positron diffraction ((T)RHEPD) was shown to be highly beneficial since the intrinsic features of this technique, i.e.\,missing exchange correlation and repulsive scattering potential, give rise to an outstanding accuracy in the determination of the atom positions at the surface.
In contrast to its electron counterpart, positron diffraction patterns are easier to calculate due to the opposite sign of the scattering potential, and  total reflexion is only present in the positron case.
Other diffraction techniques such as small angle X-ray or neutron scattering are barely sensitive to the topmost atomic layer since even at grazing incidence the penetration depth of X-rays still amounts to a few ML.
Surface scattering of He atom beams exhibit high sensitivity to the surface, but precise calculations are more complicated and information on the subsurface interspacings is not accessible.
In order to gain information on the surface composition, element sensitive techniques such as electron or X-ray induced Auger electron spectroscopy (EAES or XAES) are applied.
Positron annihilation induced AES, however, exhibits intrinsically major advantages such as suppressed secondary electron background in the energy range of Auger-transitions,
topmost layer sensitivity and less damage of the probed surface or adsorbed molecules due to the only thermal energy of the annihilating positron trapped at the surface.
Thanks to the availability of high intensity low-energy positron beams time dependent PAES could be successfully applied to observe surface segregation in situ.
Moreover, surface studies using spin-polarized positrons have been demonstrated to be particularly suitable for the observation of spin phenomena at surfaces. 
Finally, tailored surfaces for the efficient creation and release of free positronium (Ps) and the positronium negative ion Ps$^-$  play a major role in fundamental research.
In particular, a high yield of cold Ps interacting with anti-protons is required to produce larger amounts of anti-hydrogen atoms.
Apart from a well-defined surface structure, a high positron density is mandatory to pave the way for the prospective creation of a Ps Bose-Einstein condensate. 

In this paper, I will review the current status of state-of-the art positron beam setups, physical basics for surface experiments with positrons  as well as past and present surface studies with particular emphasis on diffraction, Auger-electron, polarization and fundamental experiments using low-energy positrons.
Accordingly, this review is organized in three parts: In the first more methodical part (Section\,\ref{LowEBeam}), a concise review is given on the generation of low-energy positron beams and the experimental setups realized in laboratories as well as at large-scale facilities. 
The second part (Section\,\ref{Sec:Surf}) deals with the positron's fate in matter and at the surface. In particular, the energetics at the surface is explained giving rise to the high sensitivity of the positron as surface probing particle.
The third and most comprehensive part comprises four sections of positron beam studies including introductory explanations of the basic principles and the applied techniques.
First, fundamental experiments benefiting from well-defined surfaces are presented (Section\,\ref{fundamental}), followed by an overview of positron diffraction experiments reported so far (Section\,\ref{sec:Diffraction}). 
Then PAES studies for the determination of the elemental composition at the surface (Section\,\ref{PAES}), and experiments using spin-polarized positrons are presented (Section\,\ref{sec:Polarized}).

The paper is concluded by giving an outlook with regard to ongoing and future developments of novel techniques and planned experiments.
A table of most relevant positron emitting isotopes, an overview of positron beam systems at large-scale facilities, and  calculated core-annihilation probability compiled from literature can be found in the Appendix. 

\newpage
\section{Low-energy positron beams}
\label{LowEBeam} 
Within the last years various techniques of positron annihilation spectroscopy (PAS) using low-energy positron beams became powerful tools in solid state physics and materials science for depth dependent and spatially resolved defect studies. The same holds for positron experiments in fundamental research that have benefited from the development of mono-energetic positron beams. In the context of surface studies the availability of low-energy positrons is mandatory.

There are two methods how positrons can be generated:
For laboratory positron beams usually $\beta^+$ emitters such as $^{22}$Na are used as positron source. 
At large-scale facilities such as electron linacs or nuclear reactors, which provide high-energy $\gamma$ radiation with high intensity, positrons are created by pair production.
Independent of the source, the emitted positrons show a broad energy distribution which might reach up to several MeV. 
Therefore, the produced positrons have to be cooled (moderated) to become usable for the generation of a low-energy positron beam.
In experimental setups a positron moderator is mounted close to the positron source in order to obtain a beam with narrow band width without losing too much intensity. 
For moderation purposes metals with negative positron work function, e.g.\,W, or solid rare gases (mainly Ne) are applied. 
Thus, besides a strong positron source an efficient moderator as well as an optimized positron extraction geometry are crucial to provide a maximum flux of monoenergetic positrons.

In this section, different positron sources and the moderation of positrons are briefly described leading to the principle scheme and various designs of positron beam setups.
Before the physical details are outlined how positron beams are generated the main expressions in this context will be explained in the following.

\subsection{Fundamental concepts}
\label{sec:definitions}
First, a definition is given for \textit{slow} positrons.
It is noteworthy that the SLOPOS-series of conferences deals with \textit{slow} positron beam techniques and applications (see e.g.\,most recent proceedings \cite{Hug14slopos}). 
There are several reasons why positrons with a kinetic energy below about 30-40\,keV are called \textit{slow} positrons. 
From an experimental point of view, positron beams up to about 30\,keV are usually realizable without too much efforts in more or less table top experiments. 
The according velocity of positrons with a maximum energy of 30\,keV amounts to less than one third of the speed of light. Thus, with decreasing energy relativistic effects become less relevant facilitating the physical description of the processes under study.
Another more experimental aspect is related to depth dependent defect spectroscopy of materials.
The availability of energy variable positron beams up to several 10\,keV allows the distinction between surface contribution at lower positron implantation energy and pure bulk information, since the fraction of positrons diffusing back to the surface is negligible  for implantation energies of about 30\,keV even in well annealed single crystals.
It has to be mentioned that throughout the relevant literature the expressions \textit{slow}, \textit{low-energy}, \textit{monoenergetic}, \textit{moderated} or sometimes \textit{non-relativistic} positron beams are used synonymously.
However, it depends on the physical context if other distinctions or definitions might be more suitable.
This is the case, e.g. for low-energy positron diffraction (LEPD) and reflection high-energy positron diffraction (RHEPD) similar to their electron counterparts low-energy electron diffraction (LEED) and reflection high-energy electron diffraction (RHEED), where the relevant beam energies are in the order of 10\,eV and 10\,keV, respectively.

Since facilities for the production of dense positron pulses have been developed new experiments and future applications have been discussed in an own workshop dedicated to the physics with \textit{many} positrons \cite{Dup10}.
Physics with \textit{many} positrons might occur when two or more (low-energy) positrons interact with each other and/or with various forms of ordinary matter.
In PAS and in most related theoretical considerations the precondition is fulfilled  that during its lifetime only one positron resides and subsequently annihilates inside the sample.
Even with existing high intensity positron beams providing $10^9$  moderated positrons per second focused onto a region of 1\,mm$^2$, it would be most unlikely to obtain two positrons in a time interval of about 1\,ns within a range of 10\,nm, which is in the order of the de-Broglie wavelength of thermalized positrons $\lambda_{\mathrm{dB}}=\hbar\sqrt{2\pi/m_0kT}$.
Thus, for the vast majority of positron experiments positron-positron interaction has not to be considered.\footnote{Even at the high-intensity positron source NEPOMUC, where the positron density in the converter foils is in the order of $\mathrm{10^{12} e^+/mm^2s}$, positron-positron interaction has not to be considered. }
Consequently, only in the context of experiments with \textit{many} positrons \cite{Dup10, Mil10}  one has to consider the interaction of at least two positrons.
In fundamental research (see Section\,\ref{fundamental}), large efforts have been undertaken to observe the positron-positron interaction e.g. for the formation of the positronium (Ps) molecule \cite{Cas07}.

Another question is: What is \textit{high} intensity?
A practical number for the characterization of  beams with \textit{high} intensity is $10^7$ moderated positrons per second \cite{Hug10a} to distinguish between typical  laboratory beam setups and positron beams using strong sources at large scale facilities.
The physical basis for this distinguishing mark is the maximally achievable yield of moderated positrons in tabletop beams  which is limited by the positron self-absorption in the $\beta^+$ source.

The quality of a particle beam is characterized by its \textit{brightness}, which describes the phase space density of a representative  ensemble of beam particles per time intervall.
According to Liouville's theorem, the product of the divergence $\theta=\sqrt{E_T/E_L}$ with transverse and longitudinal components of the positron energy $E_T$ and $E_L$, the beam diameter $d_+$, and the longitudinal component of the momentum $\sqrt{2mE_L}$ is constant.
For this reason, the definition of the brightness $B=\frac{I}{\theta^2 d_+^{2} E_L}$ with particle intensity (particles per second) $I$ is commonly used in the positron community (see e.g.\,\cite{Coleman2000,Mil80c,Sch88,Hug12a}).
The enhancement of the beam brightness by remoderation plays an outstanding role for the generation of (intense) beams achieving a beam spot in the micrometer range.

Positron beams with high intensity and in particular with high brightness are  desired, not only for the realization of remoderated microbeams, but also 
for the generation of positron pulses with a high repetition rate for depth dependent positron annihilation lifetime spectroscopy (PALS)\footnote{Acronyms and their meaning are listed on page\,\pageref{Acronyms}.}.
In general, most experiments profit from a high beam intensity leading to a drastically reduced measurement time and to a considerably enhanced signal-to-background ratio.
Moreover, a high positron intensity is beneficial for all kinds of coincidence techniques applied in solid state physics: ACAR measurements for the investigation of the electronic structure, CDBS for element sensitive defect spectroscopy and defect imaging in the near-surface region, and age momentum correlation (AMOC) for the study of positron and Ps states in matter.

For surface studies the maximum available positron intensity is crucial for e.g. positron annihilation induced Auger electron spectroscopy (PAES), which is applied for element selective surface studies, and for positron diffraction techniques such as LEPD and RHEPD.
In the light of the availability of high intensity positron beams with high brightness, the development of novel techniques for fundamental research and new positron applications in surface science has become possible such as \textit{time-dependent} PAES, e.g. for the observation of segregation processes, or \textit{depth-dependent} ACAR for the investigation of the electronic structure in thin layers.

\boldmath
\subsection{Positron sources}
\unboldmath 
\subsubsection{$\beta^+$ emitters}
\label{sec:BetaEmitters}
The choice of the appropriate positron source is determined by various boundary conditions such as half-life, end-point energy  and maximum available or practicable activity of the $\beta^+$ emitter. 
There is a large variety of $\beta^+$ emitting radio nuclides with a half-life of many years down to fractions of a second and  typical end-point energies $E_{\mathrm{max}}$ of the according $\beta^+$ spectra in the range between several 100\,keV and  a few MeV.
Besides nuclides showing pure positron emission, the positron yield, i.e. the branching ratio $f_{\mathrm{e^+}}$ can be much below  $100\%$ due to competing electron capture or $\beta^-$ decay as in the case of  $^{64}$Cu, which can easily bred at reactors.

Due to parity violation in the weak interaction the $\beta^+$ particles are intrinsically right-handed longitudinally spin polarized.
The helicity of the emitted positrons is given by the velocity v and the speed of light c: $H=v/c$. 
Using the notation in the review by J. Major \cite{Major2000} the helicity averaged for an ensemble of positrons can be written as 

	\begin{equation}
		H=\left\langle \frac{\textbf{s}\cdot \textbf{p}}{s\cdot p}\right\rangle 
	\end{equation}

with spin vector \textbf{s} and positron momentum \textbf{p}.
The according spin polarization component along the unit vector \textbf{e} is hence

\begin{equation}
 P=\left\langle \frac{\textbf{s}\cdot \textbf{e}}{s}\right\rangle \quad. 
\end{equation}

Using the most likely positron energy, which is almost the average energy $E_{\mathrm{av}}$ of the $\beta^+$ spectrum \cite{Major2000} one can calculate the helicity using

\begin{equation}
	H\equiv v_{av}/c=\sqrt{ 1- (1+{E_{\mathrm{av}}/{mc^2}})^{-2} } \quad. 
\end{equation}

Hence the higher the endpoint energy the higher the net-polarization of the positron beam making $^{68}$Ge a good candidate for polarization dependent positron experiments.
An overview of $\beta^+$ emitters, which are relevant for positron experiments, summarizing the characteristic energies $E_{\mathrm{max}}$, and $E_{\mathrm{av}}$,  helicity $v_{av}/c$ and, if present, the most dominant $\gamma$ transition is given in the Appendix.

Within the context of slow positron beams two nuclides $^{22}$Na and $^{68}$Ga should be highlighted.
The most prominent positron source is $^{22}$Na, which has several advantages compared to all other radio nuclides such as very high positron yield of $90\%$ and a long half-life of 2.6\,years.
In  positron experiments, which use the $\beta^+$'s directly, the emission of a prompt high-energy $\gamma$-quantum  is particularly beneficial to be used as start signal  for PALS.
$^{22}$Na is produced at cyclotrons or at accelerators by particle reactions of a Al or Mg target with suitable projectiles such as protons, deuterons or $\alpha$ particles.
The maximum  source activity is limited by the irradiation time and target cooling or by the available ion current of the accelerator.
Small  sources can be prepared easily by drying $^{22}$NaCl or $^{22}$Na$_2$CO$_3$ from its solution onto thin foils. 
 The activated material of strong sources upto 2\,GBq is usually encapsulated behind a thin Ti window allowing the emission of the $\beta^+$'s.

The nuclide  $^{68}$Ge, which decays to the positron emitter $^{68}$Ga, can be produced by irradiating a GaN target with 20\,MeV protons. 
The lifetime of the so-called generator system $^{68}$Ge/$^{68}$Ga is dominated by the half-life of the mother nucleus of 271\,days.
Due to the high energy of the $\beta^+$'s ($E_{\mathrm{max}}$= 1899\,keV, $v/c$=0.925 ) $^{68}$Ga is particularly suited for spin dependent measurements (see Section\,\ref{sec:PolarizedBeams}).  
 
\subsubsection{Pair production}
\label{sec:PairProduction}
In the last decades great efforts have been made to generate positrons by pair production using intense $\gamma$ sources at large scale facilities in order to develop low-energy positron beams with high intensity.
The present intense positron sources use either bremsstrahlung emitted from decelerating electrons or $\gamma$ radiation released from nuclear processes.
The threshold energy for positron-electron pair production in the electrical field of the nucleus corresponds to twice the electron rest mass plus an additional (usually negligible) amount of energy transferred to the nucleus which takes the recoil momentum. 
In order to produce positron-electron pairs efficiently, the $\gamma$ energy must be high enough due to the increasing pair production cross section $ \sigma_{pp} $ with larger $\gamma$ energy ($ \sigma_{pp}\propto \ln E$).
On the other hand, a too hard $\gamma$ spectrum would lead to a lower yield of slow positrons due to a lower moderation efficiency of positrons originating from high energetic $\gamma $'s.
As converter high density materials with high nuclear charge $Z$ such as Ta, W, and Pt are preferred in order to achieve high production rates since in good approximation $\sigma_{pp}\propto Z^2$.

\paragraph{Linac based positron sources}
\label{sec:LinacBasedPositronSource}

\begin{figure}[htb]
\centering
\includegraphics[width=0.5\textwidth]{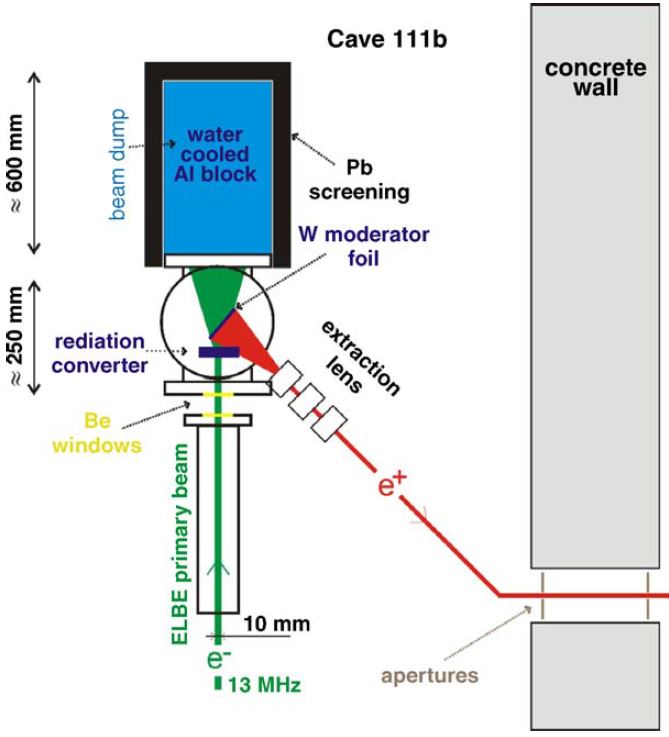} 
\caption{Generation of positrons at the electron linac ELBE (figure from \cite{Kra06}).}
  \label{linac}
\end{figure}

At electron linear accelerators (linacs), high energy brems\-strahlung is released by decelerating electrons in a beam dump. 
The heavy targets out of  Ta or W, which are applied for the conversion of the $\gamma$ radiation into positron-electron pairs, have to be cooled efficiently to dissipate the heat input.
Since a linac is intrinsically pulsed the produced positron beam shows a pulsed structure too.
At present, there are several slow positron beams in operation using brems\-strahlung from relativistic electron beams in the energy range from 10\,MeV up to a few GeV.
Figure\,\ref{linac} shows a scheme for the generation a slow positron beam at the electron linac ELBE, Germany. 
A summary of the present status of linac based positron beams and positron sources installed at research reactors is given in Table\,\ref{tabSources} in the Appendix.

\paragraph{Reactor based positron sources}
\label{reactor}
At nuclear reactors, $\gamma$ radiation is released either from nuclear fission processes or from the de-excitation of excited nuclear states after neutron absorption.
Positrons can hence be generated by the absorption of high energy $\gamma$ radiation.
At the research reactor in  Delft, an assembly of thin W tubes inside a beamtube is located close to the reactor core in order to absorb the $\gamma$ radiation from nuclear fission to generate positrons by pair production \cite{Vee01}.

\begin{figure}[htb]
\centering
\includegraphics[width=0.7\textwidth]{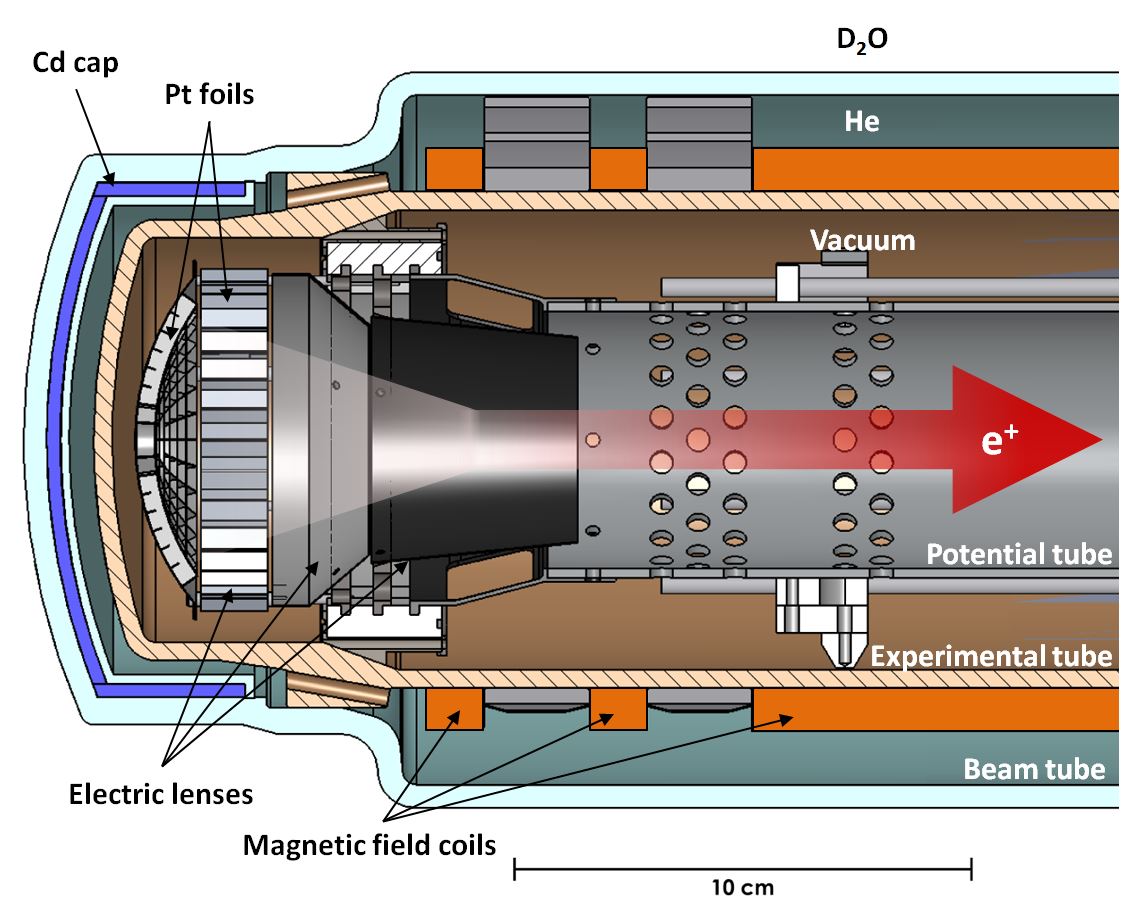} 
\caption{Cross sectional view of the neutron induced positron source NEPOMUC.
}
  \label{quellkopf}
\end{figure}

The other approach was pursued at the NEutron induced POsitron source MUniCh (NEPOMUC) where the nuclear reaction $ ^{113}$Cd(n,$\gamma)^{114}$Cd is used in order to benefit from high-energy prompt $\gamma$ rays released after thermal neutron capture \cite{Hug02a}. 
A cross sectional view of the positron source NEPOMUC is shown in Figure\,\ref{quellkopf}  (see also \cite{Hug13b}).
Due to the huge cross section for thermal neutron capture ($\sigma_{cap}(^{113}Cd)$=\,20600\,barn) a Cd cap (enriched with 80\% of $^{113}$Cd) inside the tip of a beam tube absorbs thermal neutrons very efficiently.
A structure of Pt foils is used for both the conversion of the released high-energy $\gamma$ radiation into positron-electron pairs and positron moderation (so-called \textit{self-moderation} process) \cite{Hug02d}.
The beam energy of 1\,keV is defined by the voltage applied to the Pt moderator foils. 
The positron beam is magnetically transported to the experiments connected to the positron beam facility \cite{Hug05, Hug07b}. 
At present,  NEPOMUC provides the world highest intensity of $10^9$ moderated positrons per  second  ~\cite{Hug08b, Hug13b}.
However, the brightness of the primary 1\,keV positron beam is enhanced by a positron remoderator which is operated with a W single crystal ($\Phi^+$=-3.0\,eV \cite{Coleman2000}) in back reflection geometry. 
The energy of the remoderated beam can be adjusted between 20 and 200\,eV and is presently set to 20\,eV for most experiments. 
The total efficiency of the setup is about 5\% and the beam diameter of the remoderated beam is less than 2\,mm (FWHM) in a  6\,mT guiding field \cite{Pio08}.
		
Similar to the Delft design or based on the principle of the NEPOMUC source, further projects have been initiated at research reactors. 		
Using a large positron emitting area of 900\,cm$^2$  it was demonstrated that about 5$\cdot10^8$ positrons per second can be produced at the PULSTAR reactor, USA \cite{Hat07, Haw11}.
At the Kyoto University Research Reactor a positron source was put into operation at reduced reactor power for the first time, and is currently upgraded for full power operation \cite{Sat15}.
Another reactor based positron source is currently under construction in Hamilton, Canada \cite{Mas15}. 
The operated and planned positron sources at reactors are listed in Table\,\ref{tabSources} as well.

\paragraph{Future bright $\gamma$ sources for positron production}
\label{sec:FutureGamma}

 Although nowadays positron beams reach intensities in the order of $10^9$ moderated positrons per second,  one may scrutinize whether even higher intensities could be achieved using alternative concepts for positron generation.
At present positron sources based on pair production by absorption of high energetic $\gamma$'s most of the radiation releases its energy by photo effect and Compton scattering due to their high cross sections.
Therefore, most of the radiation generates heat (so-called $\gamma$-heating) without contributing to the positron production.
In addition, at research reactors the almost isotropically emitted $\gamma$ radiation leads to the heating of the structure material close to the actual converter.
For this reason, the major technical limitation is the deposited energy in the conversion target which has to be dissipated by sophisticated cooling devices.

An alternative approach would be the positron production using a high-energy $\gamma $ \textit{beam} created by inverse Compton scattering of photons from an intense laser with a relativistic electron beam.
Due to the high brightness of such a $ \gamma $ source the $ \gamma $ radiation can be well collimated onto a small area of interaction  at the converter.
In addition, a circularly polarized laser beam interacting with a GeV electron beam would produce positrons showing a high degree of polarization \cite{Omo06}.
In principle, the $ \gamma $ energy can be chosen in the MeV range, and due to the relatively narrow band width of several 100\,keV no $ \gamma $'s hit the converter target with an energy below  the pair production threshold.
Therefore, the heat load due to $\gamma $'s not usable for pair production would intrinsically be avoided. 
After the presentation of this novel concept for such a positron source based on inverse Compton scattering \cite{Hug12a} a project for its realization was initiated recently \cite{Djo16}.

\subsection{Moderation}
\label{sec:Moderation}
\begin{figure}[htb]
\centering
\includegraphics[width=0.7\textwidth]{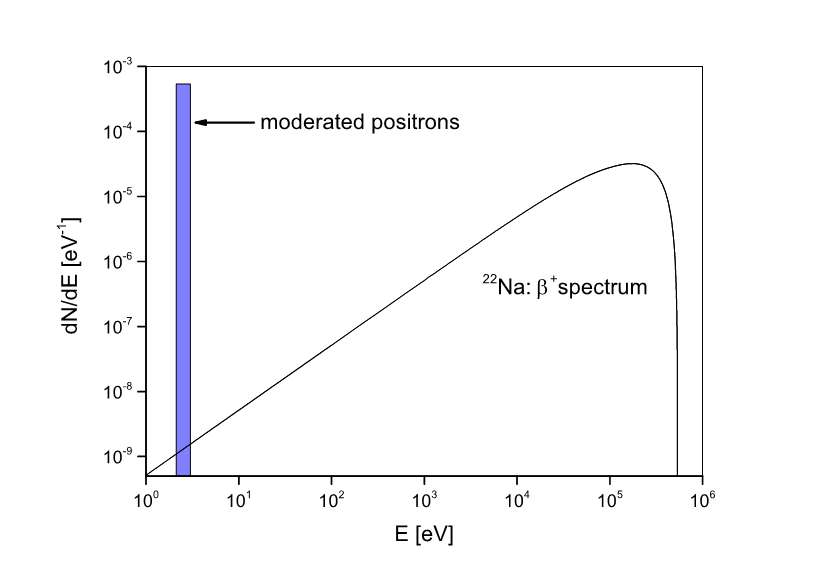} 
\caption{$\beta^+$  spectrum of $^{22}$Na and positron intensity after moderation.}
  \label{modSpec}
\end{figure}

For the generation of slow positron beams the fast positrons either from $\beta^+$ emitters or generated by pair production have to be moderated.
For this purpose,  solid state moderators using  metals with a negative positron work function $\Phi^+$ (values for $\Phi^+$are compiled e.g.\,in \cite{Coleman2000}) or solid rare gases are applied.
After implantation, a fraction of thermalized positrons diffuses to the surface where they can leave the solid perpendicular to the surface with a discrete energy $E_{0}$ corresponding to the modulus of the negative positron work function of the moderator material.
Since $E_0$ amounts to a few eV (e.g.\ $\Phi^+_{\mathrm{W}}=-3.0$\,eV \cite{Coleman2000}, $\Phi^+_{\mathrm{Pt}}=-1.95$\,eV \cite{Hug02d}) and the energy smearing is only in the order of thermal energies \cite{Gul85, Fis86} the moderation process leads to a higher phase space density if the intensity loss is not too high.   
Although the efficiency for primary moderation is typically only in the range of $10^{-4}$ to $5\cdot10^{-3} $  the obtained brightness is several orders of magnitude higher compared to conventional energy filters.
The according gain in intensity in a narrow energy window in the eV-range after positron moderation is shown in Figure\,\ref{modSpec}.
The band width of the beam, i.e.\ the positron energy distribution, is basically given by the thermal spectrum of the moderated positrons \cite{Fis86}. 
However, the spectrum is blurred by several effects such as inelastically scattered thermalized positrons, emission of epithermal positrons, local changes of the surface dipole due to surface adsorbates and the surface roughness. 
Dependent on the beam setup the band width can be additionally broadened due to different potentials applied to the moderator foils for positron extraction.
In order to minimize the positron loss due to trapping in defects such as vacancies or dislocations the moderator material has to be annealed typically at temperatures of about 80\,$\%$ of the melting temperature \cite{Coleman2000}.	
In addition, the surface should be as flat as possible and free of contamination.

Solid rare gases are applied for moderation as well where one benefits from the emission of \textit{epithermal} positrons in the eV range. 
Such a moderator can be frozen virtually free of defects and contaminations directly onto the positron source but a more elaborate setup is required including a cryostat.
The energy distribution of such a positron beam is not as narrow as in the case of metallic moderators. On the other hand its moderation efficiency is usually higher but still below 1\,\% \cite{Mil88}.
There is a large variety of moderator designs where positrons are moderated in reflexion or in transmission geometry or in a combination of both (for details see e.g.\,\cite{Coleman2000, Sch88}).
In most positron beam setups thin polycrystalline or oriented  W(100) foils (thickness $~$0.1-1\,$\mathrm{\tcmu}$m) or W meshs are used as transmission moderators (see e.g.\ \cite{Lyn85, Sch88, Str01,Sai02}). 

Repeated moderation of a positron beam, i.e.\ \textit{remoderation} \cite{Mil80c}, can be applied for further increase of the phase space density  in order to achieve a still  higher beam brightness. 
For example, at NEPOMUC a W(100) single crystal in back reflection geometry is used for remoderation and beam switches enable to toggle quickly between the primary and the remoderated positron beam \cite{Pio14a}.
Positron remoderation is of particular interest for the generation of micro beams (see Section \ref{sec:Microbeams}). 

\subsection{Positron beam setups}
In the last decades, a large variety of  low-energy positron beams  has been built either using $\beta^+$ sources or the pair production process at large scale facilities  
In the following, various lay-outs of positron beams taylored to the requirements of specific applications, will be presented.  
Based on the basic principle of a conventional tabletop positron beam (see Figure\,\ref{beamsetup}) the different setups can be categorized into micro-beams, pulsed, trap-based, and polarized positron beams. 

\subsubsection{Tabletop positron beams}
\label{sec:TabletopPositronBeams}
\begin{figure}[htb]
\centering
\includegraphics[width=0.7\textwidth]{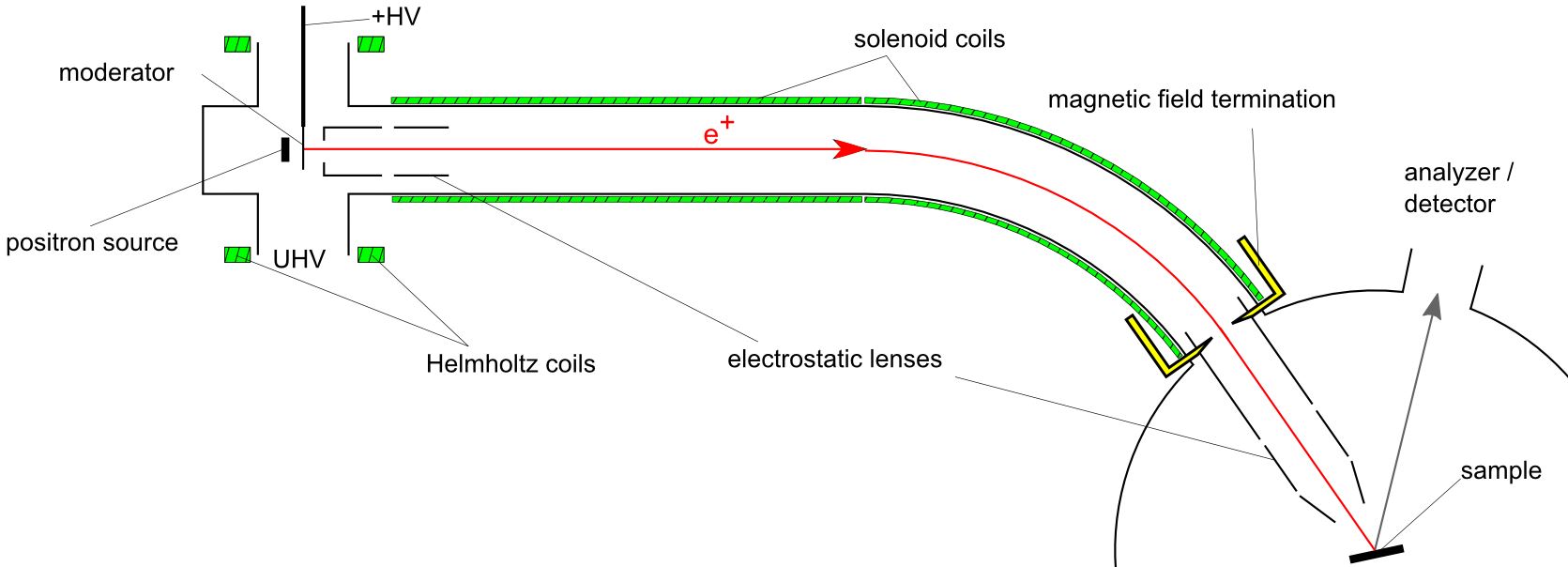} 
\caption{Scheme of a conventional tabletop positron beam setup.}
  \label{beamsetup}
\end{figure}

A conventional positron beam apparatus comprises the following main components: positron source with moderator, electrostatic extraction lenses and acceleration section, magnetic field coils for the (adiabatic) magnetic beam guidance, and a shielded bend (or a E$\times$B filter) in order to avoid fast non-moderated positrons hitting the sample. 
Similar to the positron beam presented in \cite{Str03} the basic scheme of a slow positron beam setup is sketched in Figure\,\ref{beamsetup}. 
Dependent on the application electrostatic beam transport (with or without additional magnetic fields) is also realized in particular for positron acceleration and focusing. 
The kinetic energy of the beam can be easily varied by adjusting the  potential  of the moderator or the sample.

Most tabletop setups provide beam intensities in the range of $10^4-10^6$ moderated positrons per second within a beam spot of a few mm.
Even using a very strong positron emitter in an optimized source geometry,\footnote{For example a 2\,GBq source of $^{22}$Na with a positron fraction of 0.9, back reflexion probability $<$0.4, low self absorption of 0.2, and a solid angle with respect to the moderator of $<2\pi$.} and assuming a maximum moderation efficiency of 1\% one would barely achieve a maximum intensity of $10^7$ moderated positrons per second with laboratory beams.

\subsubsection{Micro-beams}
\label{sec:Microbeams}

In order to emphasize the challenge to create a positron micro-beam providing a spot size in the micrometer  range a comparison with electron microscopy is given.
Compared to electron sources, which  are available off-the-shelf and reach current densities of more than 10$^{18}$ e$^-/$mm$^2$s, the highest available intensity of  low-energy positrons is still more than five orders of magnitude lower.  
However, more important is the achieved brightness of commercial LaB$_6$ cathodes amounting to B=4.6$\cdot10^{18}\,$s$^{-1}$mm$^{-2}$eV$^{-1}$ which is hence intrinsically ten (!) orders of magnitude higher than at high-intensity positron beams\footnote{Using the parameters reported in \cite{Hug14b} and assuming reasonable values for the transverse energy of $\approx$10\,eV and 0.1\,eV for the primary and remoderated beam at NEPOMUC, respectively, the brightness of the primary beam would be $\approx5\cdot10^{5}$\,s$^{-1}$mm$^{-2}$eV$^{-1}$. After remoderation the brightness is enhanced by almost three orders of magnitude to  $B>3\cdot10^{8}$\,s$^{-1}$mm$^{-2}$eV$^{-1}$.}.
For this reason, much more sophisticated techniques have to be applied in order to realize positron micro-beams with a lateral resolution below 1$\mathrm{\tcmu}$m and still reasonable intensity.
The minimum positron beam focus for defect spectroscopy has not to be as tiny as in electron microscopy, since  the positron diffusion length in matter, which is usually well above 10\,nm, defines the smallest volume, from which information is obtained. 

\begin{figure}[htb]
\centering
\includegraphics[width=0.7\textwidth]{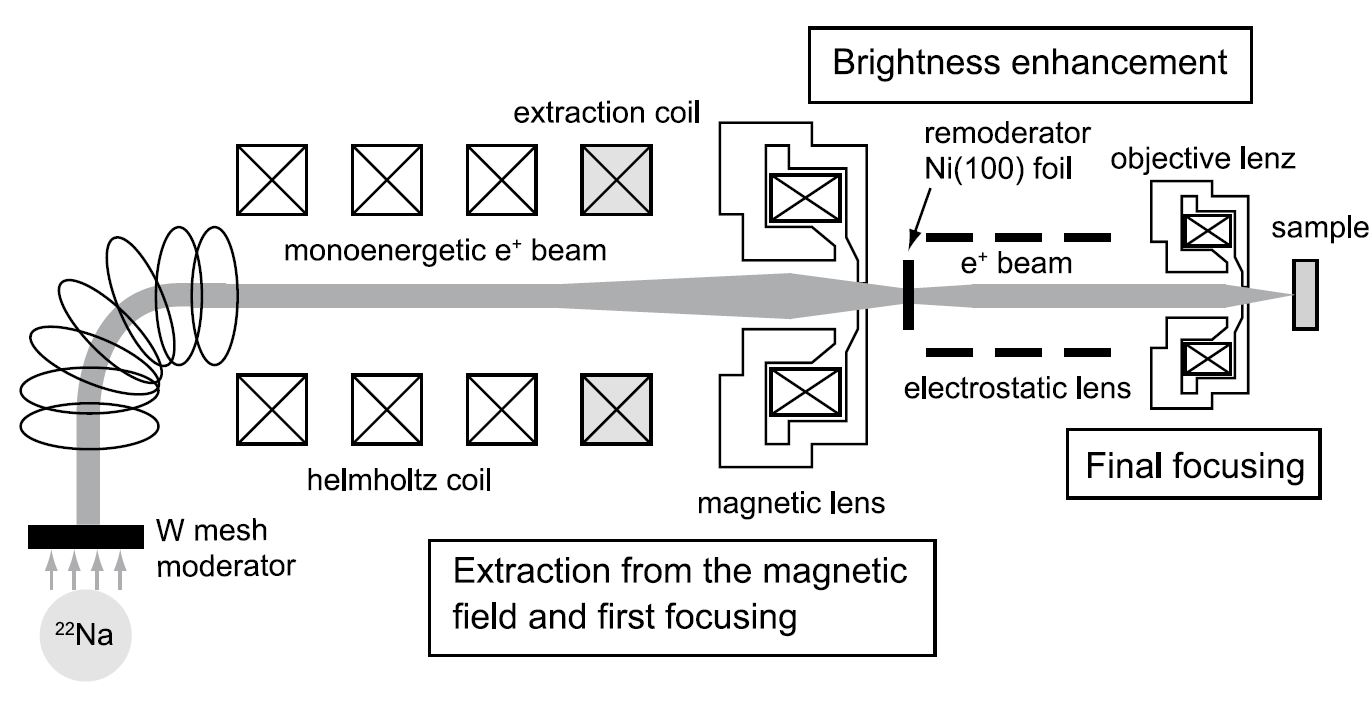} 
\caption{Scheme of the positron micro-beam system using a transmission type remoderator at Chiba university (figure from \cite{Oka09}).}
  \label{microbeam_oka09}
\end{figure}

There are basically two different approaches to build positron micro-beams. 
The first and more obvious concept is based on a positron source-moderator geometry with a tiny lateral extension and an optical device for imaging the source spot onto the sample \cite{Gre97}.
The other design benefits from the increased phase space density after remoderation. 
A	two-stage moderated scanning micro-beam using W(110) (re-)moderators was presented in the late 1980's \cite{Bra88, Bra88b, Vas95}.
A so-called scanning positron microscope (SPM) was developed to provide a pulsed positron beam for lifetime experiments with a spatial resolution below 10\,$\mu$m \cite{Koe97}. 
In order to reduce the measurement time considerably an interface with reflection remoderator for the connection of the SPM to the NEPOMUC beam line was set up, and a pulsed beam of threefold moderated positrons was successfully generated \cite{Pio07}.
 Fujinami et al.\  presented an operating positron micro-beam using a thin Ni(100) transmission remoderator allowing measurements with a spatial resolution of 80-90$\mu$m \cite{Fuj08, Osh08}.
Exemplary, the basic layout of the positron micro-beam system at Chiba university is shown in Figure\,\ref{microbeam_oka09} \cite{Oka09}. 

\subsubsection{Pulsed beams}
\label{sec:PulsedBeams}
For the measurement of the positron lifetime the Stop signal is naturally provided by the detection of the annihilation radiation. 
For PALS without depth information basically three techniques can be applied to obtain a Start signal: (i) In conventional PALS using $^{22}$Na as positron source the Start signal is provided by a prompt $\gamma$ quantum. (ii) By using relativistic positron beams, the Start signal is generated by positrons passing through a thin scintillation detector before implantation into the specimen \cite{Bau90}. (iii) It was demonstrated that  bremsstrahlung can be used to produce positrons directly in the sample \cite{Sel02}. Hence the master clock of the linac, which produces intense bursts of $\gamma $ radiation, can serve to trigger the time measurement \cite{Kra09}.

Using low-energy positrons for the investigation of the near-surface region of a sample, it is challenging to generate a Start signal to enable PALS with high time resolution in the order of 200-300\,ps\,\footnote{For the measurement of the Ps lifetime in matter the requirements are not as high, since the o-Ps lifetime, which is usually governed by the so-called pick-off process, is typically greater than 1\,ns.}.
In principle, secondary electrons generated after positron impact can be used as trigger \cite{Szp02}. However, the secondary electron yield strongly depends on material and energy, and  a time resolution $<$350\,ps is difficult to achieve.
For this reason, great efforts have been made to develop low-energy pulsed positron beams, which basically comprise a (two-stage) bunching system and a chopper for background elimination (see e.g.\,\cite{Spe97}). 
At present, three pulsed slow positron beam facilities are in operation.
Suzuki et al. demonstrated how to obtain positron lifetime spectra in a time range of 45\,ns and a time resolution of 250\,ps with about 100 counts per second (cps) using a pulsed positron beam at a linac \cite{Suz91}. 
Another linac based system is in operation at the ELBE facility yielding a peak-to-background ratio of 10$^4$ and a time resolution of 500\,ps (FWHM) \cite{Jun13}.
At NEPOMUC the pulsed monoenergetic positron beam system  provides positron pulses with a repetition rate of typically 50\,MHz corresponding to a time window of 20\,ns and a total time resolution of the system of 240\,ps \cite{Spe97, Spe08}. 
Typically, lifetime spectra can be recorded with a count rate of $>10^4$\,cps and a variable  positron implantation energy between 0.5 and 22\,keV.
	
\subsubsection{Trap-based beams}
\label{sec:TrapBasedBeams}
Trap-based positron beams have been developed to deliver tailored intense positron pulses \cite{Sur89}.
In such devices, positrons, delivered e.g. from a moderated $^{22}$Na-based positron beam, loose their energy by inelastic scattering in a buffer gas trap.
It was also demonstrated that gas cooling can be used to remoderate positrons, and to extract them continuously  from the buffer gas device \cite{Loe08}. 
The basic scheme of a so-called Penning-Malmberg trap for the accumulation of positrons comprises of a uniform longitudinal magnetic field to restrict the motion across the field and electrostatic potential wells to prevent positrons from escaping along the cylinder axis.
After a collection time in the order of seconds or less, a pulse of positrons is released allowing experiments in a quasi-pulsed mode  with higher peak flux. 
Typically, a pulse  contains about 10$^6$ positrons and its time width is restricted to $>$15\,ns  \cite{Cas06b}. 
An extremely narrow energy width of 18\,meV can be achieved by raising the potential well slowly for releasing the positrons \cite{Gre02}. 
The relevant theory of single-component plasmas and the specific applications of trap-based positron beams in particular for atomic and molecular physics have been reviewed recently by Danielson et al. \cite{Dan15}.
For fundamental studies the pulse length and repetition rate  can be synchronized with other parameters of the experiment, e.g.\ the characteristic time constant of collecting cooled anti-protons for the production of anti-hydrogen \cite{vdW02}.
Intense positron bursts can be obtained by storing almost $10^8$  positrons in an additional accumulator for a collection time of 400 seconds.
By application of an additional buncher up to $7\cdot10^7$ positrons have been compressed into an 1\,ns pulse \cite{Cas06a}.
In order to achieve a higher area density at the sample a synchronized pulsed high magnetic field coil is operated for the creation of the Ps$_2$ molecule \cite{Cas07} (see Section\,\ref{sec:PositroniumMolecule}).

\subsubsection{Spin-polarized beams}
\label{sec:PolarizedBeams}
Monoenergetic positron beams based on $\beta^+$ emitters are ideally suited for non-destructive polarization sensitive surface studies as presented in Section\,\ref{sec:Polarized}.
Most important is the fact that despite the high number of inelastic scattering events during positron moderation the depolarisation is negligible, and hence the resulting positron beam retains its axial polarization to a high degree as demonstrated by Zitzewitz et al. \cite{Zit79}.

For the realization of  an experimental setup several issues have to be addressed. 
As discussed in Section\,\ref{sec:BetaEmitters} $\beta^+$ emitters provide  intrinsically right-handed longitudinal spin-polarized positrons.
Often the positron emitter is deposited on a material with high atomic number Z in order to increase the number of usable positrons by back scattering.
Since back scattered positrons, however, reduce the net polarization a low-Z material should be used as substrate. 
For example, the positron backscattering coefficient of Be (Z=4) amounts to 10\% in contrast to Ta (Z=73) with 50\% \cite{Col08}.
Taking into account the opening angle $2\alpha$ of the emitted positrons with respect to the irradiated area of the moderator the usable polarization $P$ is 
	\[
	P=\frac{v}{2c}\cdot(1+cos\alpha) \quad.
\]
Consequently, for a polarization dependent experiment a nuclide with high average energy such as $^{68}$Ga should be chosen. 
Given e.g. $\alpha=20^{\circ}$ the positron polarization would amount to $P=90\%$ whereas with $^{22}$Na one would only obtain $P=68\%$.
For spin-dependent measurements, however, the quality factor $P^2\cdot I$, which also accounts for the intensity $I$ at the sample, is usually maximized \cite{Van84} leading to an optimum opening angle of around $\alpha=60-80^\circ$.
Since already weak magnetic fields lead to a depolarization of the beam an electrostatic guiding system has to be used in order to maintain the spin polarization during the beam transport.
Two types of beam deflectors bending the beam by 90$^\circ$ can be applied to either obtain a longitudinally polarized (magnetic deflector) or a transversely polarized beam (electrostatic deflector). 

\begin{figure}[htb]
\centering
\includegraphics[width=0.7\textwidth]{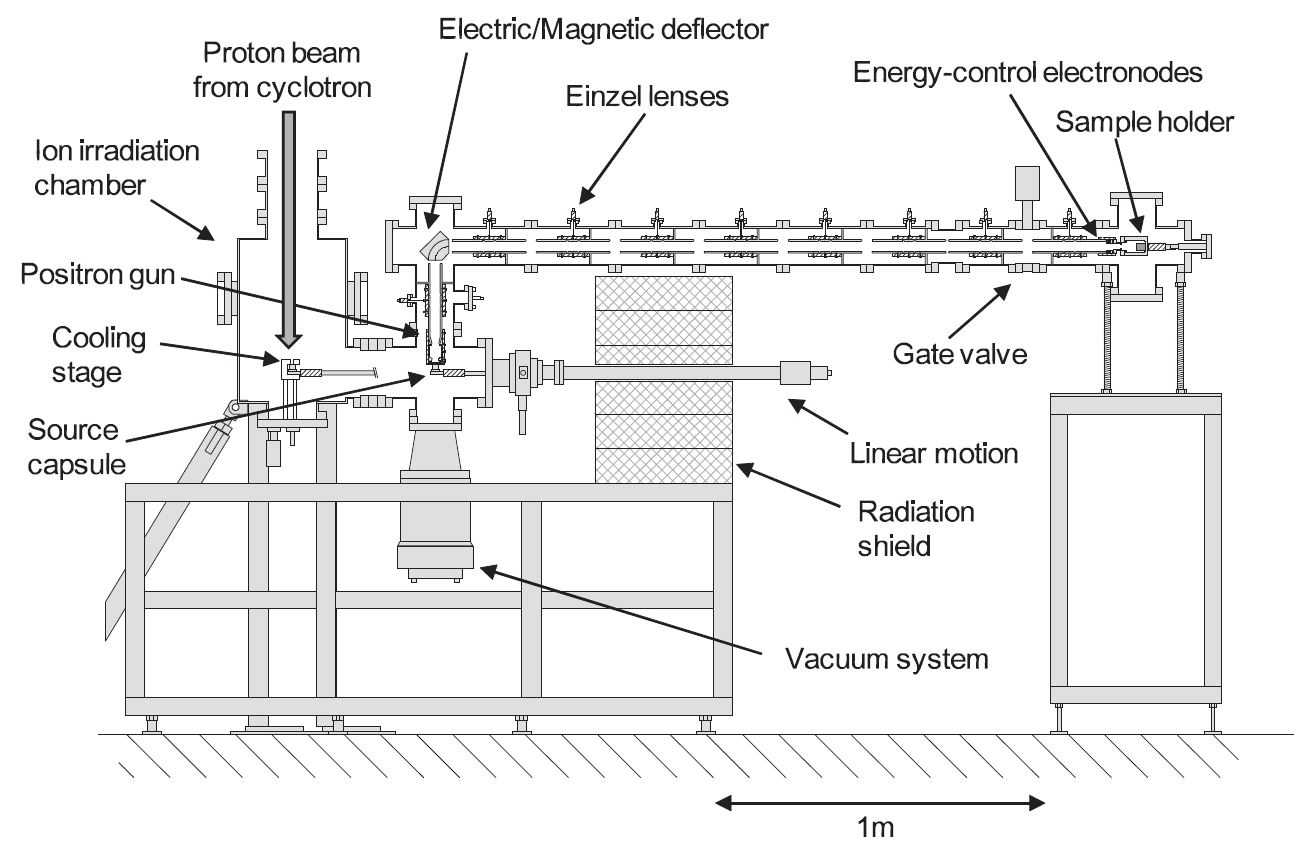} 
\caption{Scheme of the spin-polarized positron beam line at JAEA: The application of either a magnetic or an electrostatic deflector allows to toggle between longitudinal and transverse polarization (figure from \cite{Mae13}).}
  \label{PolBeam_Mae13}
\end{figure}

So far, there have been two research groups working with polarized slow positron beams. 
In the late 1970's the pioneering work of the Michigan group lead to the first polarized positron beam system.
The beam provided a positron intensity of $1.4\cdot10^4$ positrons per second with an energy between 0.3 and 1.5\,keV and a polarization of P=0.50(3) \cite{Zit79, Gid82}.
More recently, at JAEA Kawasuso et al. developed a polarized slow positron beam in close vicinity to a cyclotron used for the production of $^{68}$Ga via the $^{69}$Ga(p,2n)$^{68}$Ge nuclear reaction by irradiating a GaN substrate with 20\,MeV protons (see Figure\,\ref{PolBeam_Mae13}) \cite{Kaw13, Mae13}.
The beam is transported with an energy of 5ֱ5\,keV, and the final beam energy can be adjusted using a deceleration/acceleration tube inside the analysis chamber.
Typically an intensity of $5\cdot10^3$ positrons per second with a polarization of P=0.47(8) is achieved.

In the future, a positron source using a $\gamma $ beam is an attractive alternative to produce a spin-polarized positron beam with high brightness \cite{Hug12a}. 
In the first setup of the so-called ELI-facility, which is under construction in Bucarest, the degree of spin polarization of the positron beam is expected to be in the range of 33-45$\%$  \cite{Djo16}.

\newpage
\section{Positrons at the surface}
\label{Sec:Surf} 

\subsection{Positrons in matter}
\begin{figure}[htb]
\centering
\includegraphics[width=0.7\textwidth]{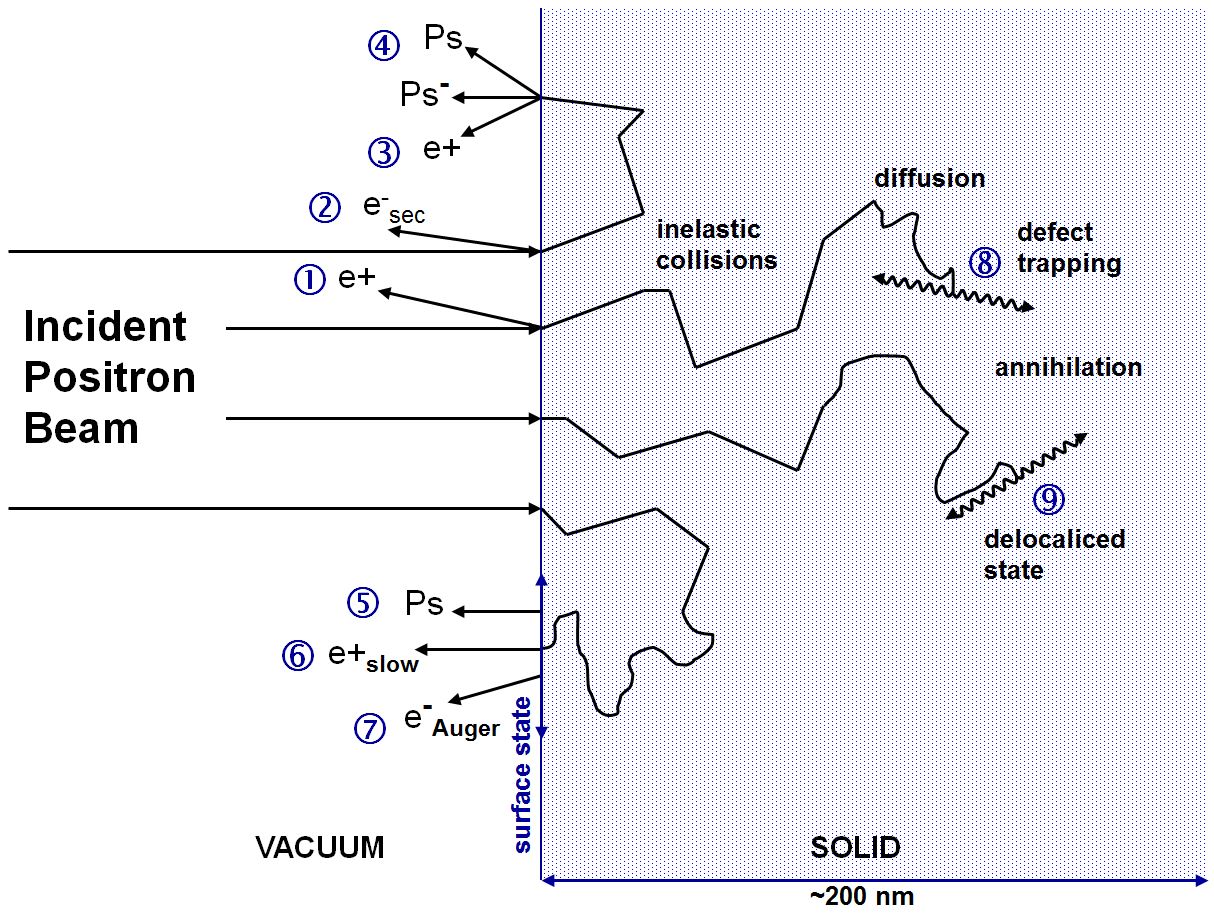} 
\caption{Positron interaction with matter:
At the surface positrons can be (1) reflected and diffracted, and (2) secondary electrons are emitted. Non-thermalized positrons can be emitted as (3) epithermal or as (4) neutral or charged positronium. After themalization and back diffusion to the surface the emission of (5) thermally desorbed Ps and of (6) moderated positrons (for materials with negative positron work function) can occur. Annihilation from a surface state can lead to the emission of (7) Auger electrons and characteristic X-rays (not shown). In the bulk, positrons annihilate either (8) after being  trapped in a defect or (9) from the delocaliced state in the lattice. } 
  \label{e+matter}
\end{figure}

After implantation into matter, the positron thermalizes within picoseconds and diffuses over hundreds of lattice spacings  until it annihilates either as a delocalized positron from the Bloch-state or from a localized state after being trapped in a crystal defect \cite{Sch88}.
In order to get an overview of the main processes of positron-matter interaction the fate of the positron in matter is visualized in Figure\,\ref{e+matter}. 
At the first encounter with the sample a fraction of the positrons is reflected.  
In the energy range of about E$_+$=2-30\,keV the reflexion probability at the surface is in the order of 7\% and only slightly energy dependent for light materials such as C.
For high-Z materials such as Au the reflexion probability increases from about 22\% at 2\,keV to $\approx$38\% for E$_+>$20\,keV \cite{Mak92}. 
Due to its importance for surface structure analysis the details of positron diffraction at surfaces are outlined in Section\,\ref{sec:Diffraction}.
		
Inside the sample the positron rapidly looses its energy by Bremsstrahlung ($\sim$MeV range), ionization and excitation of electrons (mainly below $\sim$100\,keV), plasmon excitation (in metals) and positron-phonon scattering (down to thermal energies) \cite{Sch88}.
After attaining thermal equilibrium the positron is assumed to be in the ground state of the periodic potential of the crystal lattice.
Since in the vast majority of cases  the assumption holds that during its lifetime only one positron is in the sample the Pauli exclusion principle has not to be taken into account.
At room temperature the annihilation of non-thermal positrons is usually negligible since in metals the whole thermalization process is in the order of a few picoseconds and hence much lower than the typical lifetime of about 100\,ps.
Due to the band gap in semiconductors and insulators the lower number of final electron states leads to a reduced cross section of  positron-electron scattering in the eV-range and hence to increased thermalization time. 
	
Thermalized positrons mainly interact with the lattice by phonon scattering. 
In metals the diffusion process is dominated by scattering with acoustic longitudinal phonons \cite{Bar50}. 
The positron diffusion length L$_+$ in a defect-free lattice depends on the lifetime $\tau_{e^+}$ and on the (material dependent) diffusion coefficient $D_+$: 
\begin{equation}
	L_+ = \sqrt{D_+ \cdot \tau_{e^+}} \quad .
\end{equation}
For typical values for metals of $D_+\approx $ 1\,cm$^2$/s and $\tau_{e^+} \approx$100\,ps the positron diffusion length $L_+$  is found to be in the order of 100\,nm.
Taking into account the temperature dependence of phonon scattering ($\propto T^{-3/2}$) and the thermal energy of the positron (E$_+=3/2 k_B T$) the diffusion coefficient $D_+$ is proportional to $T^{-1/2}$ leading to $L_+ \propto T^{-1/4}$ \cite{Pus94}. 
	
The positron-electron annihilation in matter is completely dominated by the emission of two 511\,keV photons in opposite direction in the center-of-mass system. 
Since the momentum of the positron in the delocalized (ground) state is much lower than the electron momenta in the crystal the annihilation parameters such as Doppler-shift and angular correlation of the emitted photons provide information on the electronic structure.
The positron annihilation rate is proportional to the local electron density. 
Therefore, the measurement of the positron lifetime allows the distinction between annihilation of ``free'' positrons and positrons trapped in lattice defects such as vacancies.
The typical core annihilation probability is in the range of a few \%  
and hence most of the positrons annihilate with valence  electrons, i.e.\,conduction electrons in metals. 
In insulators and at surfaces, however, the formation of Ps leads to a longer mean positron lifetime and in particular to a significant increase of 3$\gamma$ annihilation events arising from o-Ps self annihilation.
		
\begin{figure}[htb]
\centering
\includegraphics[width=0.7\textwidth]{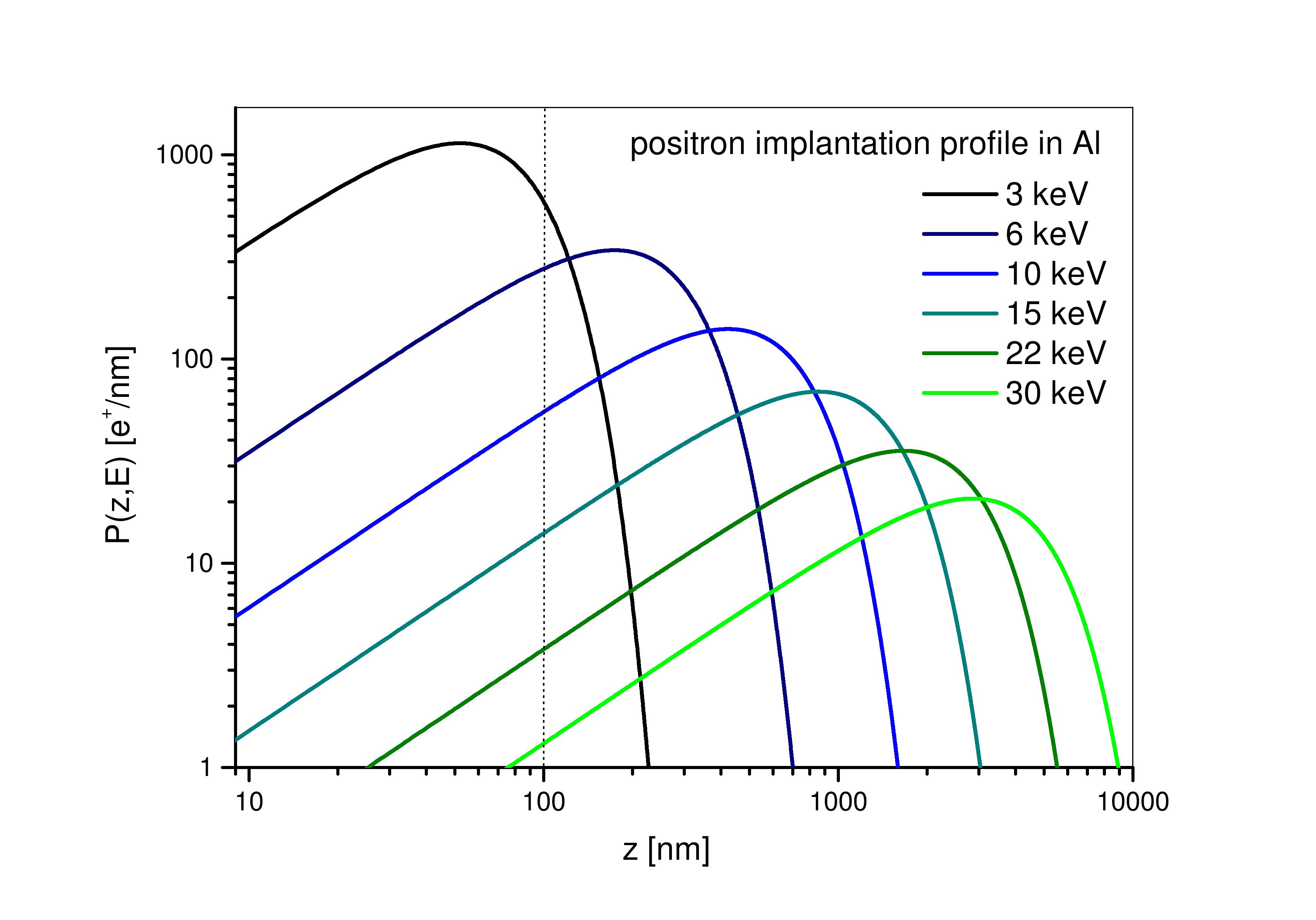} 
\caption{Positron implantation profiles in Al for positron beam energies between 3 and 30\,keV. Up to a depth of about 100\,nm (dotted line) positron back diffusion to the surface has to be considered. 
}
  \label{implprofile}
\end{figure}

The implantation process of positrons in the range 100\,keV was extensively studied by Monte-Carlo simulations in order to reveal the material dependent depth profile \cite{Mak61}.  
The depth distribution of positrons \(P(z,E_+)\) with an impinging energy E$_+$ as function of depth $z$ is described by the so-called Makhovian implantation profile
\begin{equation}
P(z, E_+)=\frac{mz^{m-1}}{z_{0}^{m}}exp\left[-\left(\frac{z}{z_{0}}\right)^{m}\right]
\end{equation}
where  $z_{0}$ 
\begin{equation}
z_{0}=\frac{\bar{z}}{\Gamma\left[(1/m)+1\right]}
\end{equation}
is related to the mean implantation range $\bar{z}$: 
\begin{equation}
\bar{z}=\frac{A}{\rho}E_+^{n}
\end{equation}
with the Gamma-function $\Gamma$, the mass density $\rho$, and  the material dependent parameters $m$, $A$ and $n$ (values are tabulated in e.g.\,\cite{Pus94}).
		Exemplary, the positron implantation profiles in Al were calculated for a kinetic energy between 3 and 30\,keV as plotted in Figure\,\ref{implprofile}.
	In general, the probed depth region, where the information is gained from, is defined by the positron implantation profile blurred by the positron diffusive motion.	
		Therefore, back diffusion to the surface has to be considered for positrons implanted in a depth corresponding to the typical positron diffusion length, which is in the order of 100\,nm for annealed crystals.

\subsection{Positrons reaching the surface}

\begin{figure}[htb]
\centering
\includegraphics[width=0.7\textwidth]{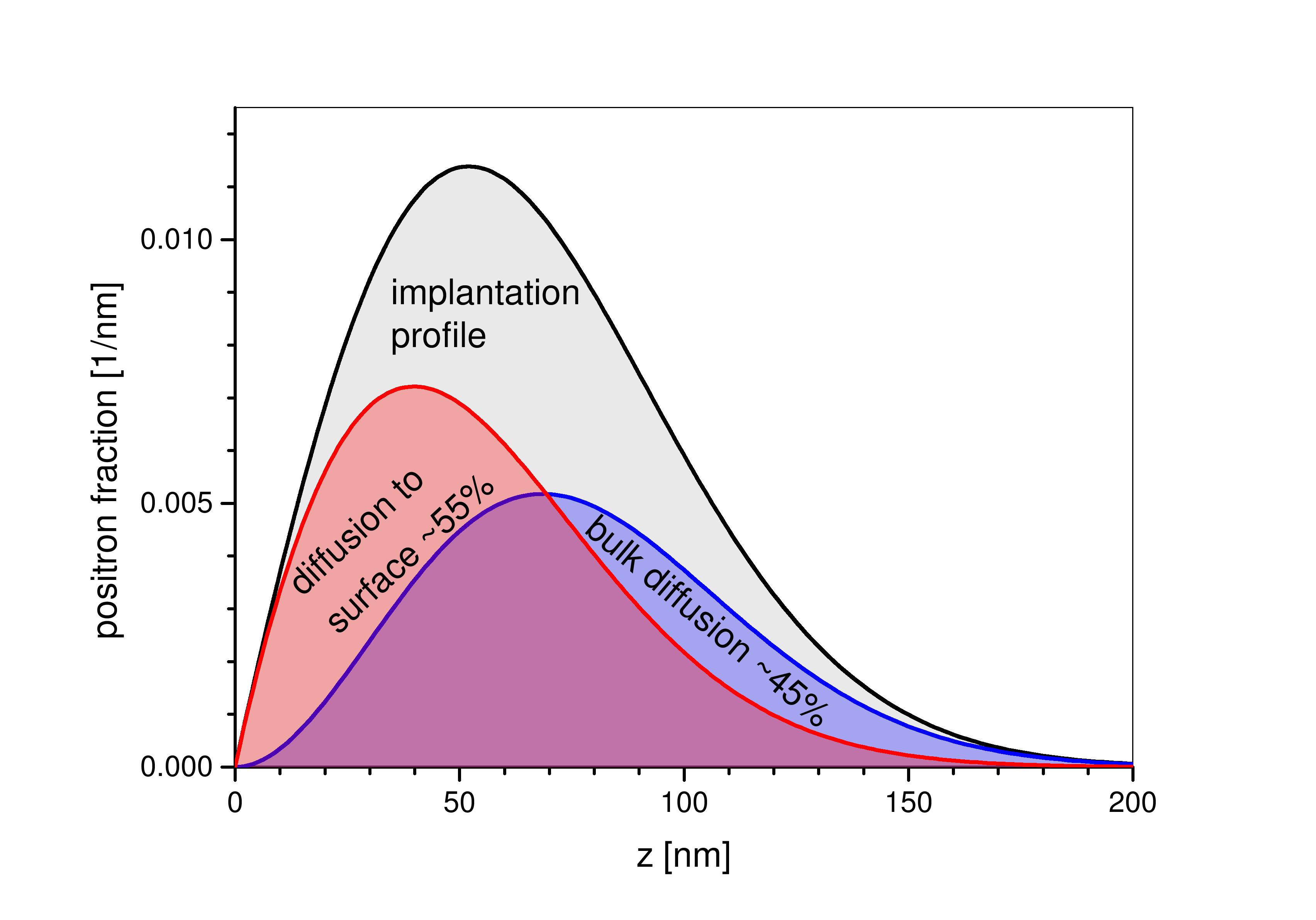} 
\caption{Fraction of positrons diffusing back to the surface calculated for Al using E$_+$=3\,keV and a diffusion length of L$_+$=100\,nm.}
  \label{BulkSurfdiffusion}
\end{figure}

For surface studies, the fraction of positrons diffusing back to the surface after implantation into the specimen is of highest importance.
Accounting for the material dependent positron diffusion length L$_+$ the back diffusion probability can be calculated by 
\begin{equation}
	\label{J(E)}
	J(E) = \int^{\infty}_{0}exp[-z/L_+]P(z,E_+)dz \quad .
\end{equation}
For positrons implanted in well-annealed Al with E$_+$=3\,keV the fraction of positrons diffusing back to the surface amounts to about 55\,\% as shown in Figure\,\ref{BulkSurfdiffusion}.

As depicted in Figure\,\ref{e+matter}, several processes can occur, which are most relevant for surface studies with positrons.
 The diffusion  of thermalized positrons to the surface with subsequent emission as so-called moderated (slow) positrons for materials with negative positron work function plays a central role for the development of efficient positron moderators.
Positrons implanted close to the surface might reach the surface prior complete thermalization and can be emitted as so-called epithermal positrons. 
The emission of (thermally activated) Ps as well as the formation of the positronium negative ion Ps$^-$ play a major role for fundamental studies on bound leptonic systems. 
Therefore, the underlying processes are discussed in Sections\,\ref{sec:ColdPositronium} and \ref{sec:PsMinus}.
In principle, after formation at the surface the Ps momentum distribution contains information about the electronic density of states at the surface \cite{Col02}.
There have been few  Ps formation experiments to demonstrate that information can be gained on oxide growth on metals \cite{Che87, Che89} and semiconductors \cite{Che88}. 
An overview of Ps formation studies on various surfaces is given in \cite{Coleman2000}.
Due to its high surface sensitivity experiments with low-energy Ps are anticipated when monoenergetic Ps beams will become available \cite{Nag14}.
At present, however, the measurement of the different decay probabilities of the triplet and singlet Ps is used to perform unique measurements of the electron polarization at the surface (see  Section\,\ref{sec:Polarized}).
Independent from the sign of the positron work function the positron might be trapped in a surface state \cite{Hut90}. 
Annihilation of a surface trapped positron with core-electrons leads to the emission of Auger electrons and hence allows elemental analysis with topmost layer sensitivity (Section\,\ref{PAES}).
		
\subsection{Consideration of the energetics of surfaces}
\begin{figure}[htb]
\centering
\includegraphics[width=0.7\textwidth]{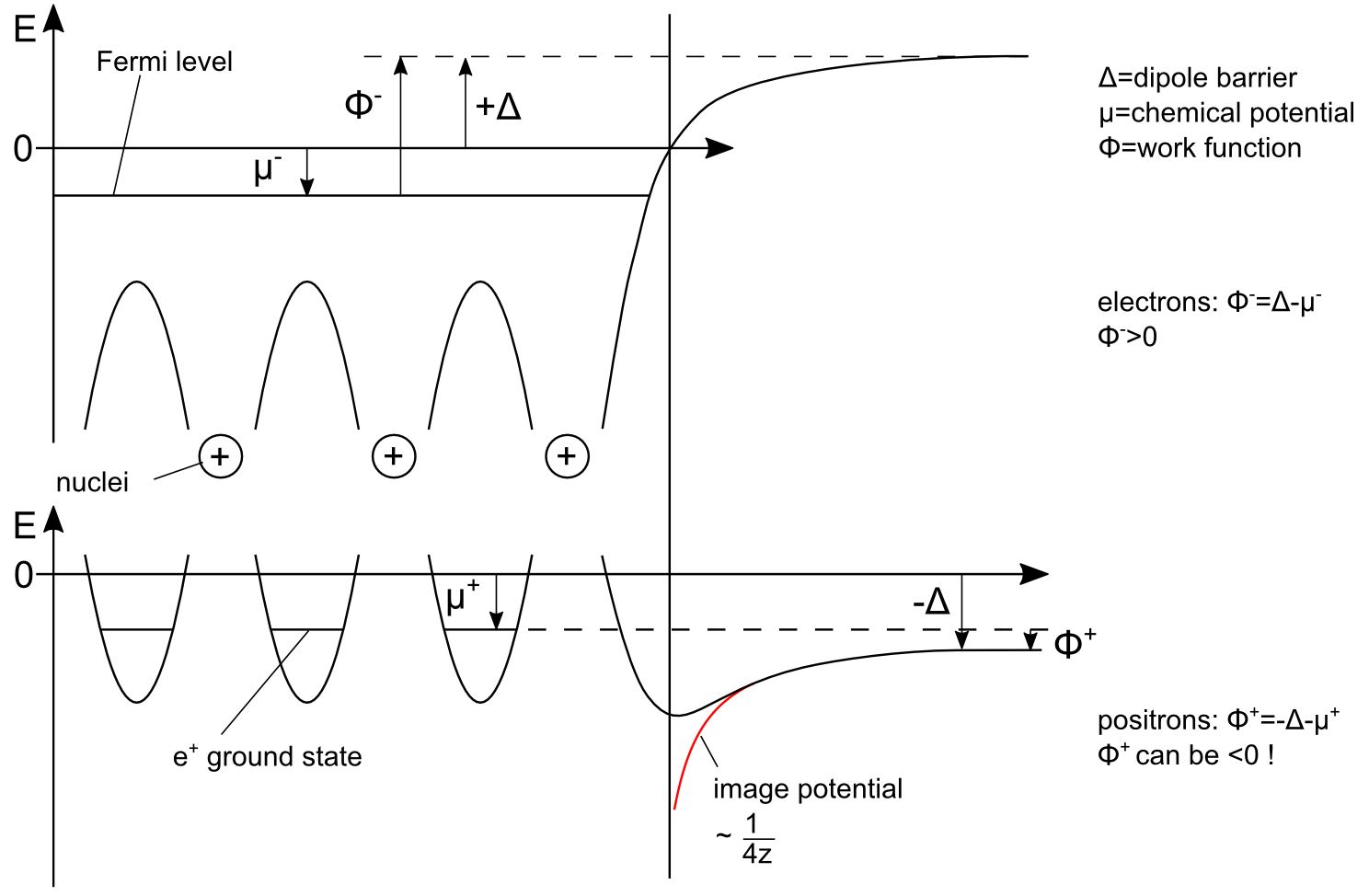} 
  \caption{ Scheme of the bulk and surface potential for electrons (top) and positrons (below):	
$\Phi^-$ and $\Phi^+$ denote the work function, $\mu^-$ and $\mu^+$ the chemical potential  of electron and positron, respectively, and 
$\Delta$ corresponds to the surface dipole potential.
Compared to electrons, for positrons the opposite sign of $\Delta$ gives rise to small or even negative $\phi^+$ as depicted here.
The solid line represents the effective potential comprising the electrostatic and the positron electron correlation potential.
	}
  \label{fig:Potential}
	\end{figure}

The transition of the potential energy from the bulk via the near-surface region to the vacuum leads to characteristic features.
The different potentials affecting the positron are best described by comparison with its counterpart the electron.
Figure\,\ref{fig:Potential} shows schematically the potential energy for an electron and a positron with respect to the so-called ``crystal zero''. 
Contrary to electrons for positrons the crystal potential is always positive leading to major  consequences for surface diffraction experiments (see Section\,\ref{sec:Diffraction}).
In the bulk the chemical potential $\mu^-$ of the electron corresponds to the Fermi level, i.e.\,to the highest occupied electron state in the crystal lattice.
Since usually only one positron resides in the specimen at the same time the  thermalized positron in the ground state with the chemical potential $\mu^+$ can be described by a delocalized Bloch wave  in the periodic potential of the lattice. 
Taking into account the surface dipole barrier $\Delta$ one obtains the work functions $\Phi^+$ and $\Phi^-$ for the positron and the electron, respectively:
	\begin{equation}
	\Phi^{\pm}=\mp \Delta - \mu^{\pm}  \quad.
	\end{equation}
Note that for positrons the opposite sign of $\Delta$ gives rise to small or even negative values of  $\Phi^+$.
As a consequence, at surfaces with a negative positron work function the positron can be spontaneously (re-)emitted into the vacuum with a discrete energy of E$_+$=$-\Phi^+$. 
For this reason,  metals such as W or Pt  ($\Phi^+ = -3.0$\,eV and -1.95(5)\,eV, respectively \cite{Coleman2000, Hug02}) are applied as so-called positron moderators.
At the surface the potential well is formed by both the outermost atomic layer and the surface image potential of the positron (see Figure\,\ref{fig:Potential}). 
The efficient positron trapping at the surface potential of metals and semiconductors gives rise to the outstanding surface selectivity of diverse positron experiments.
The positron trapped in the surface state together with an electron might also form Ps.
However, Ps can only be formed in the region of low electron density outside the surface due to the screening effect of the valence electrons in metals.
The threshold energy $\Phi_{Ps}$ to form and emit Ps from the surface is given by the work functions $\Phi^+$ and $\Phi^-$ of the positron and electron, respectively, and the Ps ground state binding energy of 6.8\,eV (see Equation\,\ref{Eq:PsForm} and discussion in Section\,\ref{sec:ColdPositronium}).

The positron affinity $A^+$ of a material is defined as $A^+= -\Phi^+ -\Phi^-$ \cite{Pus89}. 
This quantity is particularly helpful for the description of the potential step formed at an interface of different materials and for the potential well of buried layers or precipitates \cite{Pus89, Hug08a, Pik11}.
The different positron affinities $A^+$ of the respective materials give rise to an attractive potential towards the material with lower $A^+$. 
	
\begin{figure}[htb]
\centering
\includegraphics[width=0.5\textwidth]{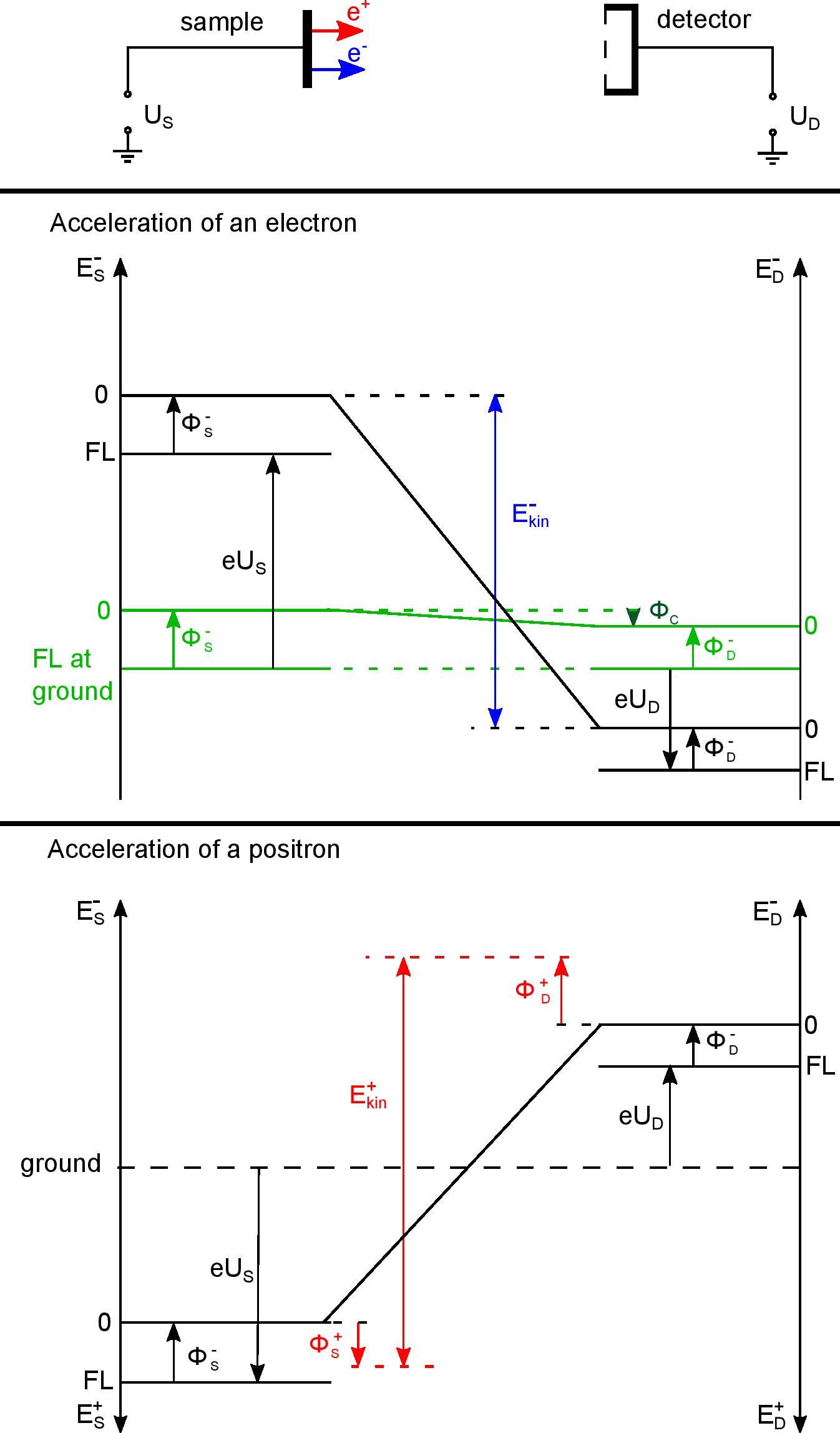} 
	\caption{Simplified measurement scheme of the kinetic energy of electrons or positrons respectively (top). 
	The gain in kinetic energy E$^{\pm}_{kin}$ of an electron (middle) and a positron (bottom) is determined by the respective work functions $\Phi^{\pm}_{S,D}$ and the voltages $U_{S,D}$  applied to the sample and detector.
	The axes indicate the potential energy E$^{\pm}_{S,D}$,	FL denotes the Fermi levels and the zeros correspond to the vacuum levels right at the surface of the sample and detector, respectively. 
		The contact potential $\Phi^-_{C}$ might lead to an acceleration of electrons (plotted for $U_{S,D}$=0 in green).
	The polarity of the external voltages $U_{S,D}$ is chosen to accelerate the respective particle (black levels, middle and botom).
		}
	\label{Emission}
	\end{figure}	

In the experiment, where the kinetic energy of an electron or a positron is detected, the work functions of both  the sample and the energy analyzer (with grid or entrance slit) have to be considered. 
Note that for electrons $\Phi^-$ is always positive and hence energy is required to release a particle whereas for positrons $\Phi^+$ can be negative leading to the emission of positrons into the vacuum.
Before the measured kinetic energy of a positron is discussed the energy landscape for an electron leaving a sample will be recapitulated first.
The acceleration of an electron right from the surface of the sample, i.e. the gain in kinetic energy from the so-called vacuum zero just above the sample, to the vacuum zero right at the detector is considered as depicted in Figure\,\ref{Emission}.
Without biasing the sample or the detector 
the remaining contact potential $\Phi^-_{C}=\Phi^-_{D}-\Phi^-_{S}$ could  lead to an acceleration of an electron right at the surface of the sample (at the so-called vacuum zero). 
By applying voltages to the sample (U$_{S}$) and detector (U$_{D}$) the kinetic energy E$^-_{kin}$ can be deduced from the formula 
	\begin{equation}
	E^-_{kin}=e U_{D}-e U_{S} +\Phi^-_{S}-\Phi^-_{D} \quad .
	\end{equation}
In the case of positron acceleration from the vacuum zero of the sample to that of the detector the contact potential acts oppositely.
For positron emission from the bulk of the sample the positron work functions of both the sample ($\Phi^+_{S}$) and the detector ($\Phi^+_{D}$) have to be taken into account as well.
In general, if acceleration potentials are applied, i.e.\,$U_{S}>0$, and $U_{D}<0$, the kinetic energy E$^+_{kin}$ of the positron amounts to 
	\begin{equation}
	E^+_{kin}= e U_{S} -e U_{D} + \Phi^-_{D} - \Phi^-_{S} + \Phi^+_{D}-\Phi^+_{S} \quad .
	\end{equation}
As exemplarily shown in  Figure\,\ref{Emission} $\Phi^+_{S}<0$ and $\Phi^+_{D}>0$ lead to an increase of the total kinetic energy of the positron measured by the detector. 
Using the definition of the positron affinities $A^+= -\Phi^+ -\Phi^-$  the measured E$^+_{kin}$ can be rewritten as
	\begin{equation}
	E^+_{kin} = e (U_{S}-U_{D}) +A^+_{S}-A^+_{D} \quad .
	\end{equation}

\subsection{Positron density on Al(100) as a show-case}
\label{sec:PositronDensityOnAl100AsAShowCase}

For a theoretical description of the surface potential with density functional theory (DFT) the metal surface is usually represented  by a slab model.
A correct calculation of the positron surface state, however, has to consider the surface image potential as well as the electron-positron correlation potential in the vacuum region.

\begin{figure}[htb]
\centering
\includegraphics[width=0.5\textwidth]{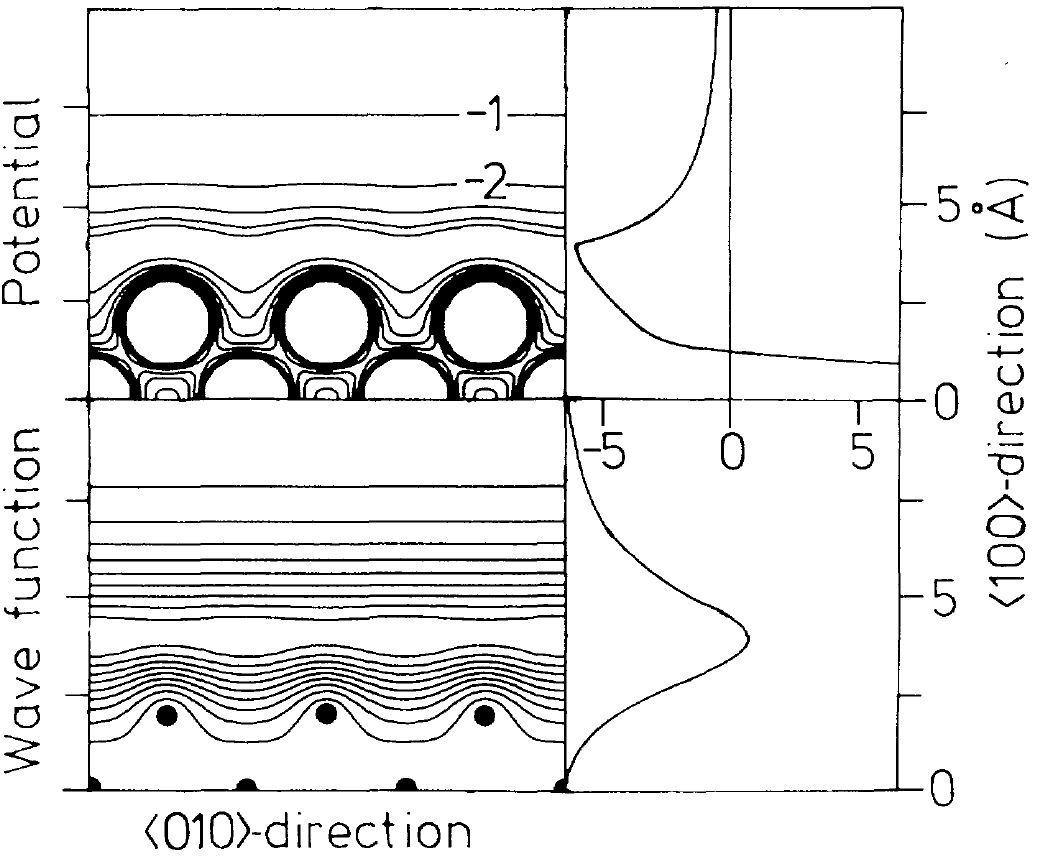} 
\caption{Positrons at an Al(100) surface: Potential and positron wave function calculated with two-component DFT using the corrugated mirror model (figure from \cite{Nie83}). }
  \label{SurfState_Al_Nie83}
\end{figure}

In the following, the positron localization at a clean Al surface and after coverage with Li atoms is exemplarily discussed.
Nieminen et al. used the so-called corrugated mirror model (CMM) to calculate the positron potential and the wave function at an Al(100) surface \cite{Nie83}.
The pronounced localization of the positron wave function at the surface and the according surface potential is shown in  Figure\,\ref{SurfState_Al_Nie83}.
The resulting positron annihilation with electrons from the topmost atomic layer leads to the outstanding surface selectivity e.g.\,for the element analysis with PAES (see Section\,\ref{PAES}) or for the measurement of the polarization of surface electrons (Section\,\ref{sec:Polarized}).

\begin{figure}[htb]
\centering
\includegraphics[width=0.7\textwidth]{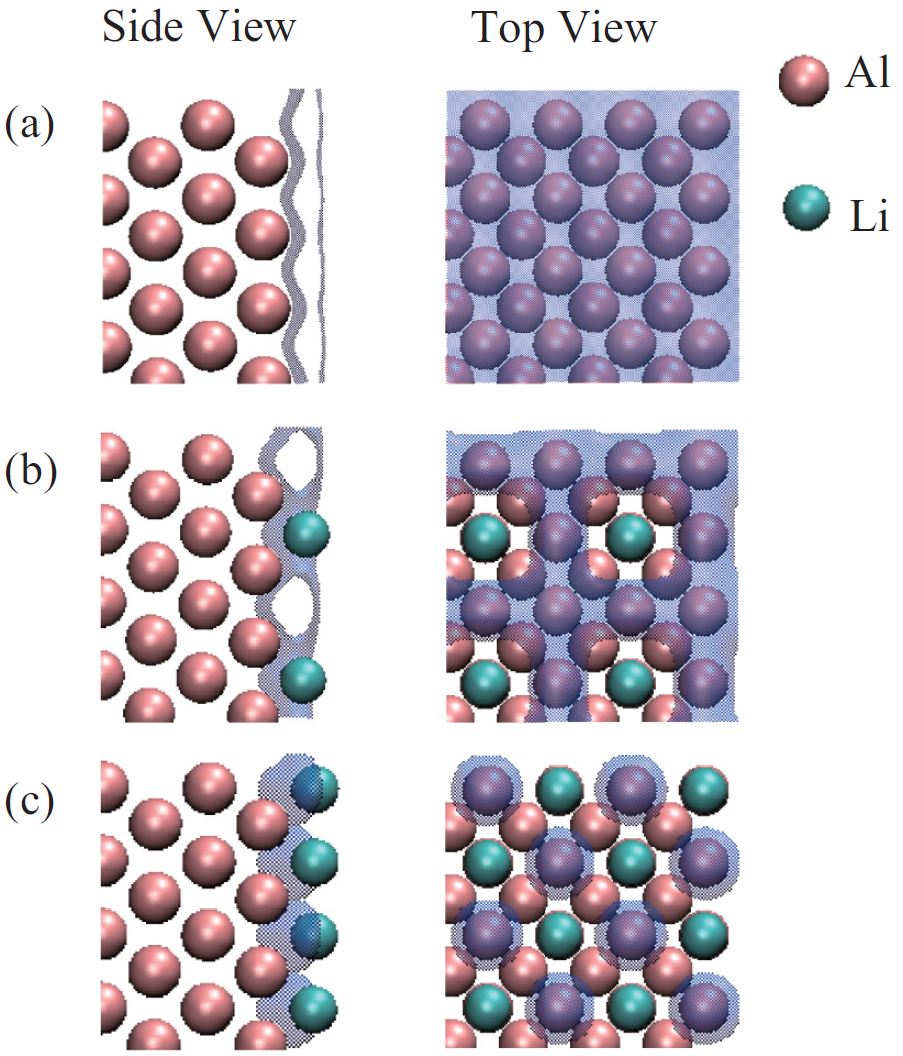} 
\caption{ Positron density distributions on a (a) clean Al(100) surface and with a Li coverage of (b) 0.25\,ML, and (c) 0.50\,ML. 
The positron density is shown as a blue transparent isosurface with an isovalue of 0.0015 a.u. (figure rom \cite{Hag15}). }
  \label{SurfState_AlLi_Hag15}
\end{figure}

More recently, the effect of Li covering the Al(100) surface on the work functions of electrons, positrons, Ps, and the positronium negative ions (Ps$^-$) was investigated in a theoretical study.
For this purpose, the positron potential and the positron wave function on Al(100) surfaces covered with fractions of a monolayer (ML) of Li have been calculated with two-component DFT using the CMM  as shown in Figure\,\ref{SurfState_AlLi_Hag15} \cite{Hag15}.
The observed initial decrease of the electron work function $\Phi^-$ from 4.39\,eV for Al(100)  to 2.18\,eV after covering with 0.25\,ML Li is attributed to the reduction of the surface dipole barrier due to electronic charge transfer from the Li adatoms to the Al substrate which has a higher electron affinity than Li.
With higher Li coverage $\Phi^-$ increases smoothly since the bonding between the 2s-electrons of Li starts to suppress the electron transfer from the Li adatoms to the Al substrate.
For a Li coverage between 0.25 to 1\,ML $\Phi^+$ decreases in the same manner (from 2.15 to 1.01\,eV) as $\Phi^-$ increases.
This behavior   corresponds  to  the expectation that the variation of the work functions results solely from the change of the modulus of the dipole barrier.
It is noteworthy that the modulus of the change in the work functions for positrons and electrons agrees within about $1\,\%$ for various theoretical models applied by Hagiwara et al. \cite{Hag15}.

\section{Tailored surfaces for fundamental experiments}
\label{fundamental}
\subsection{Cold positronium}
		\label{sec:ColdPositronium}

Positronium is a hydrogen-like bound state of an electron and its anti-particle the positron.
According to their relative spin orientation one distinguishes the singlet state as para-positronium (p-Ps, S=0) and  the triplet state ortho-positronium (o-Ps, S=1).
Due to conservation of angular momentum p-Ps with a vacuum lifetime of 125\,ps \cite{Ram94} decays into two $\gamma$ quanta (higher even numbers of photons are largely suppressed) whereas the long-lived o-Ps ($\tau_{o-Ps}$=142\,ns \cite{Val03}) decays into three $\gamma$'s.

As a purely leptonic system Ps has been subject to many fundamental  studies in particular for experimental tests of quantum electrodynamics.
In contrast to the H-atom, the energy of the spin-orbit and the spin-spin interaction are of the same order at Ps.
The hyperfine transition of the ground state Ps was directly measured by Yamazaki et al. \cite{Yam12}.
The fine structure of the 2S and 2P levels of Ps, which are non-degenerated in contrast to the H-atom, was determined by Mills et al. \cite{Mil75}.

In the fundamental research on the matter anti-matter symmetry, e.g. in gravitation experiments, the efficient formation of anti-H-atoms ($\bar \mathrm{H}$) is of outstanding importance.
For the $\bar \mathrm{H}$ production low-energy positrons have to recombine with cold $\bar p$, which are usually stored in a Penning trap.
However, the opposite charge of positron and $\bar p$ leads to the principal difficulty to let the recombination take place in a single trap. In addition, the probability to create $\bar \mathrm{H}$ in the ground state is low due to the high stability of the formed high Rydberg states. 
In order to overcome these constraints, the efficient production of cold Ps became highly important.  
Due to its neutrality a dense Ps pulse can drift into the electromagnetically trapped $\bar p$ cloud to eventually form large numbers of $\bar \mathrm{H}$ atoms  in a well-defined state by charge exchange reactions.

\begin{figure}[htb]
\centering
\includegraphics[width=0.5\textwidth]{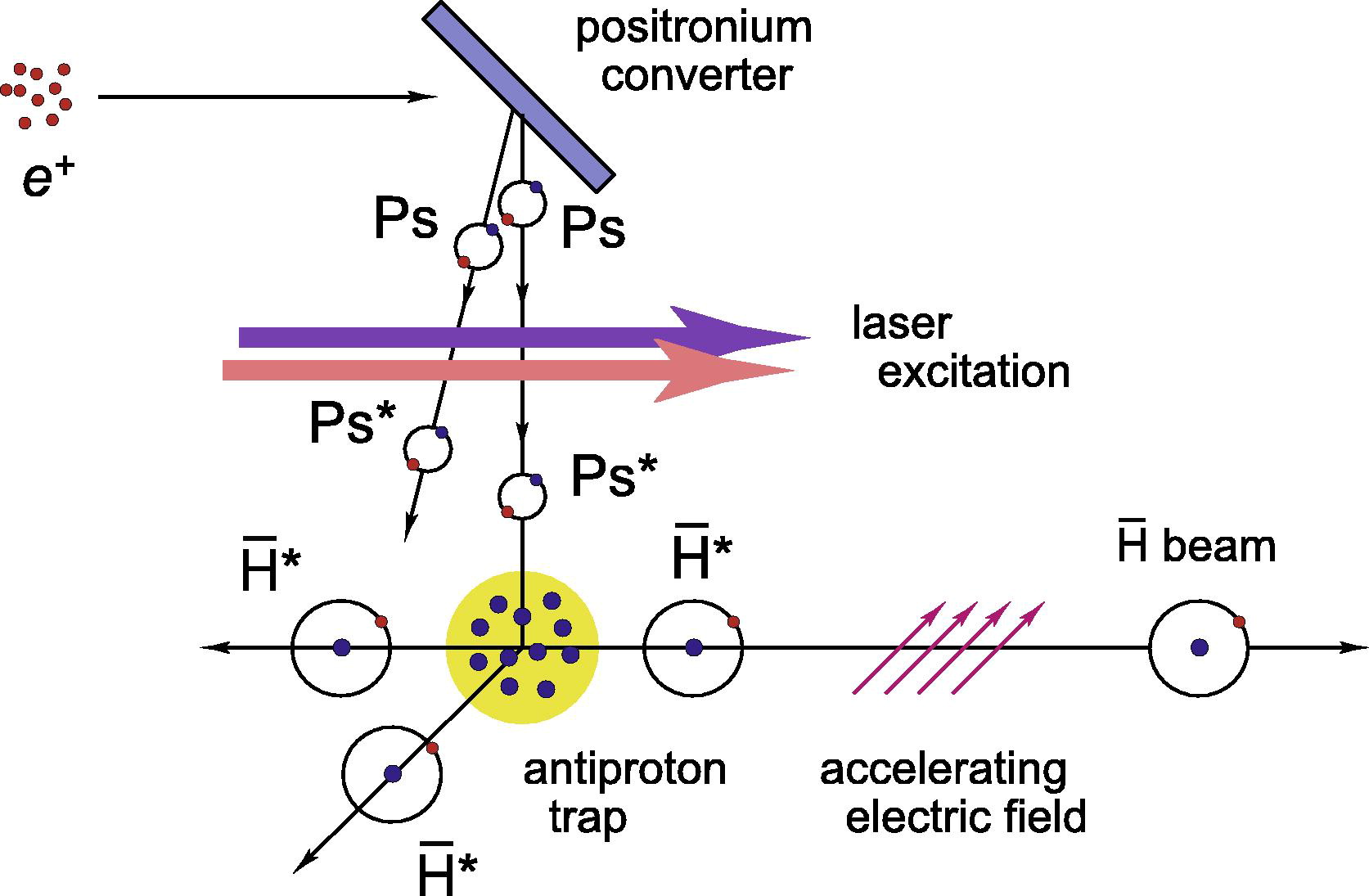} 
\caption{After hitting a converter positrons form Ps which will be excited to a high Rydberg state with principal quantum number n $\geq$30 (Ps$^{\ast}$). Resonant charge exchange between Ps$^{\ast}$ and trapped $\bar p$ generates excited $\bar \mathrm{H}^{\ast}$, which are accelerated by the Stark effect in inhomogeneous electric fields (figure from \cite{Kel08}).
} 
  \label{Anti_H_Kel08}
\end{figure}

For this reason, the aim is to prepare the surface of a target material in such a way that (i) a maximum number of o-Ps is formed in the surface near region, (ii) a high Ps diffusion length is achieved,  and (iii) Ps is emitted with a well defined (low) energy. 
Moreover, in order to allow the particles to drift a macroscopic path length the lifetime of the emitted o-Ps is increased by laser excitation to high Rydberg states.
Among several research groups with the aim to produce and study $\bar \mathrm{H}$ the principle of the experiment developed within the Aegis collaboration is exemplarily presented.
The main idea is to produce an energy variable $\bar \mathrm{H}$ beam for the test of gravity using anti-matter \cite{Kel08}. 
As sketched in Figure\,\ref{Anti_H_Kel08} the $\bar \mathrm{H}$ beam production follows  a three-step procedure:
(i) the produced Ps will be excited to a high Rydberg state with principal quantum number n $\geq$30 (Ps$^{\ast}$), (ii) resonant charge exchange between Ps$^{\ast}$ and cold $\bar p$ stored in a Penning trap generates excited $\bar \mathrm{H}^{\ast}$, and (iii) the large electric dipole moment of $\bar \mathrm{H}^{\ast}$ enables the creation of an energy tunable $\bar \mathrm{H}$ beam by Stark acceleration using inhomogeneous electric fields.

Free Ps can be either formed by  positron scattering with atoms or molecules (see \cite{Gar98} and references therein) or it is released from the surface of a solid after positron impact.
In insulators such as silica or polymers Ps can also be formed in the bulk where it might diffuse to the surface and leaves the solid as so-called thermally activated Ps.
In metals the Ps formation is prevented due to thee screening	effect of the conduction electrons \cite{Kan65}.
However, at a metallic surface several processes lead to Ps emission.
(i) Positrons can form non-thermal Ps directly by picking up an electron from the surface.
(ii ) Surface trapped positrons together with an electron may also be thermally desorbed as Ps with an energy distribution given by the temperature of the solid. Hence heating of the sample leads to an increased Ps emission rate.
(iii) Positrons diffusing to the surface may capture an electron and be ejected as Ps with an energy of a few eV. 
The kinetic energy of the Ps depends on the density of states of the electrons at the surface and the Ps formation potential $\Phi_{Ps}$ 
\begin{equation}
		\Phi_{Ps} = \Phi^+ + \Phi^- -E_{Ps}
		\label{Eq:PsForm}
\end{equation}
with $\Phi^+$ and $\Phi^-$ the positron and electron work functions, respectively, and the Ps binding energy of $E_{Ps}$=6.8\,eV.
Therefore, when slow positrons impinge on a metal, Ps atoms are produced from thermalized positrons and emitted from the surface spontaneously since the Ps formation potential $\Phi_{Ps}$ is negative.

Using a time-of-flight technique Mills et al.\,measured the velocity of Ps formed after implantation of 1-2\,keV positrons, which diffuse back to the surface of Al(111) \cite{Mil83}.
A sharp step in the Ps intensity at a an energy of 2.62(4)\,eV was attributed to the expected value of the Ps formation potential $\Phi_{Ps}$=-2.60(3)\,eV obtained from the according values of the known electron and positron work functions.
Furthermore, the Ps yield was found to be proportional to the density of states at a distance of about 0.1\,nm outside the metal surface averaged over a region corresponding to the Ps size ($\sim$0.1\,nm).
Further analysis of the Ps emission spectrum revealed that about 12\% of the positrons reaching the surface form thermal Ps \cite{Mil91}.

\begin{figure}[htb]
\centering
\includegraphics[width=0.5\textwidth]{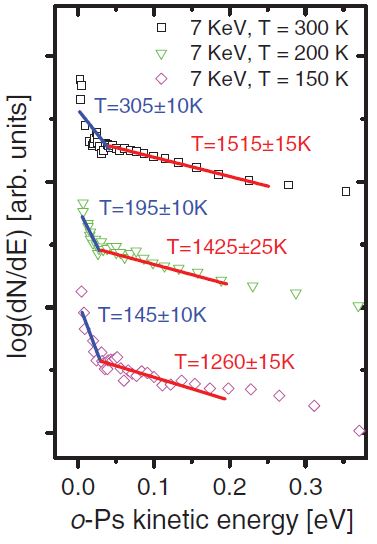} 
\caption{
Measured o-Ps energy spectra obtained after implantation of 7\,keV positrons in silica nano-channels with diameters of 5-8\,nm at various temperatures.
The two-exponential fits for each spectrum point out the presence of two different Maxwellian beam distributions at the indicated temperature
(figure from \cite{Mar10b}).
} 
  \label{PsCold_Mar10b}
\end{figure}

Higher yields of cold o-Ps are achieved by implanting positrons at low temperatures in a target material with an open porosity connected to the surface.
For this purpose, oxidized Si, i.e. thin layers of SiO$_2$ on Si, is applied where the surface of the target material is maximized by applying a geometry of nano-pores or nano-channels.
The reason to use silica is the high Ps formation probability in the bulk amounting to  72\%, and the Ps diffusion length of about 11-15\,nm \cite{VPe04} allowing a high number of formed Ps to escape from the sample.
However, since quantum confinement of Ps limits the minimum Ps energy the size and the geometry of the nano-channels is crucial for Ps emission at very low temperature.
For a detailed description of Ps formation in such materials and surface preparation see e.g. \cite{Mar10a, Mar10b} and references therein.
Mariazzi et al.\,investigated the Ps formation and diffusion in oxidized nano-channels in Si dependent on both positron implantation energy and size of the nano-channels ranging from 5 to 100\,nm \cite{Mar10a}.
The nano-channels with a length of about 2\,$\mu$m were produced in Si by electrochemical etching in hydrofluoric acid and subsequent oxidation of the inner surface.
 At 1\,keV positron implantation energy a Ps yield up to 45\% was observed, and a fraction of about 42\% of them is estimated to be emitted into the vacuum.
Temperature dependent measurements on oxidized Si nano-channels with diameters of 5-8\,nm revealed that at 150\,K about 27\% of the implanted positrons (E$_+$=7\,keV) form Ps \cite{Mar10b}. 
Figure \,\ref{PsCold_Mar10b} shows the contribution of cooled Ps to the measured Ps spectra at different temperatures.  
A fraction of 9\% of the emitted Ps following a Maxwellian distribution with a characteristic temperature of 150\,K is attributed to Ps which is cooled by collisions with the walls of the nano-channels . 

		\subsection{Positronium negative ion}
		\label{sec:PsMinus}

The positronium negative ion Ps$^-$ is a bound system consisting of two electrons and a positron.
About 50 years ago, Wheeler discussed the stability and bound states of ``polyelectrons'' such as Ps$^-$  \cite{Whe46}.
The ground state of Ps$^-$ is stable against dissociation but, naturally, unstable against annihilation into photons.
Since the constituents are point-like leptonic particles of equal mass Ps$^-$ is an ideal object to study the quantum mechanics of a three-body system. 
Furthermore, the production,  acceleration of Ps$^-$ and subsequent photo-detachment would pave the way for the creation of an energy-variable Ps beam.
Details of experiments and techniques for efficient Ps$^-$ formation, photo-detachment and the development of Ps beams can be found in the review by Y. Nagashima \cite{Nag14} and references therein.

The Ps$^-$ binding energy of E$_B$=7.13\,eV corresponds to that of the neutral Ps E$_{Ps}$ and  to that of the additional bound electron ($\approx$0.33\,eV). 
At surfaces Ps$^-$ is produced and emitted if the formation potential $\Phi_{Ps^-}$
\begin{equation}
\Phi_{Ps^-} = \Phi^+ + 2\Phi^- -E_B
\label{Eq:Ps-}
\end{equation}
is negative.
Hence, Ps$^-$ emission is energetically allowed from surfaces of diamond-like C, W and Ta \cite{Nag14}.  

\begin{figure}[htb]
\centering
\includegraphics[width=0.4\textwidth]{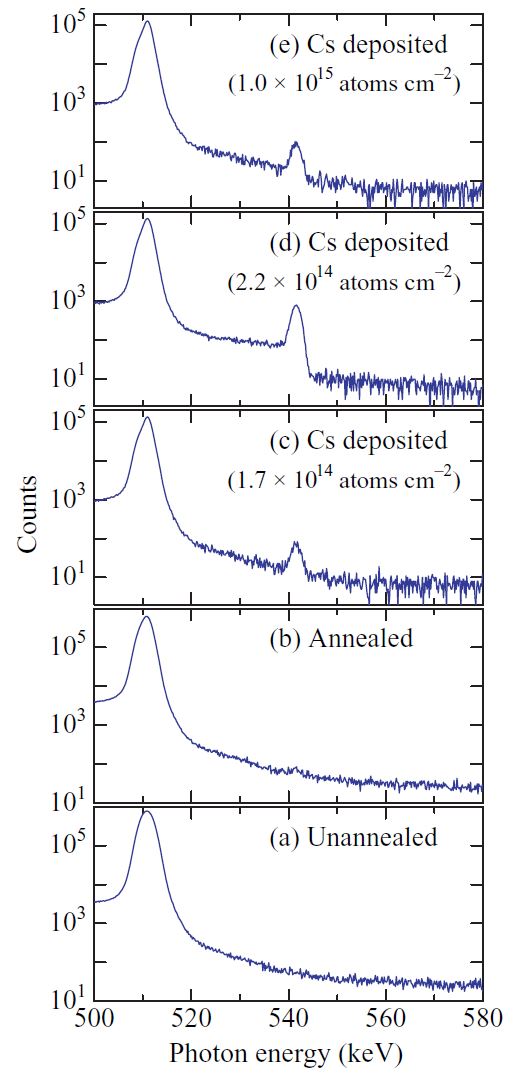} 
\caption{Spectra of the annihilation photons obtained for (a) the untreated W(100) surface, (b) after annealing  and (c)--(e) after Cs deposition.
 The appearance of the second peak at about 541\,keV corresponds to the in-flight decay of Ps$^-$ leading to a Doppler-shift of 30\,keV
(figure from \cite{Nag08}).} 
  \label{CsW_Nag08}
\end{figure}

In the experiment, Ps$^-$  was firstly observed  by Mills in 1981 \cite{Mil81}. 
For the Ps$^-$ production positrons with a kinetic energy of E$_+$=400\,eV were used which partially pass through a  (Ni mesh supported) 4\,nm thin C film. 
Dependent on the acceleration voltage at the production foil the in-flight decay of Ps$^-$ generates a distinct Doppler-shifted annihilation peak which can clearly be identified in the $\gamma$ spectrum (see also Figure\,\ref{CsW_Nag08}).
After the development of tabletop tandem accelerator setups using two 5\,nm thin diamond-like C foils for production and acceleration of Ps$^{-}$ \cite{Fle06} its ground-state decay rate could be determined with unprecedented accuracy to 2.0875(50)\,ns$^{-1}$ \cite{Cee11b} in agreement with recent theoretical predictions \cite{Puc07}.

The formation efficiency, defined as the fraction of the number of emitted Ps$^-$ to that of the incident positrons, is usually in the order of 10$^{-4}$.
By rewriting Equation\,\ref{Eq:Ps-} using the chemical potentials of positron and electron $\mu^+$ and $\mu^-$, respectively, one obtains 
\begin{equation}
	\Phi_{Ps^-} = -\mu^+-2\mu^--7.13\,eV + \Delta \quad.
\end{equation}
Hence it becomes evident that the formation potential $\Phi_{Ps^-}$ can be lowered by manipulating the surface dipole barrier $\Delta$.
An overlayer of oxygen on W would increase the value of $\Delta$ and hence suppress Ps$^{-}$ emission whereas $\Delta$ would decrease by adsorbed alkali atoms due  the depolarization of the surface dipole \cite{Nag14}. 
Therefore, great efforts have been made to improve the conversion efficiency for Ps$^-$ production by covering the W surface with alkali metals.
	
\begin{figure}[htb]
\centering
\includegraphics[width=0.5\textwidth]{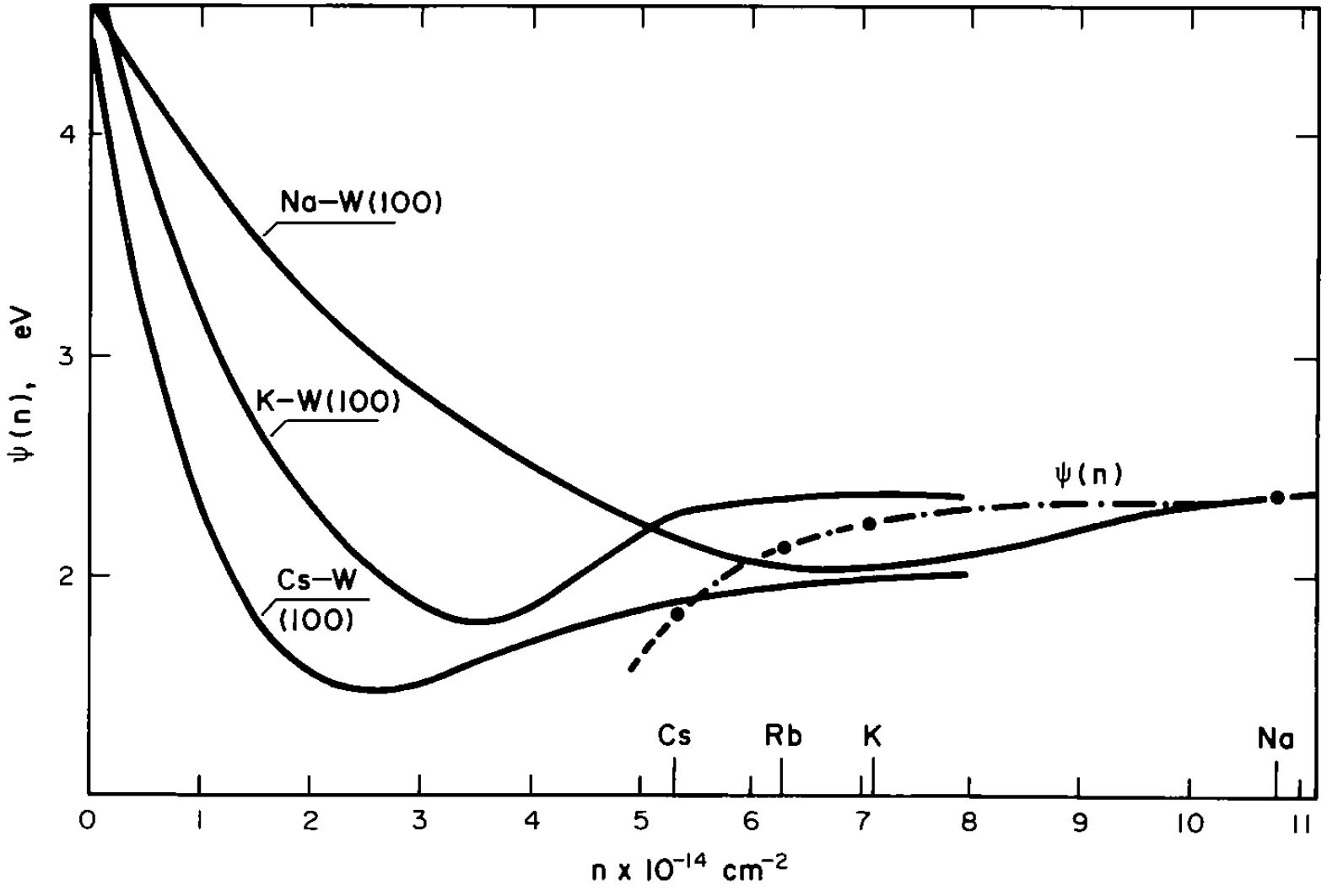} 
\caption{Dependence of the work function $\Phi^-$ of W(100) on the adatom concentration $n$ of Na, K and Cs, and $\Phi^-(n)$ of bulk alkali metals (dashed-dotted line; figure from \cite{Kie81}).
} 
  \label{workfunction_Kie81}
\end{figure}

For instance, the electron (positron) work functions for clean W(100), which amount to 4.63\,eV \cite{Lid2003} (−3.0\,eV \cite{Coleman2000}), will be shifted by $-(+)$3.10\,eV when covered with 2.2$\cdot 10^{14}$ Cs atoms cm$^{-2}$\cite{Nag08}.
The dependence of the electron work function $\Phi^-$ of W surfaces on the degree of coverage with alkali atoms was studied  by Kienja et al.\,\cite{Kie81}.
The variation of $\Phi^-$ of W(100) dependent on the surface concentration of adatoms of Na, K, and Cs is shown in Figure\,\ref{workfunction_Kie81}.		
In the same figure, the dependence of $\Phi^-$ of the alkali metals on their bulk electron density is represented in terms of surface concentration by taking into account the respective lattice constants. 
It is noteworthy that already 1\,ML of alkali atoms on the W(100) surface corresponds nearly $\Phi^-$  of the bulk alkali metal.

The energy spectra of the annihilation photons obtained for clean W(100) and after Cs deposition are shown in Figure\,\ref{CsW_Nag08}.
After annealing of the bare W crystal, a small Doppler-shifted peak appearing at about 541\,keV indicates the in-flight decay of Ps$^-$.
Adsorption of Cs atoms at the W surface results in a stronger peak, which becomes clearly distinguishable from the background photons. 
A maximum intensity is reached at a Cs surface density of about 2.2$\cdot 10^{14}$ atoms cm$^{-2}$ as expected from the change of the surface dipole barrier \cite{Kie81}. 
By covering W(100) with Cs atoms the conversion efficiency for Ps$^-$ production could be increased to 1.25\% \cite{Nag08}, which is two orders of magnitude higher compared to a clean W(100) surface.
The observed decrease of the conversion efficiency to a constant value of 0.1\% after about 16\,hours was attributed to oxygen accumulation from the residual  gas in the UHV chamber and structural change of the Cs layer on the W(100) surface \cite{Nag08}.

		\subsection{Molecules and Bose-Einstein condensate of Ps}
		\label{sec:PositroniumMolecule}

As discussed by Wheeler in 1946 \cite{Whe46} two Ps atoms may combine to form the Ps$_2$ molecule, with a binding energy of 0.4\,eV \cite{Hyl47}.
However, the experimental confirmation of its existence is demanding since the spatial density of o-Ps within the nano-second time scale must be high enough to let the Ps interact with each other.
For experiments with \textit{many} positrons -- e.g. observation of the formation of  the Ps$_2$ molecule or the creation of a Ps Bose-Einstein condensate (BEC)-- a positron density as high as possible is required.
A buffer gas trap (as presented in Section\ref{LowEBeam}) is well suited to provide a dense low-energy positron pulse.  
After accumulation, a high number of positrons is spatially focused onto a thin film sample of porous silica for Ps formation.
	
In 2007 Cassidy et al.\,\cite{Cas07} succeeded in observing the formation of  Ps$_2$ on the internal pore surfaces of porous silica.
Fur this purpose, pulses of 10$^7$ positrons with a sub-nanosecond time width were implanted into a silica film containing interconnected pores with a diameter below 4\,nm capped with a 50\,nm thick non-porous layer.
The overall probability of two Ps atoms interacting with each other was estimated to be about 10\%.
 By applying the so-called single-shot PALS technique the characteristic temperature dependence of the Ps$_2$ formation probability could be revealed hence demonstrating the existence of this exotic leptonic molecule.
 Details of the experiments and the discussion of competing processes such as spin exchange quenching and pick-off annihilation due to interactions with the pore walls can be found in \cite{Cas07}.
In subsequent experiments an excited state of the Ps$_2$ molecule was studied via optical spectroscopy.
The 1S-2P transitions in free Ps and in Ps$_2$ correspond to a wavelength of 243\,nm and 251\,nm, respectively.
By resonant excitation with the appropriate lasers single Ps atoms and Ps$_2$ molecules could be clearly distinguished \cite{Cas12}. 

The experiments with interacting Ps pave the way to create exotic  molecules such as positronic water \cite{Jia98} but also to enable the condensation of a large number of Ps to a BEC.
The formation of a Ps-BEC has attracted some attention due to the much lower mass of Ps compared to ordinary atoms.
Since the lower the mass the higher the BEC transition temperature (note that m\,$\propto$\,T$_{BEC}^{-1}$) dense Ps would be very well suited to form BEC at not too low temperatures as shown in the phase diagram for electron-positron many-body systems presented in Ref.\,\cite{Yab04}.
As proposed by Platzman et al. the formation of a Ps-BEC might become feasible by trapping 10$^5$ o-Ps atoms in a volume of about  0.1$\mu$m$^3$ at a temperature of 20-30\,K within a time of the order of nanoseconds \cite{Pla94}.
By considering the temperature dependent Ps$_2$ and Ps yield, Mills \cite{Mil02} pointed out that a high positron surface density forming a dense Ps gas would allow the near room temperature Bose-Einstein condensation of Ps.

\newpage
\section{Positron diffraction}
\label{sec:Diffraction}

\subsection{Basic principles of positron diffraction}
\label{sec:BasicPrinciplesOfPositronDiffraction}
Since decades electron diffraction techniques are applied as standard tools in surface science for a vast number of investigations such as the determination of surface structures by LEED or in-situ monitoring of layer-by-layer growth using RHEED.
In the following the underlying physics of the particle ion-core interaction, which results  in a number of differences between positron and electron diffraction, will be reviewed. 

The main difference between positron and electron scattering with a solid is the different inner potential, which is the averaged electrostatic field felt by the incoming fast charged particle \cite{Ton92}.
In contrast to electron-ion core scattering, the scattering potential between the positron and the atomic nucleus is repulsive due to the same sign of their electric charge.
In addition, since there is no contribution due to the Pauli exclusion principle in positron-atom scattering no exchange-correlation term has to be considered \cite{Fed80}.
Since the positron correlation with core electrons is small and dynamic effects are estimated to be less relevant for positrons, the static approximation, similar to LEED calculations, can be used for LEPD calculations \cite{Rea81}.
In general, the elastic scattering of positrons with atoms is considerably weaker than in electron scattering (see e.g. \cite{Duk90}).

Classically, the repulsive Coulomb potential of the ion-cores keeps a positron at distances larger than the turning point R$_t$ = Ze$^2$/E$_+$, where E$_+$ is the kinetic energy of the positron. 
By contrast, the attractive potential allows the electron to approach the core at distances r\,$\leq$\,R$_t$ as sketched in \sFF{scatterfactor_Ton92}.
Consequently, for positrons relativistic effects are less relevant since they are decelerated when approaching the ion cores. Therefore, spin-orbit coupling, which governs the Mott-scattering of electrons, and hence spin-dependent scattering is much weaker for positrons interacting with surfaces containing elements with high atomic number \cite{Fed80}.
Since at positron scattering the interaction with the ion-cores and the centrifugal barrier are both positive the classical R$_t$ is larger for positrons than the turning point for electrons  \cite{Duk90}.
The scattering factor for electrons show resonances whereas for positrons the scattering factors turn out to be similar for all elements.
Because positrons are repelled by the atomic nuclei, the scattering of slow positrons by atoms resembles the Born approximation \cite{Ton92} as depicted in \sFF{scatterfactor_Ton92}.

 \begin{figure}[htb]
\centering
\includegraphics[width=0.5\textwidth]{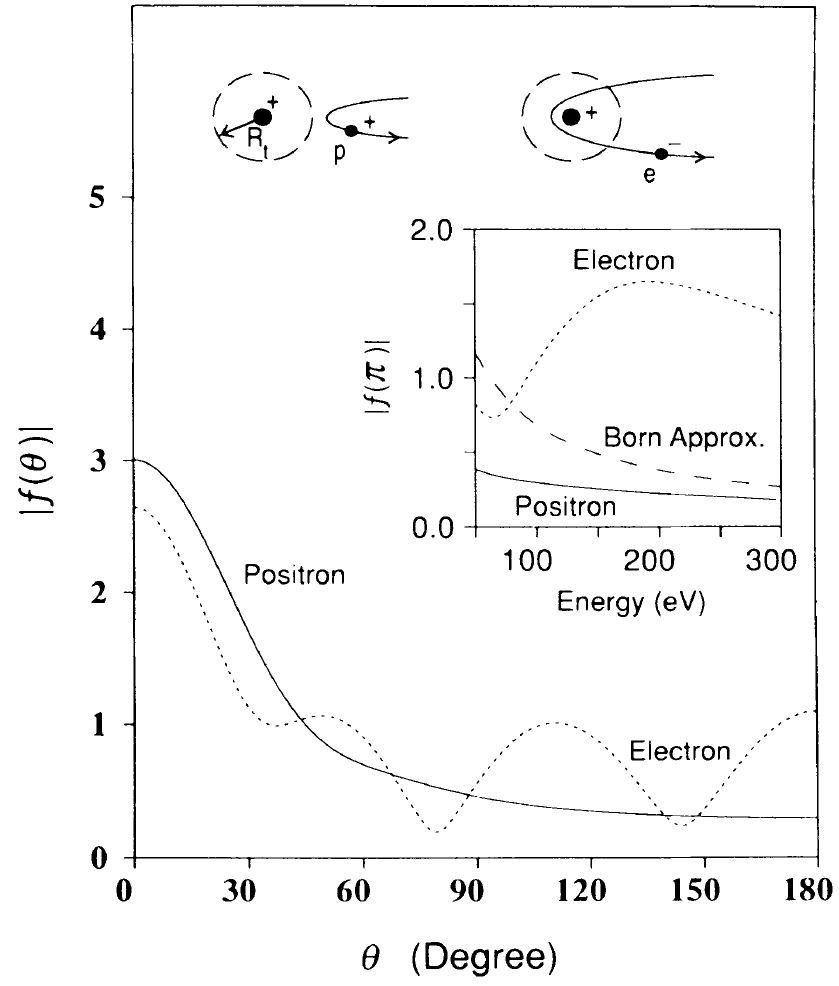} 
\caption{The differential scattering factor of positrons and electrons at 100\,eV for Cu. Insert: energy dependence of $\left|f(\pi)\right|$ and the Born approximation (figure from \cite{Ton92}). } 
  \label{scatterfactor_Ton92}
\end{figure}

The scattering factors of positrons and electrons approach each other when particle-hole excitation becomes less probable than plasmon excitation \cite{Rea81}.
Consequently, the inelastic mean free path (IMFP) of positrons and electrons  are similar above about 200\,eV  \cite{Ton00}.
Below particle energies in the order of 100\,eV the behavior of electrons and positrons experiencing inelastic collisions reveals significant differences.
Since there are no excluded final states for the positron in a solid the IMFP of a positron is shorter than that of an electron leading to an increased  surface sensitivity of positrons.
The damping for positron scattering relative to electrons is further enhanced due to other loss channels such as positron annihilation and Ps formation.
Tong et al.\,stated that the combination of weak elastic scattering (similar to photons) and strong inelastic damping (high surface sensitivity) makes positrons appealing for LEPD holography. 
Positron holography, however, would require positron densities as high as 10$^{10}$ positrons per second on a beam spot of 1\,$\mu m^2$  \cite{Col02}.  

\begin{figure}[htb]
\centering
\includegraphics[width=0.5\textwidth]{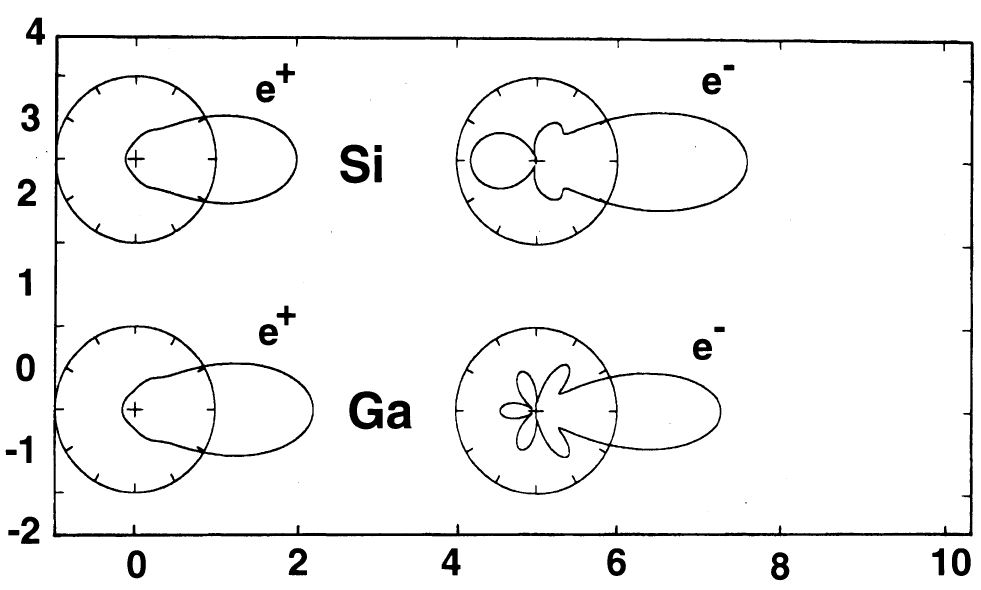} 
\caption{Radial plots of the differential scattering factors of positrons and electrons at 100\,eV for Si and Ga. 
The positive horizontal direction corresponds to $\theta$=0$^o$;  the unit of $\left|f(\theta)\right|$ is {\AA}
(figure from \cite{Ton00}). } 
  \label{scattere+e-_Ton00}
\end{figure}

Electron scattering factors are highly anisotropic in both amplitude and phase whereas the scattering factor of positrons more closely resembles that of photons as pointed out by Tong et al.\,\cite{Ton98} .
These differences are related to the fact that a positron, like a photon, is not bound by an atomic nucleus, while an electron is \cite{Ton00}.
Exemplary,  the positron differential scattering factor is compared with that of the electron for Si and Ga   in \sFF{scattere+e-_Ton00} illustrating the high anisotropy in the electron case. 
In particular, the scattering factor of positrons varies slowly and smoothly.  
Details of the calculation of the respective scattering factors can be found in elsewhere \cite{Ton98, Ton00}.
In LEED,  for a reliable determination of the surface structure the diffraction intensities are computed by so-called dynamical calculations, i.e., including multiple scattering \cite{Duk93}.
In contrast to LEED, the absence of resonance scattering effects and less multiple scattering greatly facilitate simulations for LEPD.
Furthermore, the improved description of positron-atom scattering cross sections using the simplified potential model of elastic positron diffraction yields significantly better agreement between calculated and measured intensities in LEPD than in LEED \cite{Duk90}.
As pointed out by Joly  \cite{Jol94} LEED calculations become more complex and more expensive since a nonuniform lattice of denser grid points in the ion core area has to be used in order to account for the attractive electron-core interaction resulting in a high kinetic energy of the electrons close to the ion core. 

As visualized in \sFF{scattere+e-_Ton00} the positron scattering cross sections depend only barely on the specific element. 
Therefore, using LEPD, the reduced contrast between different elements as scatterers should result in an improved structure determination of surfaces even with highly unlike atomic species. 
Taking advantage of this  feature, a comparative study on the surfaces of InP and GaAs has been perfomed by Chen et al.\,\cite{Che93} (see Section\,\ref{sec:GaAsAndInP}).

The inner potential of the crystal lattice leads to a reduction of the kinetic energy of the positron whereas it is increased for an electron. 
In the following, Si(111) surface is used as a show-case to discuss the consequences in a diffraction experiment using positrons or electrons.

  According to Snell's law of refraction the vacuum de-Broglie wavelength $\lambda$ and the glancing angle $\theta$ are related with the wavelength inside the crystal $\lambda_{s}$ and the refraction angle $\theta_s$ by 
\begin{equation}\label{snellius}
\frac{\lambda}{\lambda_{s}}=\frac{\cos \theta}{\cos \theta_{s}}  \quad.
\end{equation}
The well-known Bragg equation 
\begin{equation}\label{bragg}
2d \sin \theta_{s} = n \lambda_{s}
\end{equation}
describes the diffraction of a particle in a crystal with the lattice spacing $d$ under the actual glancing angle inside the crystal $\theta_{s}$ and the resulting  order of the constructive interference  $n$ as a positive integer.

Accounting for the crystal potential U$_{0}$ the wavelength of the electron or the positron inside the crystal  $\lambda_{s}$ amounts to
\begin{equation}\label{wavelength}
	\lambda_{s}= \sqrt{150.4 \mathrm{eV}/ (E_{kin}- qU_{0})} \,\mathrm{\AA} \quad.
\end{equation}
Finally, by replacing $\lambda$ using Equations \ref{bragg} and \ref{wavelength} one obtains the Bragg condition 
\begin{equation}
E_{kin}\cdot\sin^2 \theta = \frac{n^2}{d^2} 37.5 \mathrm{\AA}^2\mathrm{eV} + qU_{0}  \quad.
\end{equation}

For Si surfaces the mean inner potential energy of the crystal amounts to qU$_{0}=\pm$12\,eV for positrons and electrons, respectively \cite{Ich87}.
In general, according to Snell's law the refractive index for positrons (electrons) is smaller (larger) than one due to the different sign of the crystal potential for positrons and electrons.
The values taken for electrons impinging on Si(111) (qU$_{0}$ = -12\,eV, d = 3.14\,\AA) for $n=1$ one would obtain:
\begin{equation}
E_{kin}\sin^2 \theta <0 \quad .
\end{equation} 
Consequently, the first order Bragg peak from surface parallel planes is not visible in electron diffraction experiments.
By contrast, due to the opposite sign the first order reflection is observable in positron diffraction e.g.\,as demonstrated for the respective rocking curves calculated for Si(111) by dynamical diffraction theory \cite{Ich92}.

The different sign of the  potential energy  leads to an additional important consequence: 
For positrons total reflection occurs for small incident angles whereas for electrons it is not observed (see \sFF{Totalrefl}).
The critical angle for total reflection of positrons at  a surface is 
\begin{equation}
\theta_C=\arcsin\sqrt{eU_0/E_+} \quad.
\end{equation}
For the Si surface $\theta_C$ amounts to 2$^o$ for 10\,keV positrons. 
The total reflection can also be expressed by the vertical momentum component of the incoming beam with an energy of E$_{+,\bot} = E_+\sin ^2\theta_C$.
Hence, total reflection becomes experimentally accessible without too much effort. 
The benefits from total reflection high-energy positron diffraction (TRHEPD) will be discussed in Section\,\ref{sec:HREPD}  .

\begin{figure}[htb]
\centering
\includegraphics[width=0.5\textwidth]{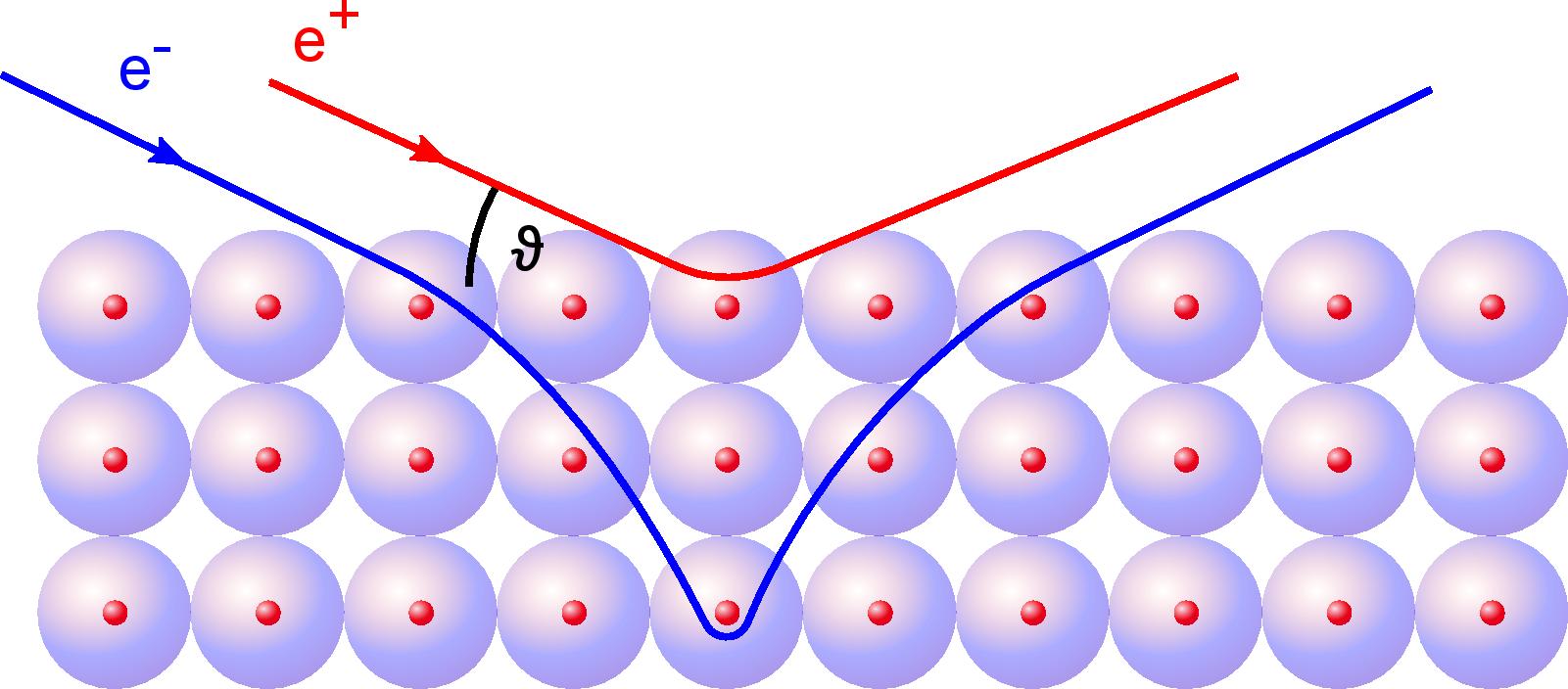} 
\caption{
For positrons total reflection occurs for small incident angles as they are repelled by the positively charged nuclei. 
} 
  \label{Totalrefl}
\end{figure}

In the past almost exclusively LEPD and (T)RHEPD have been applied for surface structure analysis using positrons. 
It is worth mentioning that positron diffraction patterns of the bulk of thin crystalline foils can also be obtained in transmission geometry.
Using a transmission positron microscope with 30\,keV positrons Matsuya et al. demonstrated that  positron diffraction patterns up to the 044 reflexes of a 10\,nm thin Au(100) single crystal foil could be observed \cite{Mat11}. 
However, compared to conventional electron diffraction using a state-of-the-art TEM it is not expected to benefit from the lower IMFP using positrons in transmission geometry.

In principle, due to its electrical neutrality Ps diffraction would offer an alternative surface sensitive tool.
The generation of mono-energetic Ps beams, however, is extremely demanding. 
Therefore, only one Ps reflection experiment has been reported so far.
By observing Ps reflected from the surface of LiF  it could be demonstrated that the Ps mean free path is extremely low ($<0.1\,$nm) \cite{Web88}.

In the following, LEPD and (T)RHEPD will be reviewed in more detail. 
Due to the aforementioned features of positron core-ion scattering, LEPD was expected to at least complement conventional LEED due to additional information leading to a higher level of accuracy in quantitative surface structure analysis.
Most recently, TRHEPD studies have attracted much attention due to the unique feature of total reflection of positrons and the availability of high-brightness positron beams of high intensity leading to short measurement times in the order of hours or below.

\subsection{First LEPD experiments}
\label{sec:LEPD}
From the beginning of the 1980's until the mid of 90's LEPD has attracted some attention since low-energy positron beams became available with sufficient brightness.
In the pioneering work by Rosenberg et al.\,Bragg peaks of elastically scattered positrons from a Cu(111) surface have been observed for the first time using a low-energy positrons \cite{Ros80}. 

Similar to LEED, at LEPD so-called I-V-curves are recorded, i.e.\,the intensity of a particular Bragg spot is measured by varying the  kinetic energy of the incoming positron beam.
Usually experimental and theoretically calculated  I-V-curves are compared for various diffraction beams in order to deduce the relative positions of the surface atoms. 
A calculation using the same code as for LEED but with a negative Coulomb contribution and without exchange term satisfactorily reproduced the measured I-V-curve \cite{Jon80}.
Although the deviation was found to be small, a comparison of three different potentials (i) keeping the exchange and correlation terms as applicable to electrons, (ii) no inclusion of exchange and correlation, and (iii) eliminating the exchange but retaining the correlation term yielded a slightly better agreement of the calculated and measured I-V-profiles using the latter model \cite{May87}.
Weiss et al.\,extended the study on Cu(111) reported in \cite{Ros80} to the Cu(100) surface and presented a comprehensive comparison of calculated and experimental I-V-curves obtained by both LEED and LEPD \cite{Wei83}.
By modifying the dynamical calculations, which are usually applied for LEED, the calculation procedure for LEPD comprises essentially four steps:  
(i) computation of the  atomic charge densities and atomic potentials, 
(ii) construction of the muffin-tin potential by superposition of atomic potentials, reverse the sign of the potential and eliminate the exchange correlation term,
(iii) calculation of the phase shifts for positron scattering 
inside an atomic sphere, and 
(iv) application of a multiple-scattering formalism for the calculation of the I-V-curves.
For further reading, the details of the theoretical calculations are described in-depth in Ref.\,\cite{Wei83} and references therein.

In 1985, first two-dimensional positron diffraction pattern could be recorded using a brightness enhanced positron beam  \cite{Fri85}.
It was demonstrated that at the W(110) surface the absolute scattering probability for the specular positron beam yields 2\%, and hence a factor of two higher than for electrons. 
Applying LEPD to ionic crystals,  the atomic plane spacings at air-cleaved and vacuum cleaved (100) surfaces of NaF and LiF were shown to be  equal to the bulk values within 5$\cdot 10^{-3}$ \cite{Mil85}.

\subsubsection{LEPD and LEED on CdSe surfaces}
\label{sec:CdSe}
A comparative study using LEPD and LEED for the determination of the relaxed atomic structure of the  (10$\bar 1$0) and (11$\bar 2$0) cleavage faces of CdSe was performed by Horsky et al. \cite{Hor89, Hor92}. 
CdSe belongs to the II-VI compound semiconductors exhibiting Wurtzite-structure where the two cleavage faces (10$\bar 1$0) and (11$\bar 2$0) are electrically neutral  since the topmost atomic plane  consists of equal numbers of Cd and Se atoms.
The fundamental characteristics of these relaxed surfaces is that the local coordination of the atoms is changed significantly while the bond lengths are conserved. 
The (10$\bar 1$0) and (11$\bar 2$0) surfaces relax by bond-length-conserving rotations of the topmost Cd-Se dimers 
and of the topmost Cd-Se-Cd and Se-Cd-Se triplets, respectively \cite{Hor92}.

\begin{figure}[htb]
\centering
\includegraphics[width=0.5\textwidth]{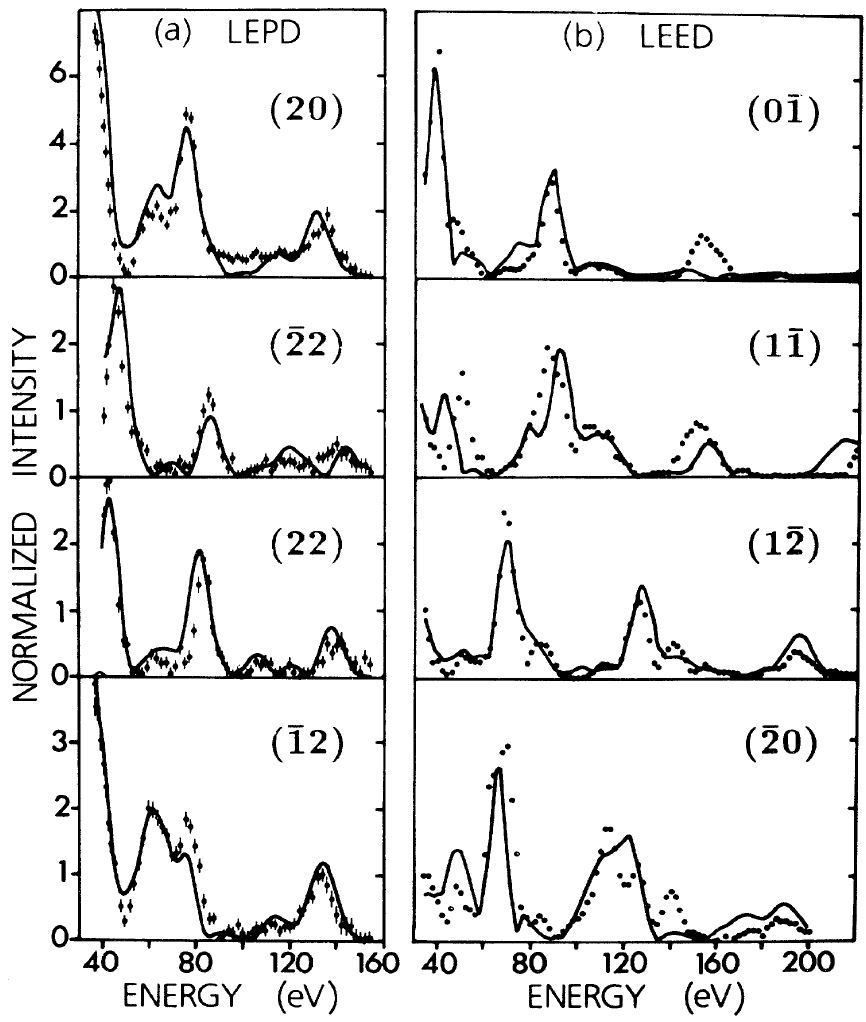} 
\caption{Experimental I-V-profiles (dots) for four beams diffracted from the CdSe(11$\bar 2$0) surface measured with (a) LEPD and (b) LEED (intensity in arbitrary units) and calculated profiles (lines) utilizing the best-fit structures
(figure from \cite{Hor89}).
 } 
  \label{CdSe_IV_Hor89}
\end{figure}

The I-V-profiles for four beams diffracted from the CdSe(11$\bar 2$0) surface obtained with LEPD and LEED are depicted  in \sFF{CdSe_IV_Hor89}. 
The reproducibility of the LEPD results was shown by using different CdSe samples \cite{Hor92}.
The agreement between calculated and experimental data was found to be significantly better for LEPD than for LEED leading to a reliable reconstruction of the atom positions at the surface \cite{Duk89}.
The goodness of the fit between calculated and experimental I-V profiles is commonly characterized by the so-called x-ray R-factor R$_x$. 
For both surfaces CdSe(10$\bar 1$0) and CdSe(11$\bar 2$0) the quality factor R$_x$  is plotted  as a function of the bond rotation angle $\omega$ in \sFF{CdSe_Rx_Hor89}.
As can be seen lower R$_x$  values are found for LEPD, i.e.\,the structure model fits better to the experimental results.

\begin{figure}[htb]
\centering
\includegraphics[width=0.5\textwidth]{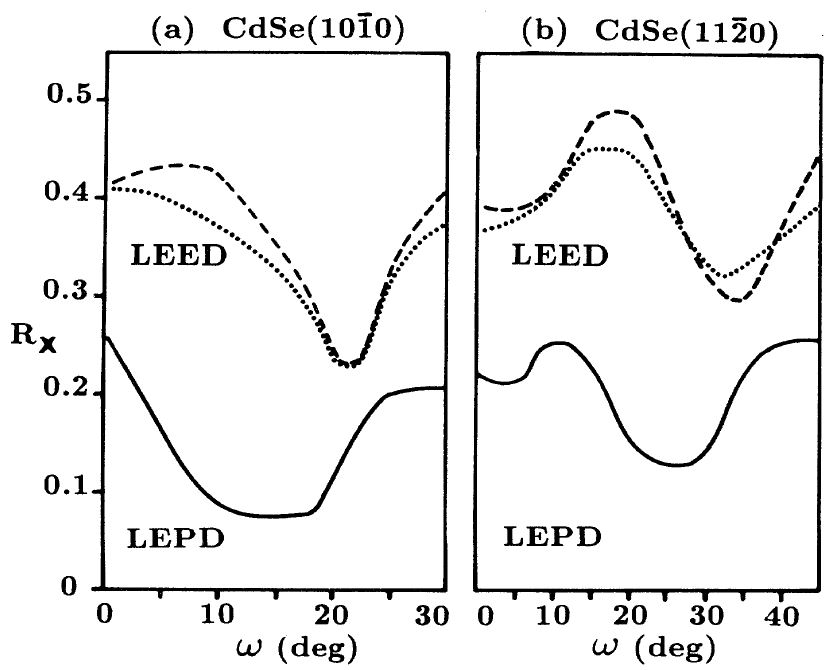} 
\caption{Quality factor R$_x$ as a function of the bond rotation angle $\omega$ for the (a) CdSe(10$\bar 1$0) and (b) CdSe (11$\bar 2$0) surfaces obtained for LEPD (solid lines) and from two different LEED measurements (dotted and dashed lines; 
figure from \cite{Hor89}).
 } 
  \label{CdSe_Rx_Hor89}
\end{figure}

It has to be emphasized that the energy-dependent elastic scattering cross section for electrons between the Cd and Se differs by up to a factor of four, whereas for positrons the difference is less than 10\%. 
The low elemental sensitivity of the positron is attributed to the positron ion-core interaction and the centrifugal barrier, which both are repulsive, whereas the effect of the two terms is opposed for electrons. 
For this reason, the correlation of the positron with core-electrons is insignificant for the calculation of diffracted positron intensities in LEPD \cite{Jon80}.

\subsubsection{Surface relaxation at GaAs and InP}
\label{sec:GaAsAndInP}
The surface structure of GaAs(110) is used as a show-case for the calculation of  the I-V-profiles by means of a dynamical multiple scattering model \cite{Les92}.
A significantly better agreement between  the intensity line shapes of theory and experiment was found for LEPD compared with LEED.
Applying this model, best-fit surface geometries for GaAs(110) and InP(110) were found by almost bond-length-conserving top-layer and second-layer rotations  \cite{Che93}.
Similar to previous studies the R-factor R$_x$ obtained in the LEPD structure calculation was a factor of two better than that in the comparable LEED analysis.
The more accurate I-V profiles calculated for LEPD is mainly attributed to the repulsive Coulomb potential experienced by the positrons when scattering from the ion cores.
However, systematic uncertainties might also influence the accuracy of the comparison of measured and calculated LEED and LEPD intensities. 
The changes of the surface composition or morphology as well as the beam parameters such as variation of the beam intensity and accuracy of  the angles of incidence have been discussed by Chen et al.\,\cite{Che93}.
 
It is important to note that the positron scattering factors are almost the same for the anions and cations leading to a negligible contrast between the according sublattices \cite{Che93}.
Hence, LEPD is particularly suited to infer the top-layer relaxations in binary semiconductors.
However, as pointed out by Joly  \cite{Jol94} LEPD is also very sensitive to the electronic configuration of semiconductor surfaces. 
The agreement between experiment and theory was found to be excellent by taking into  account the directional dependency of the covalent bonding present in semiconductors as well as empty or filled dangling orbitals at the surface. 
 At the relaxed GaAs(110) surface a charge transfer from the Ga dangling orbital to the As one could be inferred.
In general, the 3D electronic density of the GaAs(110) surface was found to be consistent with band structure calculations and ultraviolet photo-electron spectroscopy (UPS) \cite{Jol94}.

\subsection{Reflection high-energy positron diffraction (RHEPD)}
\label{sec:HREPD}

\subsubsection{Principle and features of RHEPD}
\label{sec:PrincipleAndFeaturesOfRHEPD}

The reflection high-energy electron diffraction (RHEED) is a well established tool for the investigation of surface structures. 
In most applications RHEED is used for \textit{in-situ} monitoring the epitaxial growth of thin layers. 
In contrast to LEED, the sample can easily be heat treated or manipulated taking advantage of the space available above the specimen.

The counterpart of RHEED is RHEPD by using positrons with a kinetic energy of typically 10-20\,keV to observe the diffraction pattern generated by positron small angle scattering at a surface.
In principle, diffraction patterns obtained via both,  RHEPD and RHEED, show similar features since the diffraction spots essentially result from the two-dimensional surface crystal described by the reciprocal lattice rods. 
Besides 2D diffraction patterns, the so-called rocking curve, i.e.\,the intensity of a specular reflection spot as a function of the incident glancing angle, is recorded. 
The surface structure, but also its roughness and cleanliness can be inferred from the diffraction experiment.
Usually, by taking into account the symmetry of the diffraction pattern an arrangement of surface atoms is assumed for the calculation of a theoretical diffraction pattern. 
The comparison of the experimental and theoretical results allows the deduction of the surface structure.

According to the aforementioned  distinction between positron and electron diffraction such as particle ion-core interaction and different crystal potential several  specific differences can be observed.
In diffraction patterns obtained by RHEED so-called Kikuchi lines, which arise from multiple-scattering of electrons, are  commonly observed, whereas they do not appear in RHEPD \cite{Hyo14}.
However, the most prominent feature is the total reflection of positrons at surfaces.
RHEPD near the critical angle is especially sensitive to the topmost atomic surface layer whereas at the critical angle for total reflection in X-ray diffraction (XRD), which is usually less than 0.2$^o$ \cite{Hyo14}, the penetration depth of the photons into the sample still amounts to a few nanometers \cite{Kaw03a}. 
For example, it was demonstrated that atoms in the bulk do not contribute to the diffraction pattern from a Si(111)-7$\times$7 reconstructed surface for the total reflection condition \cite{Hyo14}.
In order to highlight the uniqueness and the advantage of RHEPD in the total reflection mode Hyodo and Maekawa et al. \cite{Hyo14, Mae14} proposed to rename this method as \textit{total}-reflection high-energy positron diffraction (TRHEPD).

As pointed out by Ichimiya \cite{Ich92} the theoretical description of TRHEPD intensities can be carried out using the same dynamical diffraction theory as applied for RHEED by accounting for the opposed sign of the charged particle.
Since the surface potential is significantly influenced by adsorbates the intensity of the rocking curve (in particular in the total reflection regime) is very sensitive to the coverage as calculated e.g.\,for (sub-)ML coverage of K on Si(001) \cite{Ich92}.
In general TRHEPD is considered as a powerful tool to reveal a large variety of surface characteristics. 
Hence the potential TRHEPD applications comprise the investigation of phase transitions (including surface melting), topological irregularities,  electronic excitations and lattice vibrations in the topmost surface layer as well as tiny distortions of the surface structure, adsorbate atoms and the surface dipole potential of metals.

A drawback of RHEPD, compared to RHEED, might be the requirement of a high-intensity high-brilliant positron beam, since conventional positron beam setups with $\beta^+$ sources lead to very long measurement times.
A great step forward was the application of a positron remoderator to generate a brightness enhanced positron beam with a diameter of 0.5\,mm yielding $5\cdot 10^5$ positrons per second.
Compared to previous measurements at a beam energy of E$_+$ = 10\,keV with a spread of  $\Delta E\approx$40\,meV and $\Delta\theta\approx$12\,mrad the diffraction intensity has been enhanced by a factor of 60 enabling the observation of fractional-order spots in the higher Laue-zones with high signal-to-noise ratio \cite{Mae14, Hyo14}.
Moreover, the orientation of the sample could be performed in real time and typical recording times for a RHEPD pattern and for a full 00-spot  rocking curve could be reduced to 1\,h and 3\,h, respectively \cite{Hyo14}.
 
\subsubsection{First RHEPD experiments: H-terminated Si(111)}
\label{sec:FirstRHEPDStudies}
The first RHEPD experiment has been performed  on a hydrogen-terminated Si(111) surface by Kawasuso et al. \cite{Kaw98}.
The primary 20\,keV positron beam (3\,mm diameter, angular divergence of $<1^o$) was collimated (60\,mm long collimator with a hole of 1\,mm diameter) resulting in a final beam intensity of approximately 5000 positrons per second at the sample.
The diffraction pattern was detected by a microchannel plate (MCP) detector with a phosphor plane and CCD camera read-out within a typical data acquisition time of more than four hours.
It could also be shown that the specular intensity increases by decreasing the glancing angle, as it was expected due to the total reflection of positrons.

\begin{figure}[htb]
\centering
\includegraphics[width=0.6\textwidth]{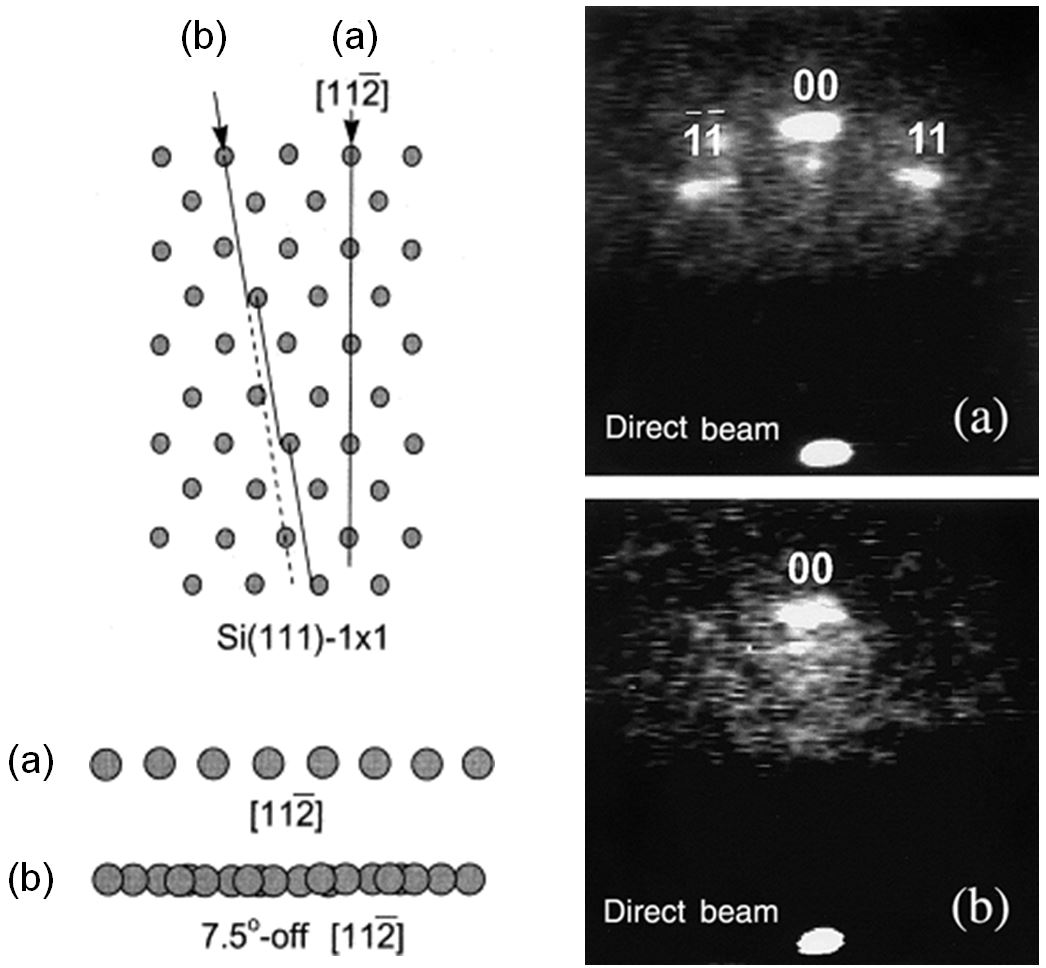} 
\caption{RHEPD on Si(111) observed with 20\,keV positrons. Left: Top view of the unreconstructed Si(111)-1$\times$1 surface and atomic positions of the surface seen from the (a) $\left[11\bar 2\right]$ direction (many-beam condition) and (b) 7.5$^o$ off from $\left[11\bar 2\right]$ direction (one-beam condition).
Right: RHEPD patterns recorded for (a) $\left[11\bar 2\right]$  incidence and (b) in one-beam condition
(figure adapted from \cite{Kaw00}).
 } 
  \label{RHEPD_Si_1_Kaw00}
\end{figure}

\begin{figure}[htb]
\centering
\includegraphics[width=0.6\textwidth]{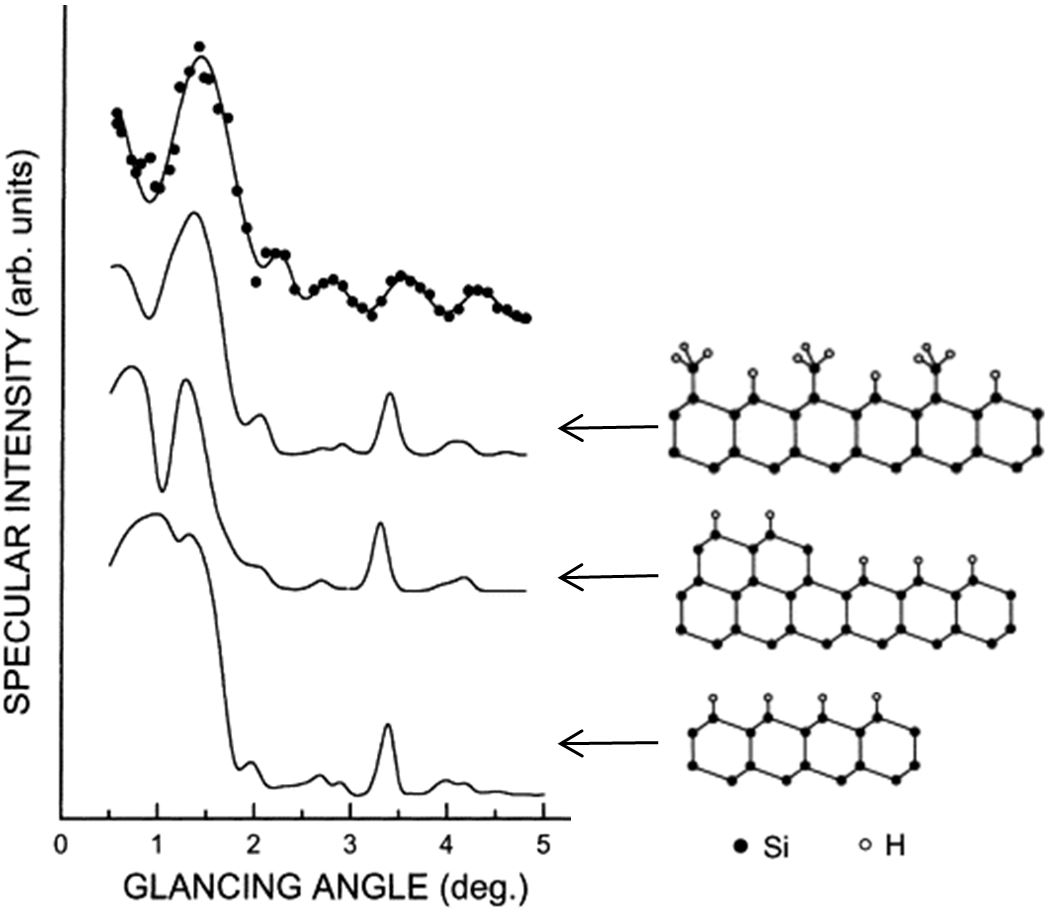} 
\caption{
Rocking curves at $\left[11\bar 2\right]$  incidence on Si(111).
Experimental data (dotted top line) and rocking curves calculated for different structural models of a hydrogen-terminated Si(111) surface (lines from bottom to top): Si(111) ideally H terminated, 
monohydride Si(111) surface containing bilayer roughness, and 
monohydride Si(111) surface with remaining SiH$_3$ molecules. 
The latter model fits the data best
(figure adapted from \cite{Kaw00}).
 } 
  \label{RHEPD_Si_2_Kaw00}
\end{figure}

In general two different geometries with regard to the incoming positron beam and the sample symmetry are considered.
In the so-called ``one-beam condition'' the incident beam points in the less symmetric direction with respect to the atoms in the horizontal lattice planes.
Since the beam direction is intentionally off any symmetric axes the low-order in-plain diffraction is suppressed, and hence the transverse symmetry virtually vanishes as sketched in \sFF{RHEPD_Si_1_Kaw00}.
Consequently, in this simple case only the diffraction between surface parallel lattice planes contributes to the diffraction spots. 

Contrary, as shown in \sFF{RHEPD_Si_1_Kaw00} by using the ``many-beam condition'', i.e. the positron beam impinges along the  $\left[11\bar 2\right]$ direction (case (a)), diffraction spots appear, which contain the full surface symmetry.
The absence of fractional order spots in the recorded diffraction patterns was attributed to the 1$\times$1 structure of the relaxed Si(111) surface.
Since the positions of the measured and calculated peaks were found to be in good agreement leading to the conclusion that the absolute value of the crystal potential is similar for positrons and electrons.

In order to reveal the structure of a H-terminated Si(111) surface the rocking curves were compared with those obtained by dynamical calculation for different structural models as plotted in \sFF{RHEPD_Si_2_Kaw00}.
By comparing the features of the respective rocking curves best agreement was found for a monohydride Si(111) surface with SiH$_3$ molecules on the surface \cite{Kaw00}.
In subsequent studies on Si(111) the first Laue zone could be observed as well by applying a positron beam with superior brightness \cite{Kaw02}.

\subsubsection{Surface dipole barrier}
\label{sec:SurfaceDipoleBarrier}
Considering  the total reflection properties of positrons it is intriguing to apply TRHEPD as a unique technique for the direct measurement of the surface dipole barrier of metal surfaces.
LEPD is assumed to be less appropriate for such a measurement due to the comparable high inelastic scattering cross section of low energy positrons (few 100\,eV).
The positron reflectivities of Au, Ni and Ir(001) surfaces have been measured  by variation of the glancing angle, i.e.\,as a function of the positron surface normal energy \cite{Kaw00}.
A smaller reflection intensity at angles significantly below the critical angle is attributed to positron capture in the surface potential and to Ps formation.
However, the sudden drop in positron reflectivity above E$_\bot >$D is attributed to the positron repulsion due the surface dipole barrier.
Although the experimental results roughly agree with the expectations deduced from the jellium model significant deviations (particularly observed for Au and Ni) have to be explained by more subtle models \cite{Kaw00}. 

\subsubsection{RHEPD on Si(111)-7$\times$7 as ideal example}
\label{sec:RHEPDOnSi1117Times7}

In general, the well-known structure of the Si(111)-7$\times$7 reconstructed surface has been intensively studied by a bunch of methods such as LEED, RHEED, STM and XRD, and hence serves as model system to benchmark surface analyses with RHEPD (see e.g.\,\cite{Hay06} and references therein).
The application of a positron beam with increased brightness allowed to observe clearly the 1/7$^{th}$ Laue zone of the reconstructed Si(111)-7$\times$7 surface at a glancing angle of 2.1$^o$ \cite{Kaw04}.
The analysis of the rocking curve associated with the Si(111)-7$\times$7 surface revealed that the mean distance between the adatom and stacking fault layers was found to be slightly larger than that expected from the electron diffraction experiments \cite{Kaw03a, Kaw03b}.
In order to obtain best agreement of experimental and calculated RHEPD patterns the surface potential was optimized leading to an increased potential of the adatoms attributed to the charge transfer from adatoms to rest-atoms \cite{Hay06}. 
RHEPD was also applied to investigate the phase transition between 2$\times$1 and c(4$\times$2) at around 200\,K  on the Si(001) surface. 
However, the small temperature dependent differences in the rocking curve could not reveal significant differences of the atom positions at the surface. This result was explained by the limitation of the beam coherence length \cite{Hay05}.

Using the Si(111)-7$\times$7 reconstructed surface as a model system Fukaya et al. extensively investigated the contribution of the scattering atoms dependent on the number of the surface layers for both RHEED and RHEPD \cite{Fuk14a}.
The diffraction patterns were calculated for two glancing angles one below ($\theta=1.3^o$) and the other ($\theta=2.1^o$) greater than the critical angle for total reflection of positrons and compared with the RHEED and RHEPD results, respectively.
As depicted in \sFF{Si(111)TRHEPD_1_Fuk14a} the calculation was carried out stepwise for the (a) adatoms, (b) adatoms and first layer, and (c) including the second layer as well. 
\sFF{Si(111)TRHEPD_2_Fuk14a} shows the goodness (R$^*$) of these results compared to the calculated diffraction pattern of the bulk crystal.
It could be demonstrated that the pattern calculated  for a two-dimensional single sheet of the Si(111)-7$\times$7 adatom configuration already displays the main features of the  THREPD pattern (case (1)).
As expected from the sparse density of the adatoms, inclusion of  the atoms in the first surface layer into the calculations increases the agreement whereas the contribution of the atoms in the second surface layer is almost negligible.
The RHEPD pattern at  $\theta=2.1^o>\theta_c$, case (2), is very well reproduced by inclusion of the second layer without further significant contribution of the bulk.

\begin{figure}[htb]
\centering
\includegraphics[width=0.7\textwidth]{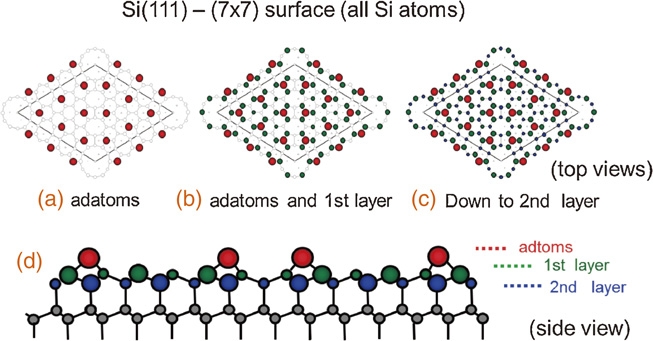} 
\caption{
Schematic drawing of the Si(111)-7$\times$7 surface structure. For clarity, Si atoms are depicted as circles of different color and size. The calculation of the diffraction pattern was performed for (a) the arrangement of the adatoms (red), (b) the adatoms and first surface layer (green), and (c) including the second surface layer (blue; figure from \cite{Fuk14a}).
 } 
  \label{Si(111)TRHEPD_1_Fuk14a}
\end{figure}

\begin{figure}[htb]
\centering
\includegraphics[width=0.5\textwidth]{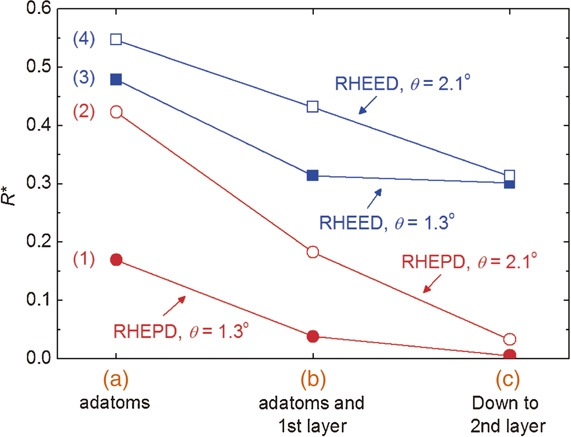} 
\caption{
Values of the residual factor R$^*$ defined as the modulus of the difference between calculated diffraction pattern for a specified number of surface layers and the bulk for RHEPD and RHEED at the glancing angles shown. 
The assumed 2D structures for the calculations are denoted by (a), (b), and (c) according to those in \sFF{Si(111)TRHEPD_1_Fuk14a}.
Note that for THREPD, case (1), inclusion of the adatoms and the first surface layer already reproduces very well the diffraction pattern calculated for the bulk crystal
(figure from \cite{Fuk14a}).
 } 
  \label{Si(111)TRHEPD_2_Fuk14a}
\end{figure}

\subsubsection{Determination of surface structures}
\label{sec:DeterminationOfSurfaceStructures}

\paragraph{Superstructures induced by metal adatoms on Si(111)-$\sqrt{3}\times\sqrt{3}$-Ag}
The Si(111)-$\sqrt{3}\times\sqrt{3}$-Ag surface  can be regarded as a model 2D metal system.
The arrangement of Ag atoms on the Si(111)-$\sqrt{3}\times\sqrt{3}$-Ag surface has been studied by recording RHEPD rocking curves in one-beam condition \cite{Fuk05}.
The atomic height of the topmost Ag triangle was found to be 0.77\,\AA{} consistent with XRD experiments and very similar to the interlayer spacing of 0.78\,\AA{} of the double layer in the bulk Si.
In this system the more complex $\sqrt{21}\times\sqrt{21}$ super-lattice structure is induced by additional adsorption not only of noble metal atoms such as Ag or Au but also by alkali adatoms such as Cs on a Si(111)-$\sqrt{3}\times\sqrt{3}$-Ag surface.
It was shown that the formation of the new superstructure is accompanied by a drastic increase of the electrical conductivity
Usually the surface structure was analyzed by  STM and surface X-ray diffraction (SXRD). 
However, various models with regard of the $\sqrt{21}\times\sqrt{21}$ superstructure have been discussed.
For further details of both, theoretical and experimental studies see e.g.\,Ref.\cite{Fuk12} and references therein.
The  complementary RHEPD experiments mainly performed by Fukaya et al., the gained insights in the structure and the underlying physics of the Si(111)-$\sqrt{21} \times\sqrt{21}$ system will be discussed in the following. 

For sample preparation, first, 1\,ML of Ag atoms was deposited on a Si(111)-7$\times$7 surface at high temperature (770\,K) in order to obtain a Si(111)–$\sqrt{3}\times\sqrt{3}$-Ag structure. 
After cooling to 110\,K additional Ag atoms were deposited until the $\sqrt{21} \times\sqrt{21}$ spot intensities measured by RHEED reached a maximum according to 0.14\,ML of Ag \cite{Fuk06a}.
A similar sample preparation on Si(111)-$\sqrt{21}\times\sqrt{21}$-Ag surfaces was carried using Au or Cs atoms, respectively, to obtain the $\sqrt{21}\times\sqrt{21}$ super-lattice structure \cite{Fuk07, Fuk12}. 

For different species of adatoms (Ag, Au and Cs) the RHEPD diffraction pattern and the measured rocking curves have been analyzed using the dynamical diffraction theory.
The determination of the atom positions at the surface is conducted in two steps.
First, the vertical component of the atomic positions is deduced from the rocking curves in one-beam condition. 
In this way, only the atomic density in the plane is considered  yielding  the vertical position of the (ad-)atoms. 
In the second step, the found vertical components are fixed and the rocking curves in many-beam condition are analyzed in order to obtain the in-plane components \cite{Fuk14b}.

\begin{figure}[htb]
\centering
\includegraphics[width=0.5\textwidth]{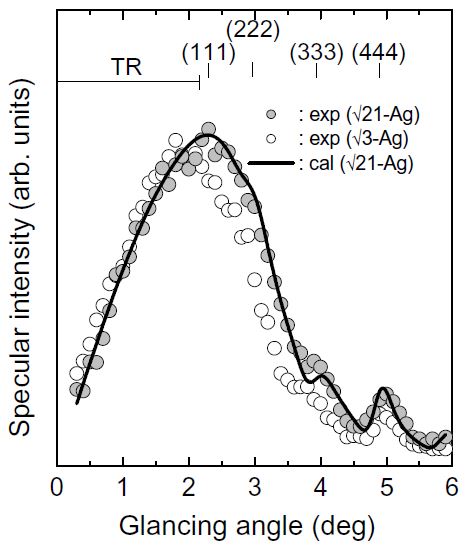} 
\caption{
Measured RHEPD rocking curves for specular spots from the Si(111)-$\sqrt{3}\times\sqrt{3}$-Ag (open symbols) and Si(111)-$\sqrt{21}\times\sqrt{21}$-Ag (gray symbols) surfaces in one-beam condition. 
A fit of the calculated rocking curve from the Si(111)-$\sqrt{21}\times\sqrt{21}$-Ag surface to the measured data yields a height of 0.53\,\AA\,of the additional Ag atoms at a coverage of 0.14\,ML. 
The positions of the Bragg reflections are labeled on the top; TR denotes the total reflection region
(figure from \cite{Fuk06a}).
 } 
  \label{SiAg_1_Fuk06a}
\end{figure}

\begin{figure}[htb]
\centering
\includegraphics[width=0.5\textwidth]{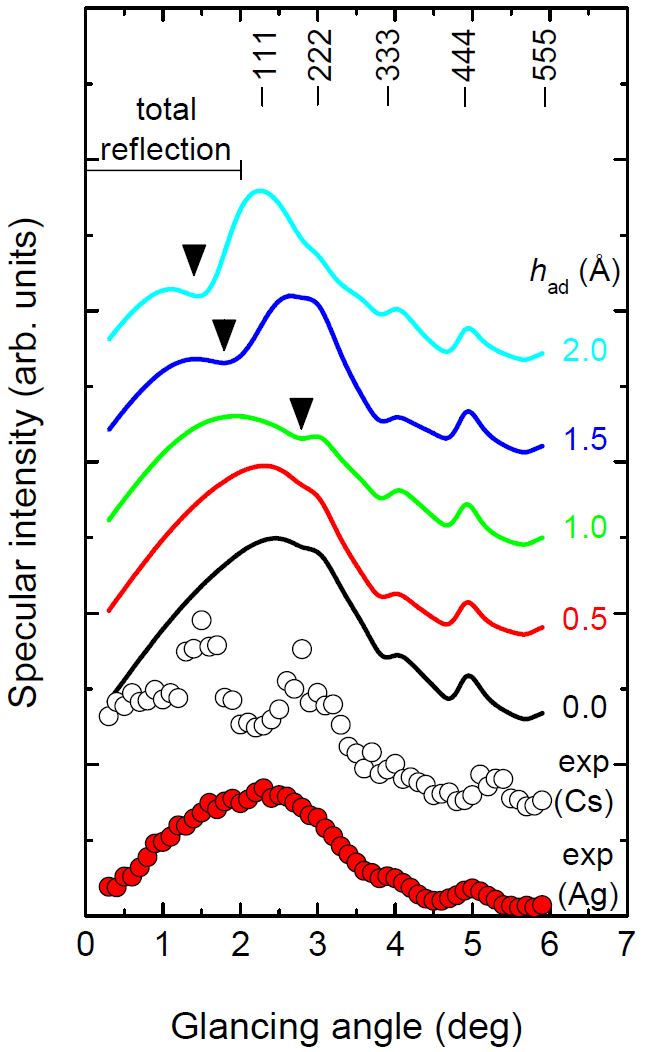} 
\caption{
RHEPD rocking curves for specular spots from the Si(111)-$\sqrt{21}\times\sqrt{21}$-Ag surface in one-beam condition. 
The experimental data (red dots) are compared with rocking curves calculated with various heights h$_{ad}$ of the  Ag adatoms with respect to the underlying Ag layer (solid lines). 
The rocking curve measured  for Si(111)-$\sqrt{21}\times\sqrt{21}$-(Ag,Cs) show a very dissimilar shape (open circles);
(figure from \cite{Fuk14b}).
 } 
  \label{SiAgCs_1_Fuk14b}
\end{figure}

In the RHEPD pattern, the formation of the $\sqrt{21}\times\sqrt{21}$-Ag structure after additional deposition of 0.14\,ML Ag could be confirmed reliably \cite{Fuk06a}.
\sFF{SiAg_1_Fuk06a} shows the RHEPD rocking curves from the Si(111)-$\sqrt{3} \times\sqrt{3}$-Ag and Si(111)-$\sqrt{21} \times\sqrt{21}$-Ag surfaces, respectively. 
The incident azimuth is set at 7.5$^o$ off from the $\left[11\bar 2\right]$ direction (one-beam condition).
Note the excellent agreement between the calculated rocking curve and the experimental data.

The results of the calculated rocking curves assuming various heights (h$_{ad}$) of the  Ag adatoms with respect to the underlying Ag layer are plotted in \sFF{SiAgCs_1_Fuk14b} illustrating the high sensitivity of RHEPD.
The shift of an appearing dip structure (marked by black triangles) towards lower glancing angles with increasing distance of adatom and the substrate Ag layer is attributed to the interference of the positron waves reflected from the adatoms and the layer below \cite{Fuk06a, Fuk14b}.
For comparison, the measured  rocking curve  for Si(111)-$\sqrt{21}\times\sqrt{21}$-(Ag,Cs) is displayed as well  featuring a very different shape (see below).
After the optimization of the calculation,  the height of the additional Ag atoms was determined to 0.53\,\AA\,, and the number of the additional Ag atoms found to be three in the $\sqrt{21}\times\sqrt{21}$ unit cell.
The value of 0.53\,\AA, being smaller than predicted by first-principles calculations, is explained by charge transfer from the additional Ag atoms to the bulk leading to a smaller radius of the ionized Ag atom and a reduced bond length \cite{Fuk06a}.

\begin{figure}[htb]
\centering
\includegraphics[width=0.5\textwidth]{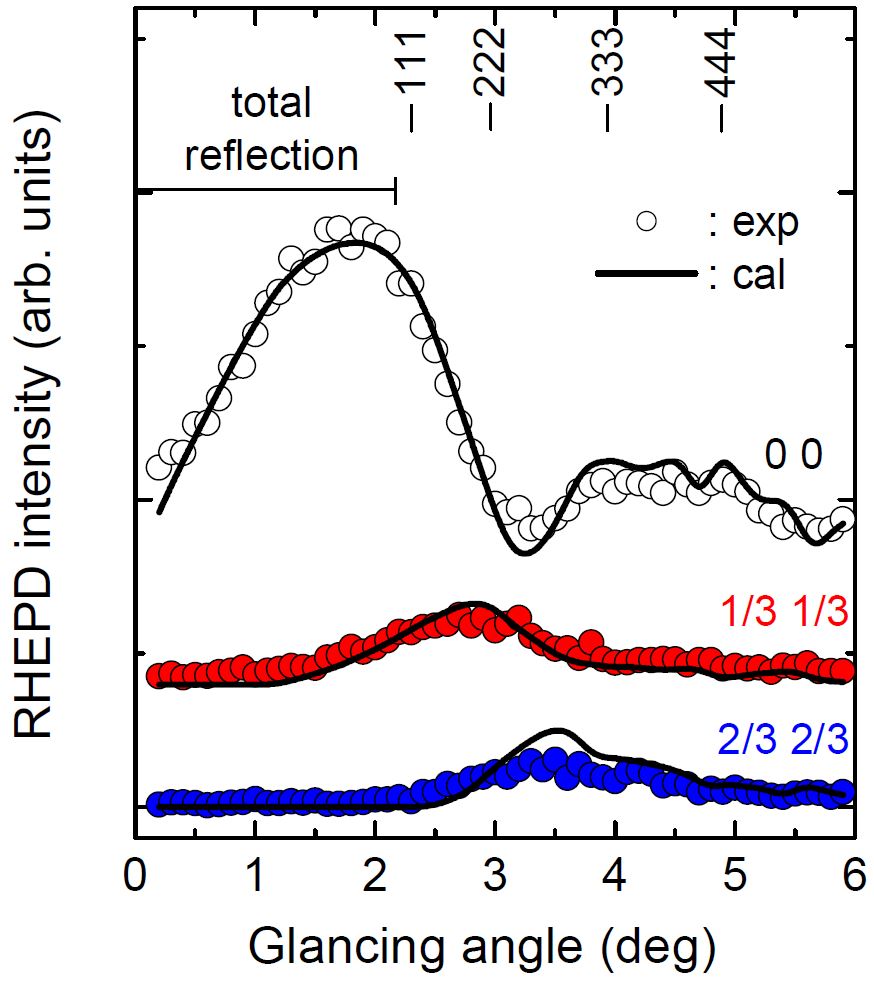} 
\caption{
RHEPD rocking curves along the $\left[11\bar 2\right]$ direction (many-beam condition) for the indicated diffraction spots from the Si(111)-$\sqrt{21}\times\sqrt{21}$-Ag surface.
The calculation with optimized structure parameters (lines) are in reasonable agreement with the experimental data (symbols) for all indicated spots on the 0$^{th}$  Laue zone
(figure from \cite{Fuk14b}).
 } 
  \label{SiAgCs_2_Fuk14b}
\end{figure}

\sFF{SiAgCs_2_Fuk14b} shows the rocking curves along the $\left[11\bar 2\right]$ direction (many-beam condition) for the Si(111)-$\sqrt{21}\times\sqrt{21}$-Ag surface.
Due to their much higher intensity only spots on the 0$^{th}$ Laue zone ((0\,0, 1/3\,1/3, and 2/3\,2/3)) were used for the analysis. 
It was found that all of the three Ag adatoms per unit cell are centrically located on top of large subjacent Ag triangles (see also \sFF{SiCsAuAg_Fuk12}).

In subsequent RHEPD studies the $\sqrt{21}\times\sqrt{21}$ super-lattice structure was stabilized by Au \cite{Fuk07} and Cs adatoms \cite{Fuk12}.
The intensity distribution in the diffraction pattern from the Si(111)$-\sqrt{21}\times\sqrt{21}$-(Ag,Au) surface was found to be very similar to that from the Si(111)-$\sqrt{21}\times\sqrt{21}$-Ag surface.
Analogous to the analysis of the Ag/Si(111) system the height of the Au adatom relative to the underlying Ag layer could be determined to 0.59\,\AA.
From that value the bond length between the Au and Ag atoms for the Si(111)-$\sqrt{21}\times\sqrt{21}$-(Ag,Au) surface was estimated to be 2.35\,\AA.
Hence, both species of adatoms Au and Ag lead to the same $\sqrt{21}\times\sqrt{21}$ super-lattice surface structure \cite{Fuk07}.
For the Si(111)-$\sqrt{21}\times\sqrt{21}$-(Ag,Cs) surface the height of the Cs adatom relative to the underlying Ag layer was determined to 3.04\,\AA\,and hence much larger than in the case of Ag or Au adatoms.
The Cs adatoms were found to arrange in a triangular lattice with an in-plane distance between the atoms of 10.12\,{\AA} \cite{Fuk12}.
The configurations obtained of the Si(111)-$\sqrt{21}\times\sqrt{21}$ superstructure stabilized either by Cs or by Au and Ag are displayed in \sFF{SiCsAuAg_Fuk12}.

\begin{figure}[htb]
\centering
\includegraphics[width=0.7\textwidth]{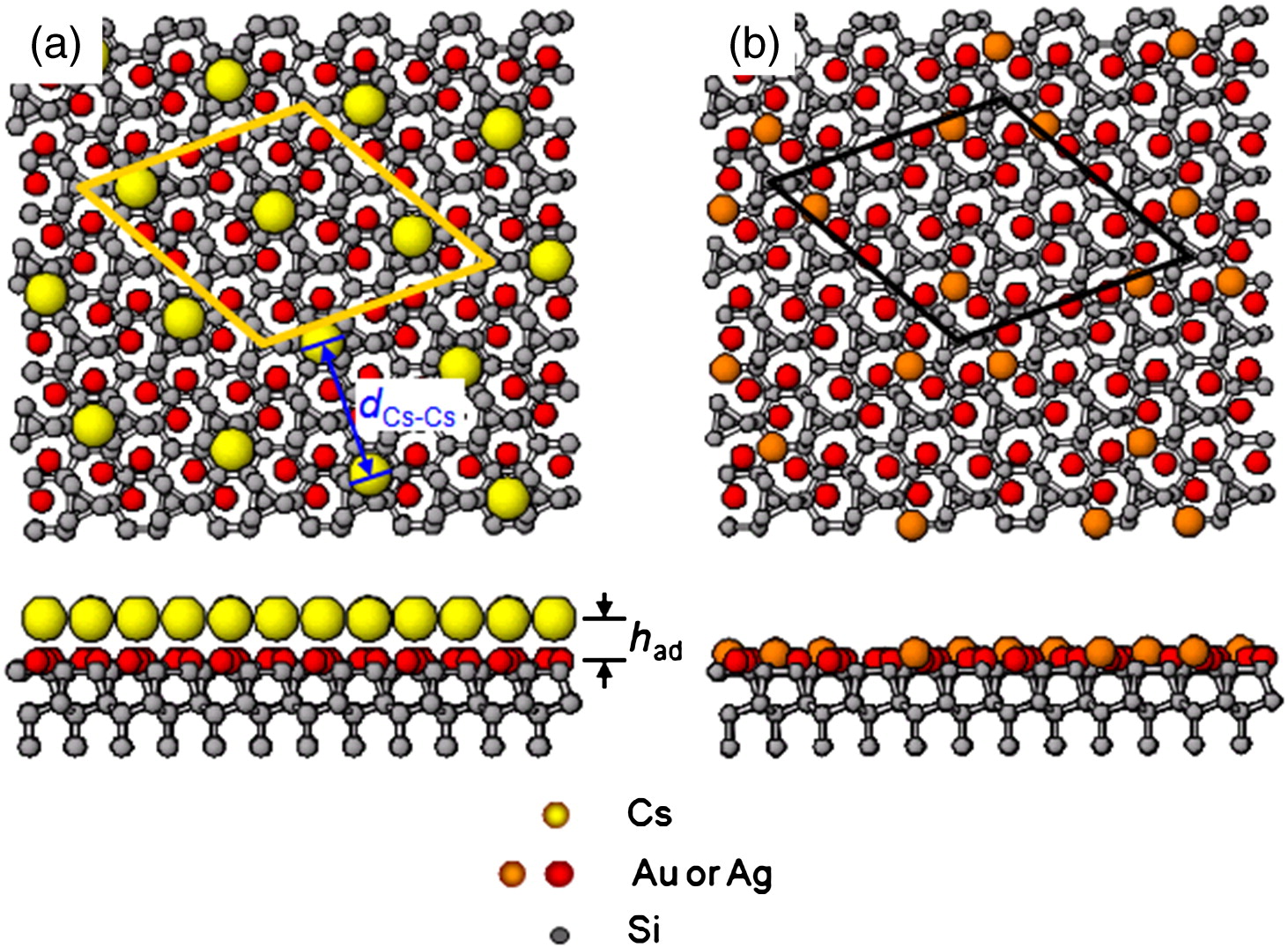} 
\caption{
Schematic structure of (a) the Si(111)-$\sqrt{21}\times\sqrt{21}$-(Ag,Cs) and (b) the Si(111)-$\sqrt{21}\times\sqrt{21}$-Ag and -(Ag, Au) surfaces. 
The rhombs represent a $\sqrt{21}\times\sqrt{21}$ unit cell, respectively
(figure from \cite{Fuk12}).
 } 
  \label{SiCsAuAg_Fuk12}
\end{figure}

Astonishingly, the $\sqrt{21}\times\sqrt{21}$ super lattice is formed for noble metals as well as for Cs adatoms although the structure of the Si(111)-$\sqrt{21}\times\sqrt{21}$-(Ag,Cs) surface differs significantly from that of the  Si(111)-$\sqrt{21}\times\sqrt{21}$-Ag and the Si(111)-$\sqrt{21}\times\sqrt{21}$-(Ag, Au) surface.
This fact is attributed to a largely different underlying physical behavior of the adatoms related to their electron
configurations of the \textit{s-d} hybridized orbitals (noble metal) or almost pure \textit{s} orbitals (alkali metal).
Following the interpretation of Fukaya et al.\,\cite{Fuk12}, first, Ag or Au adatoms are mutually and locally bound to form triangle islands.
With further increase of coverage $\sqrt{21}\times\sqrt{21}$ periodicity is eventually established due to the cohesive nature of the Ag and Au atoms. 
In the case of Cs, the adatoms are assumed to be highly mobile at low coverage.
At higher coverage the Cs atom is affected by the potentials of the neighboring adatoms that eventually leads to an ordered structure. 
This is supported by the large inter-atomic distance of the Cs adatoms of 10.12\,\AA\,being very close to the maximum
adatom distance of  10.16\,\AA\,of a triangular lattice consisting of 1/7 ML of adatoms (see \sFF{SiCsAuAg_Fuk12}).

\paragraph{Pb and Sn adsorbed on Ge(111)}
Both surfaces Pb/Ge(111)  and Sn/Ge(111) are considered as prototypical two-dimensional metal/semiconductor systems.
It is known, that the adsorption of about 1/3\,ML of Sn or Pb on Ge(111) leads to a $\sqrt{3}\times\sqrt{3}$ superstructure at room temperature.
The surfaces change from the $\sqrt{3}\times\sqrt{3}$ phase to a $3\times3$ periodicity below about 200\, K.
Although these systems have been extensively studied by STM, LEED and SXRD the underlying mechanism of the phase transition and the relative positions of the adatoms remained unclear.

In Sn/Ge(111) it was found that the positions of the Sn atoms exhibit two different heights.
Using RHEPD this height difference could be determined to 0.26\,\AA\,in agreement with theoretical calculations and previous experimental studies.
However, low-temperature RHEPD revealed that the found Ge(111)-$\sqrt{3}\times\sqrt{3}$-Sn structure is consistent with the model of one Sn atom being up and two Sn atoms being down.
In addition, it was found that the Sn adatoms lead to a large shift downwards of the relaxed first-layer Ge atoms of 0.35\,\AA\,compared to the ideal Ge bulk positions \cite{Fuk06b}.
A comparative RHEPD study at room temperature and at 60\,K on  Pb/Ge(111) revealed that the equilibrium positions of the surface atoms remain the same.
In addition, the model of one adatom being at the higher and two adatoms at the lower position was confirmed for both phases \cite{Fuk08b}.

\paragraph{Rutile-TiO$_2$(110)-1$\times$2}
Although it was known that in the system rutile-TiO$_2$(110) the 1$\times$1 phase transforms into 1$\times$2 at high temperature, different models describing the phase transition had been discussed for several decades.
Most recently, RHEPD studies could reveal the subtle atomic configurations at the topmost surface layer in this system.
Using the different modes, i.e. one-beam and many-beam condition, Mochizuki et al.\,demonstrated that TRHEPD allows to identify oxygen atoms on the topmost surface and to distinguish between the Ti atoms positioned at interstitial-vertical and at interstitial-horizontal sites.
As shown in \sFF{TiO2_Moc16} excellent agreement between the experimental rocking curve and the asymmetric Ti$_2$O$_3$ model with the interstitial-vertical sites was found.
Various structure models as well as the different relative positions and distances of the reconstructed rutile-TiO$_2$(110)-1$\times$2 surface are discussed in greater detail in Ref.\,\cite{Moc16}.

\begin{figure}[htb]
\centering
\includegraphics[width=0.5\textwidth]{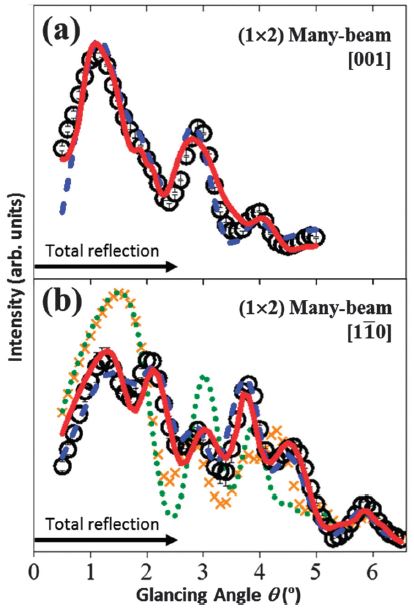} 
\caption{
TRHEPD rocking-curves for the rutile-TiO$_2$(110)-(1$\times$2) surface
(many-beam condition) along the (a) $\left[001\right]$, and (b) the $\left[1\bar 10\right]$ direction. 
The experimental data (open symbols) are compared with the theoretical results  for different models:
symmetric (blue dashed line) and asymmetric (red) Ti$_2$O$_3$ with interstitial-vertical sites;  
symmetric (green dotted) and asymmetric (orange crosses) Ti$_2$O$_3$ with interstitial-horizontal sites
(figure from \cite{Moc16}).
 } 
  \label{TiO2_Moc16}
\end{figure}

\paragraph{Graphitization of the SiC surface}
\label{sec:CSiCAndSi}
At a SiC(0001) surface it was demonstrated that flashing at 1020$^o$C yields an oxygen free and atomically flat surface using RHEPD.
Complementary AES measurements revealed that this procedure leads to the surface graphitization.
The shape of the rocking curve of the specular reflection obtained from the SiC(0001) surface was best modeled by a calculation based on the dynamical diffraction theory assuming a graphite coverage of 0.7\,ML and a interlayer distance of 3.2\,\AA.
This relatively large value was correlated to a weak van-der-Waals bonding of the graphite layer formed on the SiC(0001) surface  \cite{Kaw03a}.

\paragraph{Silicene on Ag(111)}
Analogous to graphene, silicene is a two-dimensional sheet of silicon, which can be synthesized on Ag(111).
It is noteworthy that the buckling of silicene (as shown in \sFF{Si_Ag(111)_1_Fuk13a})  is in strong contrast to the flatness observed for the case of graphene. 
RHEPD studies have been carried out to experimentally confirm the buckled structure predicted by theory.
For this purpose, RHEPD rocking curves of the specular spot were measured with the beam incident 13$^o$ off (one-beam condition) and in the $\left[11\bar2\right]$ direction (many-beam condition) in order to determine the vertical  and the in-plane components of the atomic positions, respectively \cite{Fuk13a}.
As shown in \sFF{Si_Ag(111)_2_Fuk13a} the best fit of the rocking curves calculated with the dynamical diffraction theory agree excellently with the experimental data for both the  Ag(111)-1$\times$1 surface and silicene on a Ag(111) surface.
For the Ag(111)-1$\times$1 surface no significant relaxations were found since the optimum interlayer distance of the first two layers was determined to 2.34\,\AA\,and hence being almost the same value as in the bulk (2.36\,\AA).
After the formation of the 4$\times$4 reconstructed structure of silicene on the
Ag(111) surface the measured rocking curve differs significantly. 
The calculated best-fit curve yields d=2.14\,\AA, and $\Delta$=0.83\,\AA, for the layer distances as depicted in \sFF{Si_Ag(111)_1_Fuk13a}.
The analogous analysis of the data in many-beam condition as function of the bond angles allowed their calculation to $\alpha=112^o$ and $\beta=119^o$. 
It is emphasized, that this RHEPD study not only confirmed the theoretically predicted 4$\times$4 model but also yielded accurate values of the atomic positions in the buckled configuration of silicene on the Ag(111) surface. 

\begin{figure}[htb]
\centering
\includegraphics[width=0.5\textwidth]{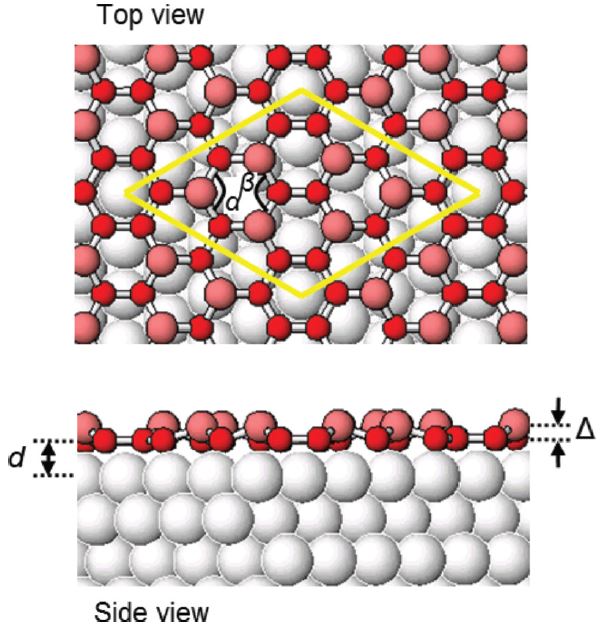} 
\caption{
Scheme of the structure of silicene (Si atoms depicted by red circles)  on a Ag(111) (Ag atoms depicted by  gray circles) surface. In the top view, $\alpha$ and $\beta$ denote the bond angles. In the side view, $\Delta$ and d denote the distances between the two top layers forming silicene, and that between the bottom layer of silicene and the Ag layer underneath, respectively
(figure from \cite{Fuk13a}).
 } 
  \label{Si_Ag(111)_1_Fuk13a}
\end{figure}

\begin{figure}[htb]
\centering
\includegraphics[width=0.5\textwidth]{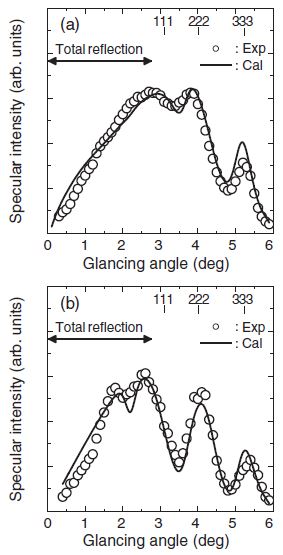} 
\caption{Experimental (open symbols) and calculated (lines) RHEPD rocking curves for (a) the Ag(111)-1$\times$1 surface and (b) silicene on a Ag(111) surface in the one-beam condition
(figure from \cite{Fuk13a}).
 } 
  \label{Si_Ag(111)_2_Fuk13a}
\end{figure}

\subsubsection{One-dimensional systems on surfaces }
\label{sec:1DSystemsOnSurfaces}

\paragraph{Surface band structure of In/Si(111)}
A novel approach to determine the electronic structure of (reconstructed) surfaces was demonstrated at the In/Si(111) surface \cite{Has07, Fuk08a}:
First RHEPD is applied to determine the surface structure with high reliability and then the found atomic positions are used as input  for the calculation of the electronic surface band structures.
Hence RHEPD should provide information on the electronic structure completely independent from results gained via  complementary techniques such as angle-resolved photoemission spectroscopy (ARPES) or scanning tunneling spectroscopy (STS). 

At room temperature, zigzag chains of In atoms are formed at the Si(111) surface displaying a Si(111)-4$\times$1-In superstructure whereas at low temperature a 8$\times$2 superstructure is formed. 
Dependent on the details of the underlying model this phase transition at  130\,K is generally attributed to be metal-insulator (semiconductor) transition.
The RHEPD measurements yielded two different vertical positions of the In atoms of 0.99\,\AA\, and 0.55\,\AA, respectively,
but the heights of the upper and lower In atoms do not depend on the temperature and are the same in both 4$\times$1 and 8$\times$2 phases \cite{Has07}.
However, significant changes are observed in the RHEPD data for the 4$\times$1 phase at room temperature and the 8$\times$2 phase at 60\,K. 
The analysis of the rocking curves showed that the zigzag chain structure of the In adatoms is transformed into a hexagonal structure at low temperatures. 
The first-principles calculations on the found surface structures revealed the appearance of a band gap at low temperature attributed to a metal-semiconductor phase transition. 
By comparing high-symmetry points in the surface band structure the values were found to be in good agreement with the ARPES and STS measurements \cite{Fuk08a}.

\paragraph{Pt-induced nanowires on Ge(001)}
Systems  consisting of self-assembled nanowires formed on a  semiconducting surface attracted much interest as they are considered to be a show-case for the investigation of fundamental physical properties restricted to one dimension. 
In a comprehensive study, RHEPD has been combined with STM and ARPES in order to investigate highly ordered Pt-induced nanowires on Ge(001) surface.
Previous STM and STS studies as well as ab-initio calculations lead mainly to three different structure models described by various arrangements of Pt and Ge dimers  (for details see \cite{Moc12}).
However, the atomic configuration and the mechanism of the phase transition in this system could  not bean clarified so far.
For the RHEPD study, the different structure models have been used to calculate the respective rocking curves in order to reveal the atomic arrangement at the Ge(001) surface.
Best agreement was found for a nanowire model with a Pt coverage of 0.75\,ML where the topmost adjacent Ge dimers are alternately buckled normal to the surface.
This surface structure is transformed into a flat symmetric structure at high temperature \cite{Moc12}.

\subsubsection{Surface excitations}
\label{sec:SurfaceExcitations}
Usually, electron diffraction is applied to investigate the thermal vibrational amplitude of surface atoms.
However, the main drawback in such experiments is that the information is gained as an average over several layers from the surface due to the large penetration depth of electrons.
For this reason, TRHEPD is particularly suited to study the vibrational amplitude of surface atoms.
Fukaya et al.\,performed temperature dependent RHEPD on a Si(111)-7$\times$7 surface to reveal both, the thermal vibrational amplitude of the adatoms and the surface Debye temperature.
The temperature dependences of the RHEPD intensities from the Si(111)-7$\times$7 surface reveal largely different Debye temperatures of the surface and the bulk which were found to be 290\,K and 600\,K, respectively.
The  change of the rocking curve with increasing temperature is explained by the enhanced thermal vibration of the adatoms at the surface. 
The thermal vibrational amplitude of the adatoms increases from  0.14\,\AA\,at room temperature to 0.23\,\AA\,at 873\,K.
This remarkable result reveals the significant softening of the adatom bonds of the Si(111)-7$\times$7 surface considerably below the 7$\times$7 to 1$\times$1 phase transition temperature \cite{Fuk04}.

The aforementioned topmost layer sensitivity of TRHEPD offers the possibility to study the excitation of phonons and electrons solely at the first surface layer. 
For this reason, the absolute reflectivity and energy-loss spectrum from a Si(111)-7$\times$7 surface have been investigated using RHEPD and RHEED for comparison \cite{Fuk09}.

The absolute reflectivities of positrons and electrons as a function of glancing angle (in one-beam condition) are shown in \sFF{Si111_Vib_1_Fuk09}.
For electrons the reflectivity is very low ($<4\%$) for all angles. 
The low reflectivity is attributed to the high penetration depth for electrons ($>$10\,\AA) due to their negative potential energy int the crystal. 
Therefore, electrons are lost due to multiple scattering inside the crystal and hence do not contribute to the specular beam. 
In contrast, the positron reflectivity is higher than 20\% in the total-reflection region ($\theta<2^o$), i.e.\,the surface normal energy of positrons is less than the crystal potential energy of Si (E$_\bot<$12\,eV).
As shown in \sFF{Si111_Vib_1_Fuk09} a calculation using the simplest case of a slab model for total reflection would yield 100\% reflectivity of positrons for small angles.
Including the 7$\times$7 surface structure in the dynamical-diffraction theory yields deviation from the total reflection from a simple flat surface, as expected. 
Furthermore, by considering both phonon and electronic excitations the experimental curve is excellently reproduced. 
Although the reflectivity of electrons is also well described by the calculation taking into account all the absorption potentials, the effect of surface excitations is much more pronounced in the positron reflectivity.
It has to be emphasized that the positrons are mainly reflected at the topmost layer where the thermal vibration amplitudes of atoms are significantly larger than those in the bulk. 
This effect explains that the absorption potential for phonon excitation is much larger (about a factor of 3.5) for positrons than for electrons. 

\begin{figure}[htb]
\centering
\includegraphics[width=0.5\textwidth]{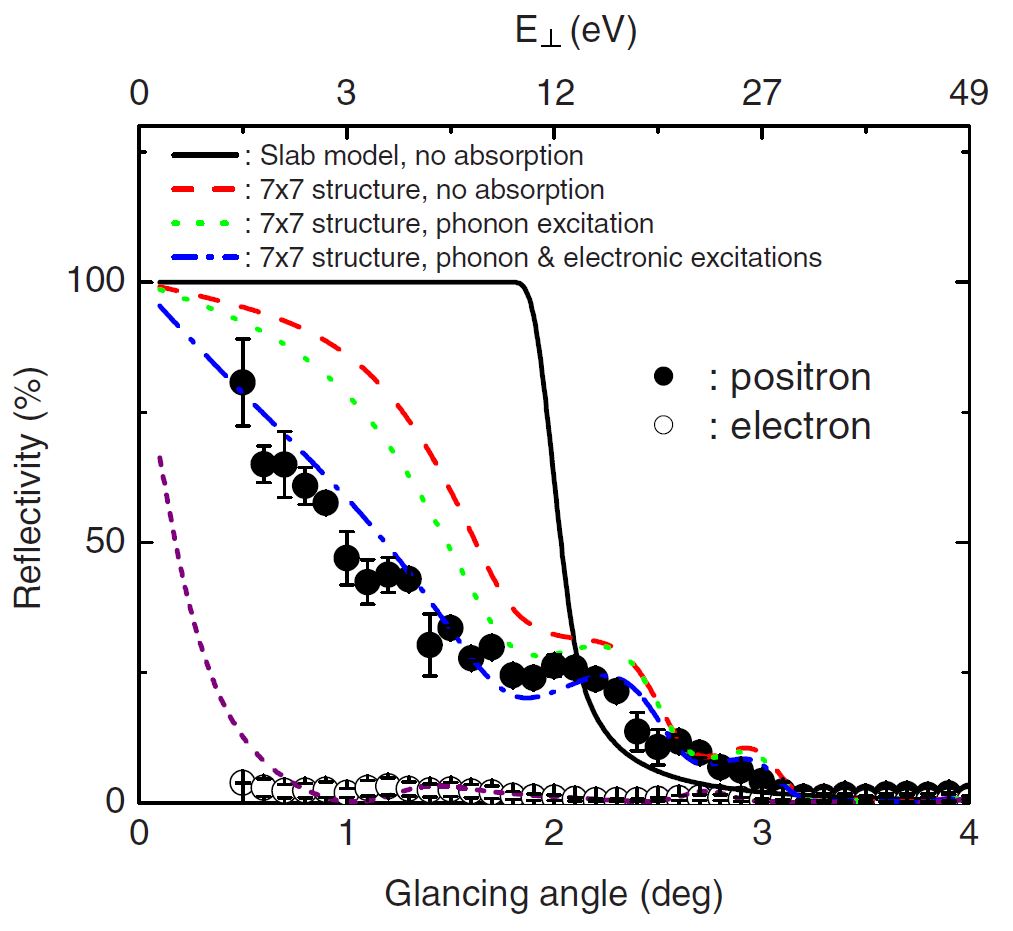} 
\caption{
Absolute reflectivities of positrons (black symbols) and electrons (open symbols) from the Si(111)-7$\times$7 surface as a function of the glancing angle. 
The lines represent the positron reflectivities calculated including relevant effects  as indicated.
For electrons the reflectivity is calculated with the 7$\times$7 structure and all absorption potentials included (short-dashed line). 
At the upper horizontal axis E$_\bot$ denotes the surface normal energy of the incoming beam
(figure from \cite{Fuk09}).
 } 
  \label{Si111_Vib_1_Fuk09}
\end{figure}

Further information about the surface excitation is gained by the energy loss spectra obtained for positrons and electrons.
Clear differences between electrons and positrons are observed by plotting the intensities of the specular beams as a function of energy loss (see \sFF{Si111_Vib_2_Fuk09}).
For electrons the intensity increases more or less continuously whereas the intensity rise of positrons exhibits distinct steps.
The differential curves reveal peaks which are attributed to elastic scattering and the excitation of upto five surface plasmons since the separation of the energy-loss peaks corresponds to the surface-plasmon energy of Si of 11\,eV.
It is striking that the energy loss is much more pronounced in the positron case (see \sFF{Si111_Vib_2_Fuk09}\,(b) and (c)). Apparently, compared to electrons, positrons excite more surface plasmons.
An analysis in greater detail revealed that this behavior can be attributed to the interaction length for positrons being twice as  high than for electrons.
This results in longer interaction times leading to multiple surface-plasmon excitations almost exclusively in the first surface layer since positrons do not penetrate the bulk \cite{Fuk09}.

\begin{figure}[htb]
\centering
\includegraphics[width=0.4\textwidth]{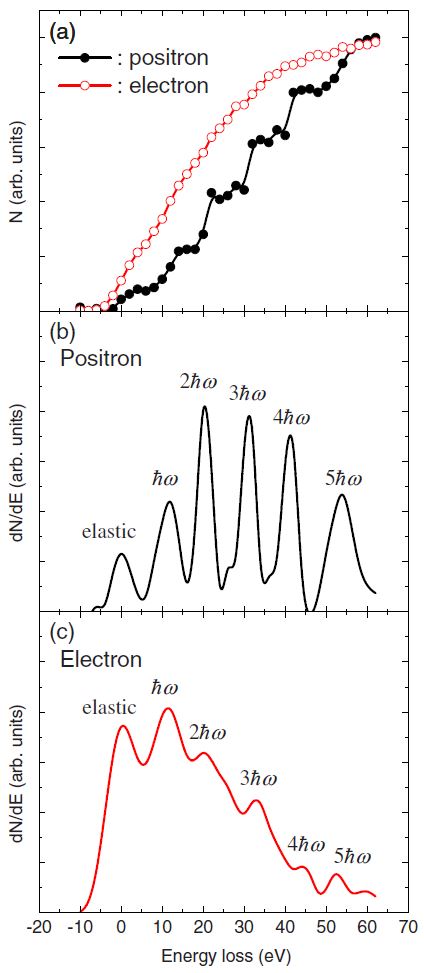} 
\caption{
(a) Intensities of the specular beams for positrons (glancing angle $\theta^+=1.5^o$) and electrons ($\theta^-=1.3^o$)
from the Si(111)-7$\times$7 surface and the resulting differential curves for (b) positrons, and (c) electrons as functions  of energy loss 
(figure from \cite{Fuk09}).
 } 
  \label{Si111_Vib_2_Fuk09}
\end{figure}

\newpage
\section{Positron annihilation induced Auger electron spectroscopy}
\label{PAES}
Positron annihilation induced Auger electron spectroscopy (PAES) is a non-destructive surface analysis tool exhibiting topmost layer sensitivity. 
In addition, the secondary electron background in the range of Auger-transition energies is intrinsically avoided that gives rise to an excellent signal-to-noise ratio (SNR) when using PAES.
In this section the main features and advantages of this technique will be discussed first. 
Then, starting from the pioneering work of Weiss et al.\,in 1988 \cite{Wei88} an overview of surface studies with PAES will be presented.
These studies comprise the investigation of metal and semiconductor surfaces covered with (sub-)monolayers of adatoms, the analysis of impurities, passivation, oxidation and H adsorption at surfaces, high resolution line shape and core annihilation measurements as well as the observation of surface segregation processes.

\subsection{Principle of PAES}
\begin{figure}[htb]
\centering
\includegraphics[width=0.7\textwidth]{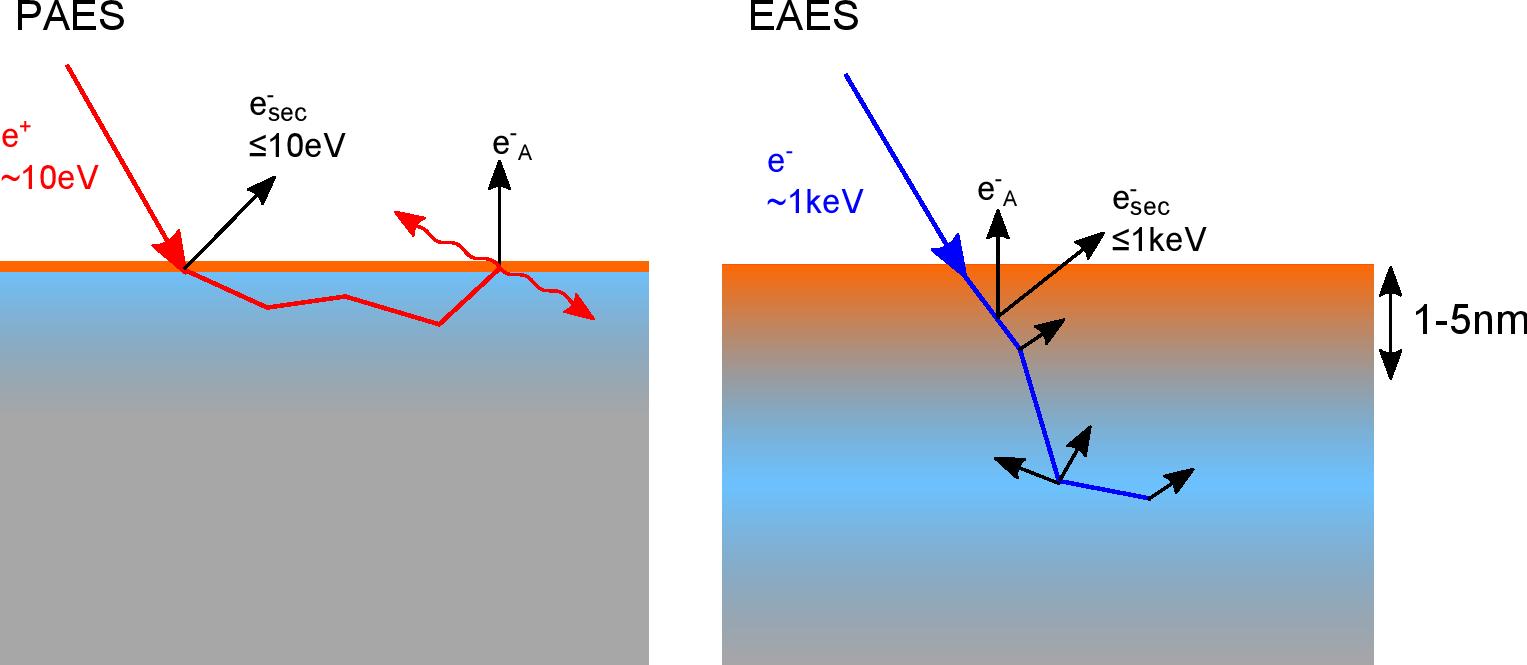} 
\caption{ 
Scheme of the emission of Auger electrons initiated by positron annihilation with core electrons (PAES) and collision induced by keV electrons (EAES). 
The positron annihilates after thermalization from a surface state leading to topmost layer sensitivity.
At EAES the information depth is averaged over several atomic layers.
}
  \label{PAES_principle}
\end{figure}

In conventional Auger electron spectroscopy (AES) the sample is irradiated with X-rays or keV-electrons to initiate the Auger process by photo- or impact ionization (XAES or EAES) of core electrons. 
In contrast to EAES or XAES, at PAES the core shell ionization is induced by the annihilation of positrons with core electrons.
For this reason, the energy of the primary positron beam can be chosen much lower (E$_+\approx$\,10\,eV)  than the typical electron energy ($\approx$\,keV) in EAES.
As a consequence of the low positron energy and the different ionization process PAES exhibits major advantages compared to conventional AES.
Independent on the nature of the primary particle ---i.e.\  electron, photon or positron--- the element information is gained from the energy of the emitted Auger electron as messenger particle.
At PAES, the positron trapping in a surface state leads to the exceptional surface sensitivity, whereas a much larger sample volume is excited at conventional AES.
Therefore, the information depth  is governed by the inelastic mean free path (IMFP) of the released Auger electrons, and hence the information gained by EAES or XAES is typically averaged over several atomic layers.
Figure \ref{PAES_principle} shows schematically the different excitation of the sample surface at PAES and EAES.
The main features of both techniques are summarized in Table\,\ref{tab:PAES}.

\subsubsection{Features of PAES}
 	\paragraph{Surface sensitivity}	
PAES allows experiments with extremely high surface sensitivity, since the vast majority of the implanted low-energy positrons is trapped in surface states and hence annihilate mainly with electrons in the topmost atomic layer \cite{Meh90, Koy92a}. 
It was demonstrated that up to 95$\%$ of the Auger signal stems from the uppermost atomic layer \cite{Wei95}.

 	\paragraph{Auger yield}	
Due to the annihilation with core electrons mainly from the topmost atomic layer the  Auger yield, defined as the number of detected Auger electrons per incident primary particle, is at PAES more than two orders of magnitude higher than at EAES.
The much lower Auger yield at EAES is mainly attributed to the larger volume, wherein the atoms are excited by the incoming primary electron beam, and the IMFP that limits the emission of non-scattered Auger electrons \cite{Cha76, Briggs90}.
For example, the Auger yield measured on a pure Cu surface turned out to be at least a factor of 200 higher at PAES than at EAES \cite{Hug07a}.
The experimentally gained results are in agreement with estimations given in \cite{Coleman2000}.

	\paragraph{Electron background and signal-to-noise ratio}
A quantitative analysis of Auger intensities is limited at EAES due to the high secondary electron background and the accordingly low SNR which  is about 1:2 at best.
In PAES no collision-induced secondary electron background  is produced in the higher energy range of released Auger electrons due to the low kinetic energy E$_+$ of the implanted positrons \cite{Wei88}. 
The maximum energy of secondary electrons E$_{\mathrm{sec}}$ is defined by the kinetic energy of the positron E$_+$ as well as the work functions of the positron $\Phi^+$ and electron $\Phi^-$ of the sample material:  E$_{\mathrm{sec}}\leq\,$E$_{e ^+}+\Phi^+-\Phi^-$.
As exemplarily shown in the PAES spectrum recorded at NEPOMUC (see \sFF{fig:compEPAES}) the  secondary electron background  ends at about 20\,eV according to the positron beam energy \cite{May10b}.
Moreover, the amount of inelastically scattered Auger electrons is lower, since most of the emitted Auger electrons are released from the topmost atomic layer \cite{Wei88, Koy92a}. 
At high-resolution measurements on metallic surfaces it was shown that the SNR is more than a factor of 20 higher at PAES than that at EAES \cite{Hug10b}.  

\begin{figure}[htb]
\centering
\includegraphics[scale=0.7]{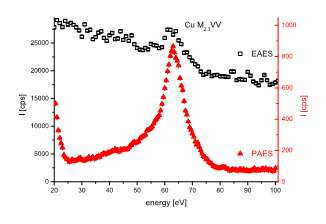}
\caption{Comparison of EAES (open black squares) and PAES (full red triangles) on Cu. The secondary electron background in the range of the Cu Auger peak is almost completely suppressed using PAES. Note the different intensity scales in counts per second (cps) for EAES (left) and PAES (right). The energy of the incident positrons amounts to 20\,eV and that of the impinging electrons to 2\,keV 
(figure from \cite{May10b}).}
\label{fig:compEPAES}. 
 \end{figure}
 
	\paragraph{Element selectivity}
As in conventional AES, the element information is gained by the detection of the discrete Auger peaks representing a fingerprint of the element distribution at the surface.
At PAES improved element selectivity is obtained due to the element dependence of the positron affinity $A^+=-(\Phi^++\Phi^-)$ \cite{Pus89}. 
For instance, the high positron affinity of Cu leads to an effective positron trapping at Cu clusters embedded in a Fe surface \cite{Faz06b}. 
For the same reason, the element selectivity of PAES leads to the observed enhanced Auger intensity of Cu in the measurements on Cu/Fe \cite{May10b}.

	\paragraph{Non-destructive surface probing}
At PAES the emission of the Auger electron is completely decoupled from the impact position of the primary particle, since the positron first thermalizes and diffuses to the surface before it ionizes an atom  via positron-electron annihilation. 
This is particularly important for the investigation of weakly bound adsorbates, e.g.\  organic molecules, which can hardly be investigated with conventional AES, since the elastically and inelastically scattered electrons lead to the capture of electrons and formation of negatively charged molecules resulting in the detachment or destruction of the object of investigation.
Moreover, due to the low positron energy the applied energy dose in PAES is considerably lower than in EAES and leads consequently to a remarkably reduced damage \cite{Wei88}, that might become essential, e.g.\  if metastable surfaces are investigated. 	

\begin{table}[htb]
\begin{small}
\begin{center}
\begin{tabular}{lcc}
{\bf Method} & {\bf EAES} & {\bf PAES}  \\
\hline 
Current  							&  $I_{\mathrm{e}^-}>\mu$A	&	$I_{\mathrm{e}^+}<$ pA \\
Setup 								&  simple					& 	elaborate \\
Beam energy 					& 	$\approx$\,keV		&	$\approx$\,20\,eV \\
 e$^-_{}$ background &	high						&		"zero" \\
Information depth			&	several at. layers	&		topmost at. layer \\
 Auger yield	(relative to EAES)		& 1							&		$>$100 \\
 SNR (relative to EAES)						& 1							&		$>$20
\end{tabular}  
\end{center}
\end{small}
	\caption{Features of EAES and PAES: Besides the drawback of a more elaborate setup and the much lower current of the primary beam, PAES exhibits several major advantages.}
   \label{tab:PAES}
\end{table}
 	
\subsubsection{Challenges}
The main drawback of PAES has been the long measurement time of typically several days for a single spectrum, if conventional $^{22}$Na based beams are used.
For instance, the time to record a PAES spectrum on annealed Cu using such a beam (E$_+$=40\,eV) was 20 days \cite{Str03}. 
Even with a stronger source and a cylindrical mirror analyzer with larger acceptance angle the shortest measurement time still amounted to about 30\,hours \cite{Wei95}. 
For this reason, both, a high intensity positron beam and the efficient electron detection are required to lower the measurement time considerably.

	\paragraph{Positron intensity}
In order to overcome the low positron intensity of laboratory beams two PAES systems were constructed at high intensity positron sources up to now, one at a linac by Suzuki et al.\  \cite{Suz96, Ohd97a} and the other in the  TUM positron research group at NEPOMUC \cite{Str01, May07}.
Suzuki et al.\ \cite{Suz96} reported a measurement time of only ten minutes for a PAES spectrum, however, at the expense of energy resolution, which was about 10\% and hence one order of magnitude lower than obtained in experiments at NEPOMUC \cite{May10b}.
Even using high intensity positron beams the positron flux is in the order of 10$^{-10}$A or even below and is hence typically more than five orders of magnitude lower than the electron current from commercially available electron guns (see Section\,\ref{sec:Microbeams}) which overcompensates the lower Auger electron yield in EAES.  
Therefore, from the experimental point of view, the magnetic and electrostatic positron beam guiding and focusing system have to be as efficient as possible.

	\paragraph{Electron detection in PAES spectrometers}
In first PAES studies cylindrical or sector electron analyzers with adjustable pass energy and equipped with a single channeltron have been applied for the energy dispersive electron detection.
The limited solid angle of a few percent and the necessity to scan the energy lead to very long acquisition times for PAES spectra.
In a completely other spectrometer design a permanent magnet behind the sample in combination with the magnetic guide field enabled  an efficient collection of the electrons onto the spectrometer axis.
As depicted in \sFF{PAES_Setup_Weiss} a Ne moderated positron beam is used with trochoidal energy filters which allow the separation of the positron transport to the sample and the detection of emitted electrons.
 Finally, the electron energy is detected by using a combination of a trochoidal monochromator and a position sensitive detector \cite{Lei89, May90}.

\begin{figure}[htb]
\centering
\includegraphics[width=0.8\textwidth]{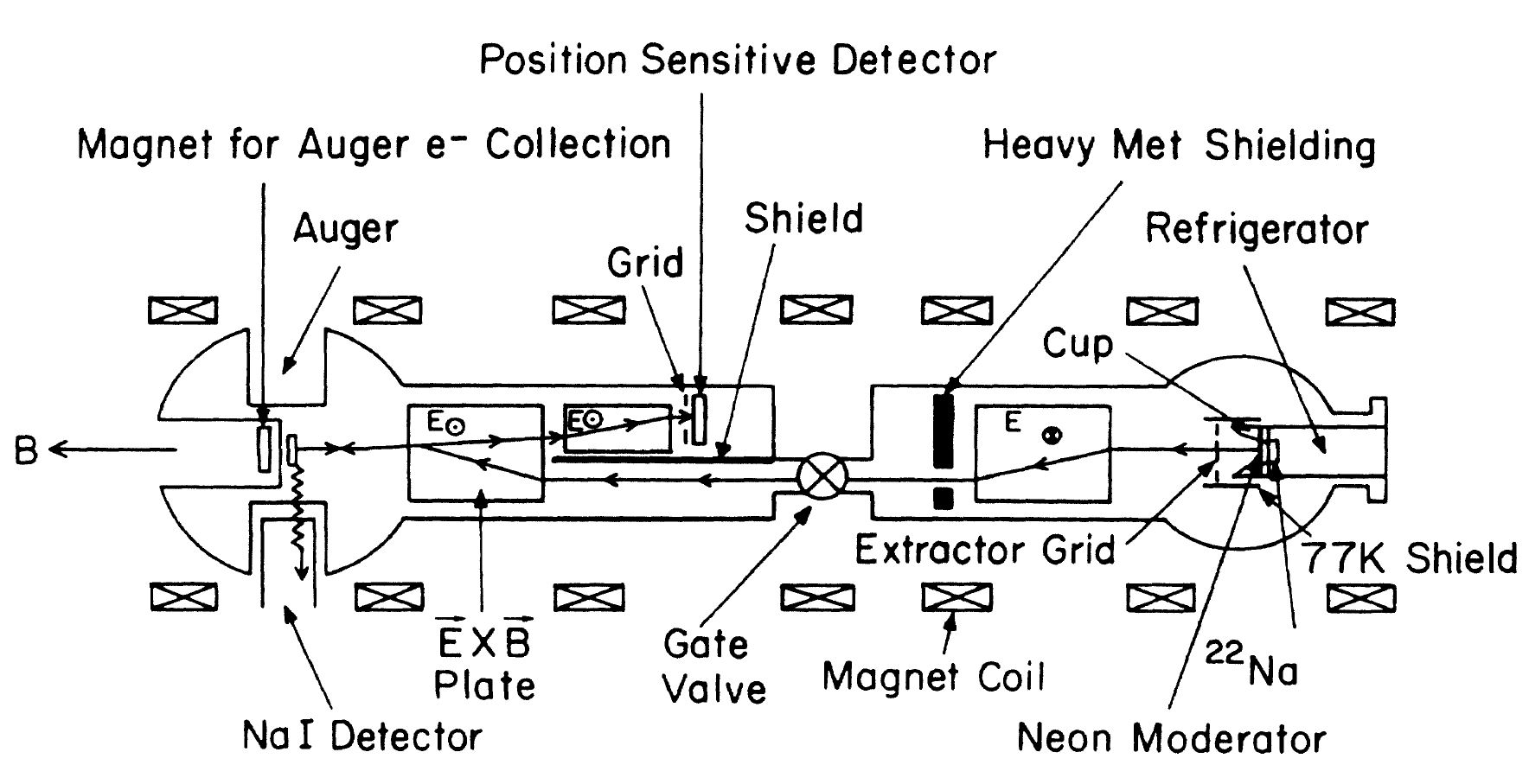} 
\caption{Setup for the detection of Auger electrons for PAES. For details see text 
(figure from \cite{May90}).
} 
  \label{PAES_Setup_Weiss}
\end{figure}

More recent devices use state-of-the-art hemispherical electron energy analyzers with MCPs and CCD read-out.
After first feasibility studies at a $^{22}$Na based beam with a low intensity of $10^4$ moderated positrons per second  \cite{Str01, Str03} the spectrometer was connected to the NEPOMUC beam line, where the measurement time was considerably reduced from about 20 days to a few hours \cite{Hug06a}.
Further improvements lead to even shorter measurement times (improved positron beam transport) and an enhanced SNR (background suppression), and hence enabled PAES studies with various positron beam energies (see \cite{Hug07a, Hug07b, May07}). 
The application of a hemispherical electron energy analyzer (PHOIBOS\,150) with a large acceptance angle of up to $\pm13^\circ$ and MCP readout allows measurements with high transmission and high energy resolution.
Both, the high intensity of the low-energy positron beam, and the efficient detection with the hemispherical analyzer lead to PAES spectrum acquisition time of seven minutes  using  the 20\,eV positron beam \cite{May10c}. 
The present setup of the surface spectrometer at NEPOMUC is shown in \sFF{NEPOMUC_PAES_FOTO}. 
Since for PAES a purely electrostatic beam guidance is required a magnetic field termination of $\mathrm{\tcmu}$-metal is mounted at the entrance of the $\mathrm{\tcmu}$-metal shielded Auger chamber in order to release the low-energy positron beam non-adiabatically from the magnetic guiding field. 
The analysis chamber is also equipped with an X-ray source and an electron gun for the surface analysis wiht complementary techniques such as  X-ray induced photo electron spectroscopy (XPS), XAES and EAES.
The features of the spectrometer including Auger analysis chamber with sample preparation chamber are presented in Refs. \cite{Hug07a, Zim14a}.

\begin{figure}[htb]
\centering
\includegraphics[width=0.8\textwidth]{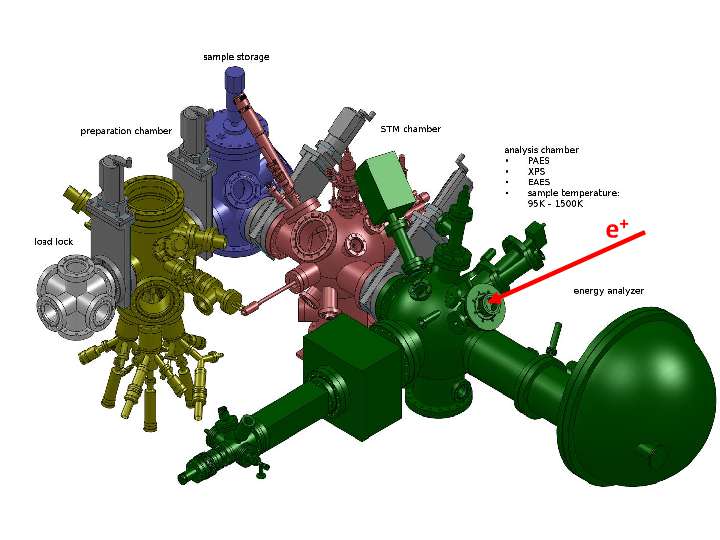} 
\caption{PAES spectrometer at NEPOMUC: The positron beam is magnetically guided to the Auger chamber, which is connected with a STM and a sample preparation chamber \cite{Zim14a}. }
 \label{NEPOMUC_PAES_FOTO}
\end{figure}

Another approach is the time-of-flight (TOF) technique to determine the kinetic energy of the emitted electrons. 
Up to now, three TOF-PAES setups have been reported so far, which are based on the design shown in \sFF{PAES_Setup_Weiss} extended by a drift section for the electrons. 
In the setup using a pulsed beam the start signal is provided by a 10\,ns pulse \cite{Ohd99}. 
In systems with  a continuous beam, the start signal is generated by the detection of the 511\,keV radiation that originates from positrons annihilating in the sample \cite{Lei89}.
The stop signal is generated by the detection of the electrons after having passed a retarding flight tube.
One disadvantage of these devices  is the lower energy resolution compared to electrostatic electron energy analyzers, which was reported to be $\Delta E/E$=\,3\,eV/30\,eV \cite{Ohd99}. 
Therefore, an improved TOF-setup, which combines a trochoidal filter and a flight tube mounted in a Faraday cage, with an energy resolution of about 1\,eV at high electron energies up to E\,$\approx$\,1000\,eV was developed \cite{Hug06b} and demonstrated to work in an energy range between 50\,eV and 400\,eV \cite{Leg07}.  

\subsection{Surface studies with PAES}
\subsubsection{Core-annihilation probability and line shape analysis}
\label{sec:PAESmeasurements}
The annihilation probability of tightly bound electrons with the positron amounts typically to a few \%.
For the 3p electrons in Ni  the value was experimentally found to be 3.7(7)\% by applying PAES \cite{Wei88}.
The so-called core-annihilation probability p$_\mathrm{core}$ is described by summing up the overlap of the positron wave function $\Psi_{+}(\vec r)$ with the wave function of the electrons in the i'th level $\Psi_{i}(\vec r)$ 
\begin{equation}
p_\mathrm{core} = \pi r_0^2 c\int d^3\vec r \cdot \left|\Psi_+(\vec r)\right|^2 \cdot {\sum_{i}^{}  \left|\Psi_{i}(\vec r)\right|^2}
\end{equation}
with classical electron radius $r_0$ and speed of light $c$.
Jensen et al.\,\cite{Jen90} have calculated  the according bulk values for a number of metals (see Appendix: \ref{tab:Core-annihilation}).
The core-annihilation probabilities can be roughly approximated by an empirical formula \cite{Coleman2000}:
\begin{equation}
p_\mathrm{core}\approx \frac{600 \cdot N(E_B)}{(E_B/eV)^{1,6} }\% 
\end{equation} 
with N$(E_B)$ the number of electrons in the respective level with electron binding energy E$_B$ in eV. 
In general, the higher electron binding energy E$_B$ the lower p$_\mathrm{core}$.
By applying the corrugated mirror model (CMM) for the description of the surface tapped positron Jensen also determined p$_\mathrm{core}$ at the surface of selected metals \cite{Jen90}. 
In addition, for the case of Cu it was demonstrated that the calculated annihilation probability with core electrons from the substrate decrease drastically when foreign atoms are adsorbed.

PAES with  high  energy resolution allows to determine experimentally both, the energy and the relative intensities of Auger transitions.
The only complementary but very time consuming technique to determine Auger line shapes is Auger photo electron coincidence spectroscopy (APECS) where, at the expense of count rate, the emitted photo electron released by X-ray absorption and the Auger electron are detected in coincidence \cite{Bar96}. 
In first PAES studies using a cylindrical mirror analyzer with an energy resolution of $\frac{\Delta E}{E}$=2.5\% the double peak structure of the CuM$_{2,3}$VV transition could be observed \cite{Yan97}.
In addition, by appropriate analyzer biases Weiss et al.\,succeeded in clearly separating the energy distribution of ``true secondaries'' from that of ``redistributed primary'' particles by bombarding Ge(100) with 370\,eV positrons and electrons \cite{Wei97}.
The application  of a positron beam allows to record the spectrum of ``true secondaries'',  i.e. electrons which are ejected but originally stem from the sample, since the positron is distinguishable from the electron. 
As shown in \sFF{Ge_emission}\,(c) the low energy secondary electron peak is clearly visible.
By measuring the positron spectrum (see panel (d) in \sFF{Ge_emission}) the distribution of the ``redistributed primary'' particles, i.e. of positrons after interaction in the sample, is obtained.
A peak at 370\,eV due to elastic and quasielastic scattering of the primary particles as well as plasmon loss peaks about 13\,eV and 26\,eV below the elastic peak are observed in both spectra recorded for (a) electrons and (d) positrons hitting the sample. 
As plotted in panel (b), it is intriguing that the linear combination of the positron induced secondary electron spectrum and the backscattered positron spectrum corresponds to the electron induced electron spectrum to a high degree.

\begin{figure}[htb]
\centering
\includegraphics[width=0.65\textwidth]{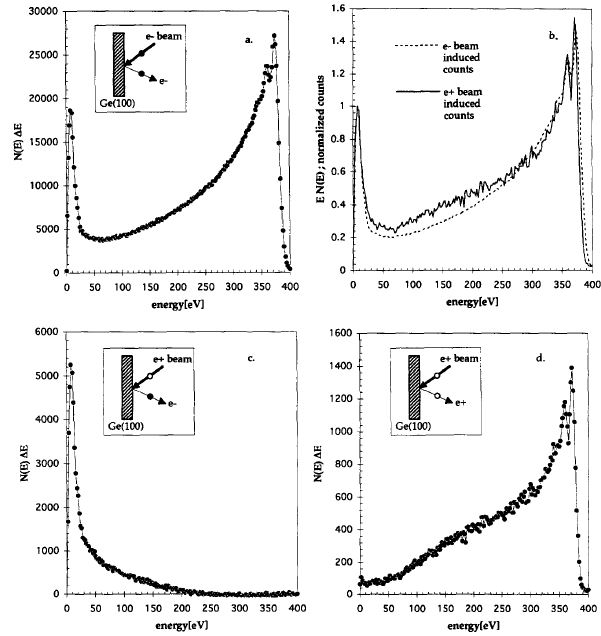}
\caption{ 
Energy spectra of electrons and positrons emitted from a Ge(100) surface bombarded with electrons and positrons:
The spectrum of emitted electrons is shown after impact of 370\,eV (a) electrons and (c) positrons.
Panel (d) shows the spectrum of re-emitted positrons as a result of bombardment with 370\,eV positrons. 
For comparison, in panel (b) the electron induced electron spectrum is plotted with a linear combination of the positron induced secondary electron spectrum and the backscattered positron spectrum
(figure from \cite{Wei97}).
}
  \label{Ge_emission}
\end{figure}

\begin{figure}[!htb]
\centering
\includegraphics[scale=0.7]{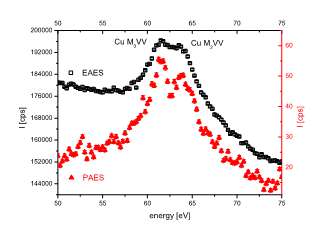} 
\caption{High resolution spectroscopy of the $\mathrm{Cu\,M_{2,3}VV\,}$-transition with EAES (open black squares) and PAES (red triangles). The energy resolution was  set to $\frac{\Delta E}{E}=1\%$. The double peak structure is clearly resolved  with PAES. The measurement time for EAES amounts to 10\,min and for PAES to 5.5\,h
(figure from \cite{May10b}).}
\label{fig:Double}. 
 \end{figure}

The improved energy resolution of $\frac{\Delta E}{E}<1\%$ and the excellent SNR allows the almost background-free analysis of line shapes of Auger transitions using the NEPOMUC setup.
The line shape analysis of the Cu\,M$_{2,3}$VV transition recorded with PAES revealed a peak separation of 2.3(1)\,eV according to EAES and as expected from theory (see Figure\,\ref{fig:Double}) \cite{May10b}.
Subsequent studies have been performed on the surfaces of annealed polycrystalline samples of Fe, Ni, Cu, Zn, Pd, and Au \cite{Hug10b}.
The contributions of the sub-levels p$_{1/2}$ and  p$_{3/2}$, e.g.\ for the M$_{2,3}$VV transition of Ni, as well as three distinct Auger peaks of the O$_{3}$VV, O$_{2}$VV and N$_{6,7}$VV transitions of Au could clearly be resolved.
Consequently, the relative contribution of  the respective electron orbital participating in the Auger process could be determined quantitatively \cite{Hug10b}.
Although the calculated values of the core-annihilation probabilities show a large straggling \cite{Jen90}, they can be very roughly approximated by the formula given above.
In the experiment the same trend was observed. The respective core-annihilation probability is lower for deeper bound electrons due to the repulsive Coulomb potential of the nucleus \cite{Hug10b}.

\subsubsection{Adsorbed layers on Cu}
\label{sec:CuSurfaces}
After the first observation of positron annihilation induced Auger-electrons emitted from Ni(100) and Cu by A. Weiss et al.\ \cite{Wei88} various experiments were performed on Cu substrates covered with S \cite{Meh90}, Au \cite{Lee94, Wei95, Hug07a}, Cs \cite{Koy92a, Faz94}, and Pd \cite{Koy93} as well as on CuO \cite{Nad07a}. 

On the surface of clean polycrystalline Cu the inelastically scattered Auger electrons and the temporal dependence of the Auger intensity of non-scattered electrons were studied to observe the  degradation of the surface cleanliness \cite{May08}.
The temperature dependence of the Auger intensity recorded with PAES was studied on clean Cu(100).
In contrast to collisional induced AES, which is independent on temperature, a strong decrease of the PAES intensity was observed between 300$^o$C and 700$^o$C. 
Simultaneously, an increase of Ps annihilation was measured by detecting the 3$\gamma$/2$\gamma$ ratio. 
Hence, the found decrease of the Auger intensity could be attributed to competing thermal emission of Ps formed at the surface. 
This in turn confirms the  exceptional surface sensitivity of PAES in the region ($\approx$1\AA) where thermal Ps is formed \cite{May90}.

At a Cu(110) surface covered with 0.5\,ML of S a c(2$\times$2) structure is formed leading to a higher positron annihilation rate at the S adatoms. 
Accordingly, the L$_{2,3}$VV transition of S could clearly be observed whereas the Cu PAES intensity was reduced by a factor of four \cite{Meh90}.

The PAES intensities have been studied dependent on the fraction of Au deposited on polycrystalline Cu in order to demonstrate the surface selectivity of PAES. 
In the Au/Cu system, even at an amount of only 0.5\,ML of Au on Cu the decrease of the Cu Auger intensity was clearly observed \cite{Hug07a}.
After a coverage of 1\,ML of Au only Auger electrons from Au were detected whereas the Auger peak of the Cu\,M$_{2,3}$VV transition disappeared \cite{Wei95}.
Covering Cu(100) with sub-ML films of Au showed a steep increase of the PAES intensity of Au already at a Au coverage of only 0.07\,ML. This behaviour found at 173\,K is attributed to positron localisation at Au islands formed at the Cu surface.
At higher temperature (303\,K) intermixing and formation of a surface alloy lead to a change in the relative Auger intensities of Cu and Au \cite{Lee94}. 

At a coverage of 0.7\,ML Cs on Cu(100) a sharp drop in the Cu\,M$_{2,3}$VV Auger intensity almost to zero was observed at two temperatures (163 and 303\,K).
 This intensity decrease is explained in terms of a phase transition of the surface structure from a disordered state of the Cs overlayer  to adsorbate metallic islands ordered in a hexagonal close-packed structure.
This in turn leads  to a migration of positrons trapped at the Cs/Cu interface (low Cs coverage) to the surface state on the vacuum side of the Cs overlayer at high Cs coverage \cite{Koy92a, Faz94}. 

In the system Pd/Cu a Cu(100) surface was covered with 0.55\,ML of Pd.
At low temperature (173\,K) Auger emission from the substrate could not be observed (see \sFF{PAES_PdCu_Koy93}).  
With increasing temperature up ot 423\,K the decreasing Pd signal obtained by PAES was attributed to surface alloying of Pd and Cu \cite{Koy93}.

\begin{figure}[htb]
\centering
\includegraphics[width=0.4\textwidth]{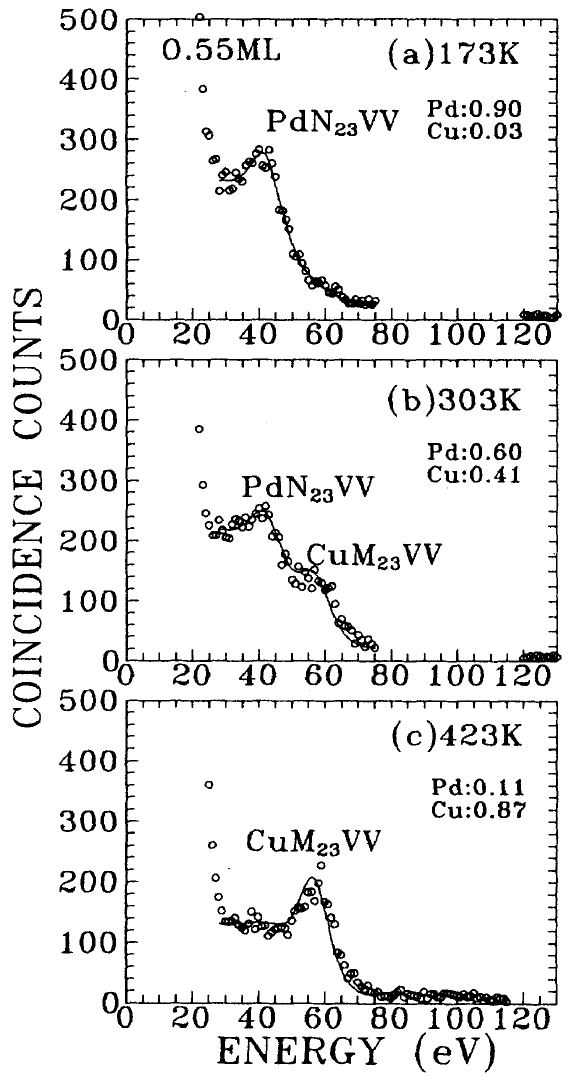} 
\caption{PAES spectra recorded for Cu(100) covered with 0.55\,ML of Pd at different temperatures 
(figure from \cite{Koy93}).}
\label{PAES_PdCu_Koy93}. 
 \end{figure}

The oxidization and desorption of O from an oxidized Cu(100) surface at different temperatures was studied by PAES \cite{Faz10}.
The initial increase of the Cu M$_{2,3}$VV  transition up to 300$^o$C was attributed to O desorption or diffusion into the bulk.
At higher temperature the reoxidation of the Cu surface leads to a reduction of the Cu Auger signal.
The core-annihilation probabilities of surface-trapped positrons were calculated taking into account electron positron correlation as well as charge redistribution at the surface and surface reconstruction.
As a result the annihilation probability for the Cu M$_{2,3}$VV  transition was found to decrease by more than a factor of four after adsorption of 1\,ML of O \cite{Faz10}.

PAES studies were performed at the surface of an annealed Fe-1.0\,wt.\%Cu alloy revealing a strongly enhanced  Auger intensity of Cu.   
This effect suggests that positrons are trapped at Cu aggregations at the Fe surface.
The calculations of positron surface states showed that the positron wave function at the Fe(100) surface with Cu nanoparticles embedded in the top atomic layers is localized on the vacuum side of  the Cu atoms in agreement with the experimental results \cite{Faz06b}.

\subsubsection{Cover layers in the systems Cu/Fe, Cu/Pd and Ni/Pd}
Thin membranes of Pd and Pd based alloys play an important role for catalysis, for H storage and H purification.
Since their mechanical stability is important for its industrial application a deeper understanding of the mechanical properties of Pd based alloys is crucial, where among others PdCu belongs to one of the most promising materials \cite{Lov05}.
For this reason, on a fundamental level the system Cu/Pd was investigated by PAES and EAES on two systems Cu/Pd and Cu/Fe due to the different positron affinity of the respective substrates \cite{May10b}.
In a subsequent experiment, time dependent PAES was applied for the first time in order to gain information of the stability of the Pd surface covered with (sub-)ML of Cu \cite{May10c}.

\begin{figure}[htb]
\centering
\includegraphics[width=0.55\textwidth]{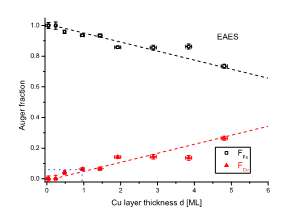}
\includegraphics[width=0.55\textwidth]{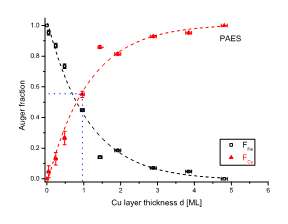} 
\caption{Fraction of the Auger signal from Cu and Fe, respectively, as a function of the Cu layer thickness, measured with EAES (left) and PAES (right; figure from \cite{May10b}).}
  \label{CuFe}
\end{figure}

As exemplarily shown for different surface coverage of Cu on Fe the raw spectra obtained by  EAES and PAES clearly demonstrate the superior surface sensitivity of PAES (see \sFF{CuFe_raw}). 
After the quantitative analysis of the spectra the high sensitivity to the topmost layer is impressively seen by plotting the respective Auger fractions as a function of the Cu layer thickness for both techniques (in Figure\,\ref{CuFe}).
For example, the observed Cu Auger intensity at PAES amounts to more than 5$\%$ at only 0.03\,ML of Cu on Fe whereas Cu is not yet detected with EAES. 
Compared to Cu/Pd, the sensitivity to the Cu cover layer is enhanced at Cu/Fe due to the higher positron affinity to Cu than to the substrate \cite{May10b}.   
In the Cu/Pd system, the pure Cu signal is obtained at a coverage of $3.33\,\mathrm{ML}$ Cu on Pd. 
Therefore,  the Pd surface is assumed to be completely covered with at least 1\,ML of Cu since the positron affinity to Pd is higher than to Cu \cite{May10b}.

\begin{figure}[htb]
\centering
\includegraphics[width=0.55\textwidth]{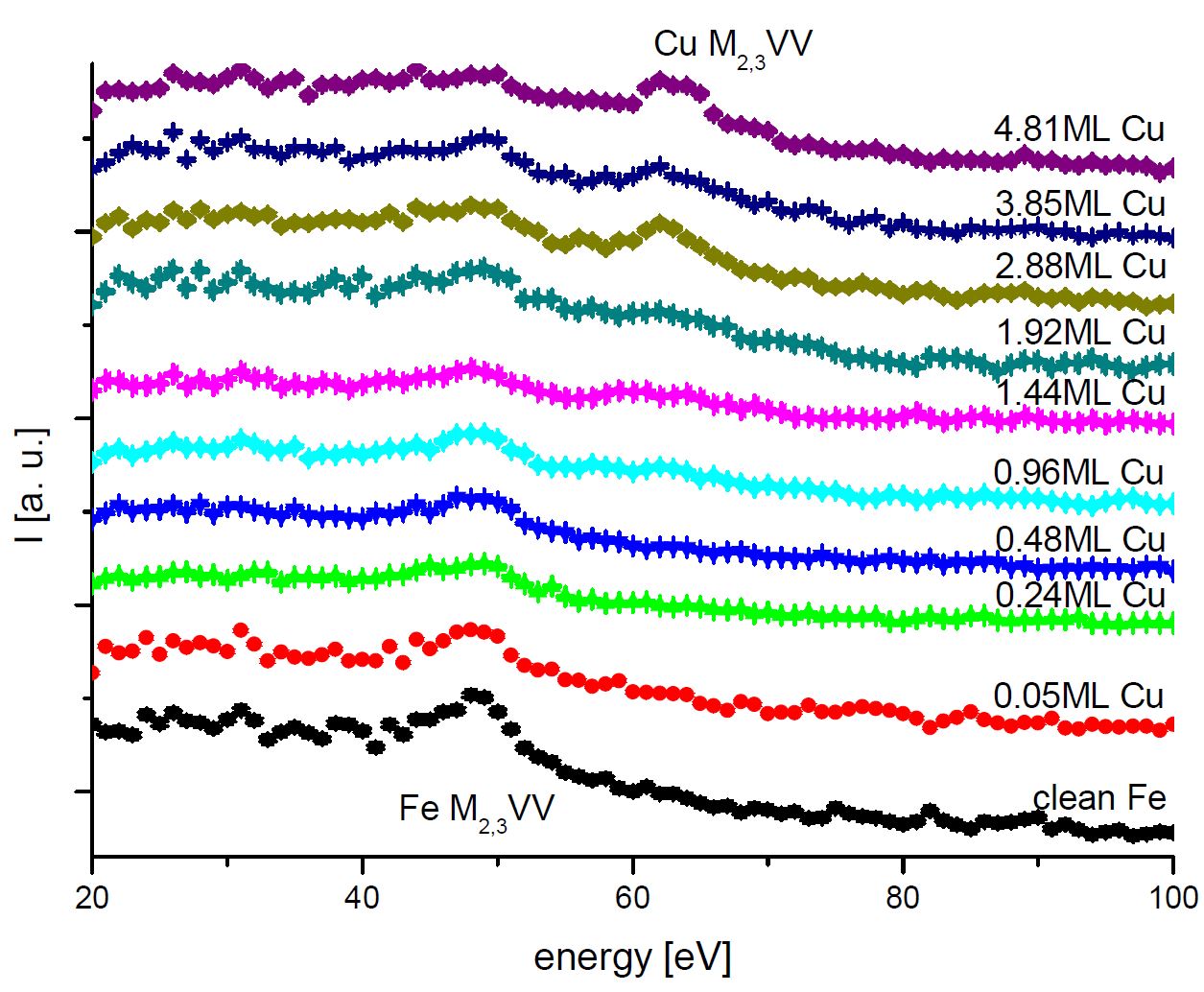}
\includegraphics[width=0.55\textwidth]{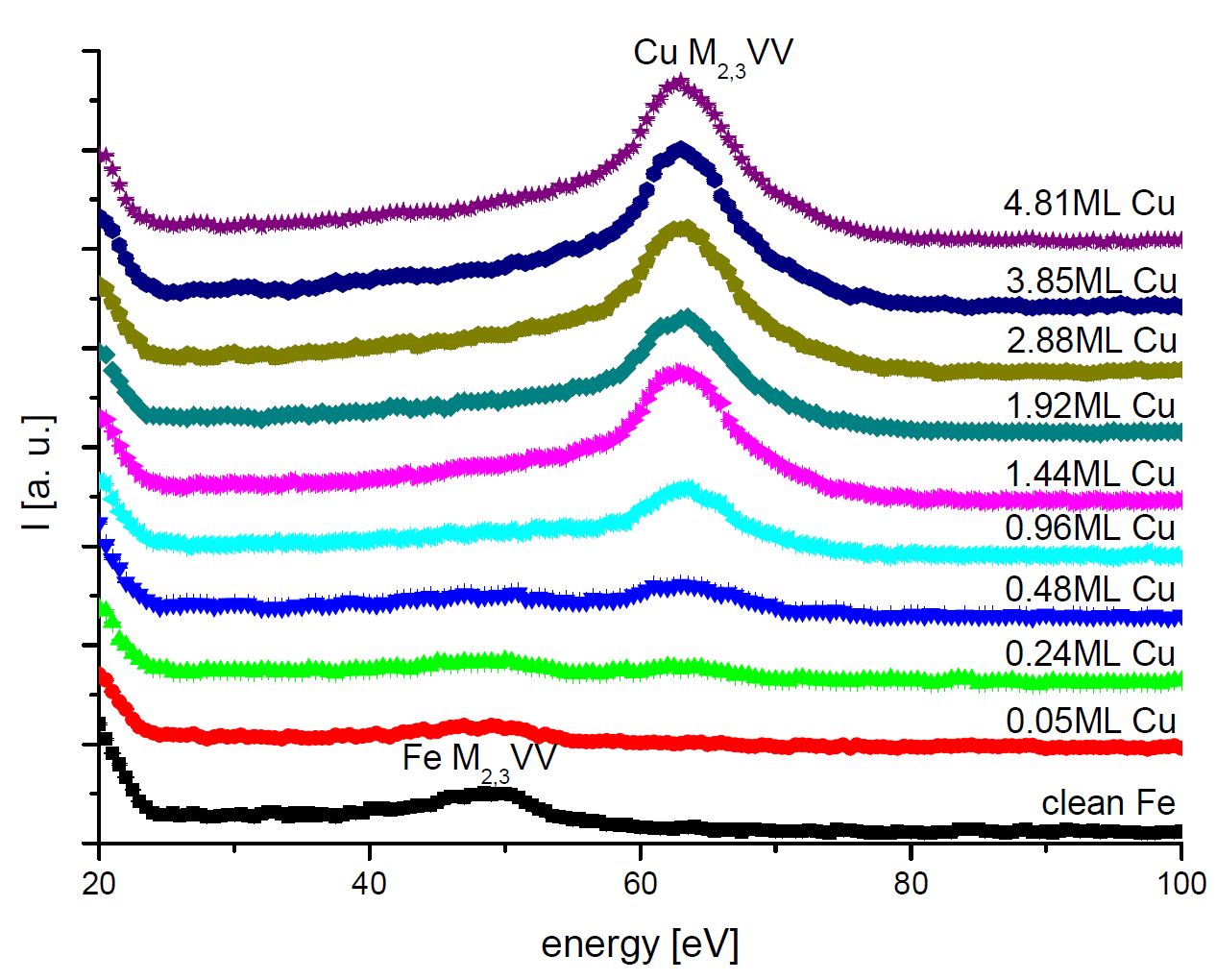} 
\caption{ 
EAES and PAES spectra of clean Fe with Cu cover layers of various thicknesses (data are offset for clarity).
}
  \label{CuFe_raw}
\end{figure}

Recently, a comparative study using AES induced by positron annihilation, electrons and X-ray as well as XPS was carried out on the system Ni/Pd. 
As shown in \sFF{PAES_NiPd_Zim16a} the PAES spectra exhibit negligible background and reveal clearly the Auger peaks of the adsorbed Ni atoms.
In addition, time-dependent PAES accompanied by XPS was performed on Pd covered with 0.5\,ML of Ni in order to investigate changes in the elemental composition of the surface at room temperature. 
The strong decrease of the Auger emission from Ni observed with PAES was attributed to the migration of Ni atoms into the Pd substrate.

\begin{figure}[htb]
\centering
\includegraphics[width=0.5\textwidth]{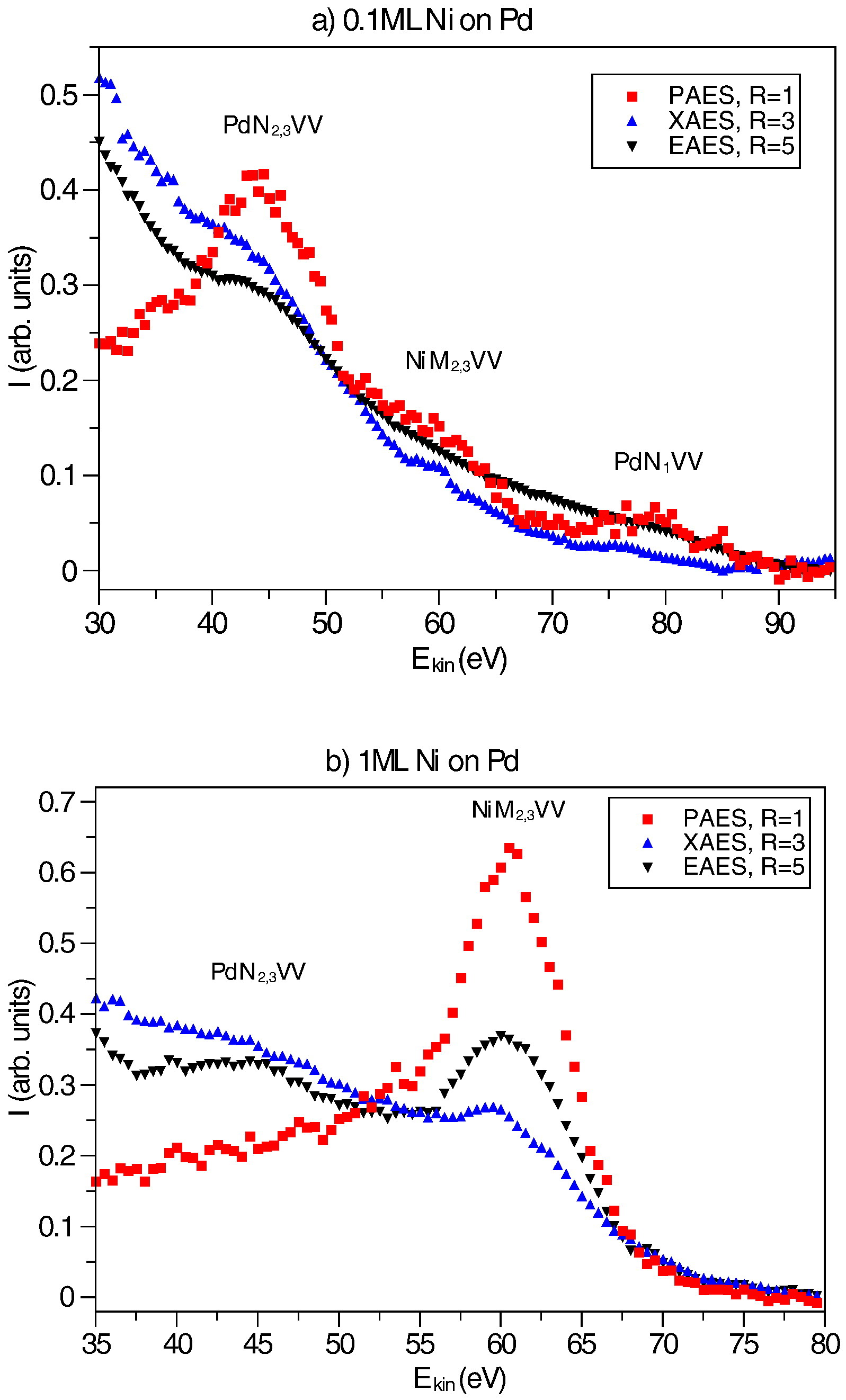} 
\caption{
Comparison of PAES with XAES and EAES on a Pd  surface covered with (a) 0.1\,ML Ni and (b) 1\,ML Ni. 
In the PAES spectra at a coverage of 0.1\,ML Ni the Ni\,M$_{2,3}$VV transition is clearly visible, whereas at 1\,ML Ni almost no Pd of the substrate is detected. 
Note the signficantly lower electron background at PAES. 
The spectra are normalized and the intensity was set to zero at E=90\,eV and 80 eV\,in (a) and (b), respectively. The retarding ratios R for the recording mode are given in the legend
(figure from \cite{Zim16a}.}
  \label{PAES_NiPd_Zim16a}
\end{figure}

\subsubsection{Surface segregation of Cu in Pd}
\begin{figure}[htb]
\centering
\includegraphics[width=0.55\textwidth]{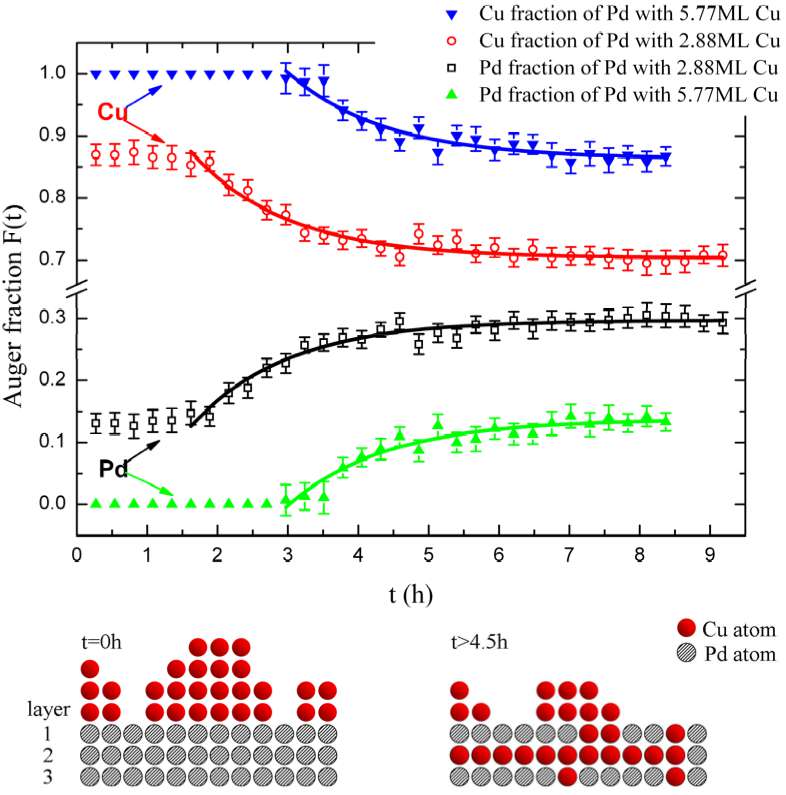} 
\caption{Segregation of Cu in Pd: The fraction of the relative Auger intensities from Pd and Cu, respectively, as a function of time. The characteristic time for the segregation process is determined to 83 minutes by exponential fits (solid lines). Scheme of the arrangement of the surface atoms (below; figure from \cite{May10c}).}
  \label{Segregation}
\end{figure}

Mayer et al.\,succeeded to observe the segregation process in the Cu/Pd system in situ with \textit{time dependent} PAES. 
For this purpose, a Pd surface was covered with almost 3\,ML of Cu in order to study the evolution of the PAES intensities of Cu and Pd.
By plotting the Auger intensities as function of time an increase of the Pd Auger intensity at the expense of the Cu Auger peak could be observed after about two hours. 
A saturation value was reached after about six hours.
This behavior could be reproduced with a Cu covered Pd sample where no Pd signal at all was detected at the beginning, i.e. the Cu amount at the beginning was twice as large as at the previous sample. 
From the time dependent slope of the respective Auger intensities, several explanations could be excluded such as surface contamination, surface diffusion, or Cu diffusion into the Pd bulk \cite{May10c}.
For both samples the time dependent PAES intensities of Cu and Pd exhibit the same characteristic time constant of the segregation process  (see fits in Figure\,\ref{Segregation}) which was found to be 83 minutes.
It is noteworthy, that these  first \textit{time dependent} PAES investigations became only feasible due to the unprecedented low measurement time for recording high quality PAES spectra as exemplarily shown in Figure\,\ref{fig:compEPAES}.

\subsubsection{Auger-mediated positron sticking}
\label{sec:AugerMediatedPositronSticking}
An efficient mechanism for positron sticking to surfaces of Cu and Au was observed by a using positron beam with very low energy of 1.5-7\,eV \cite{Muk10}.
In particular, a  narrow secondary electron peak was still observed at incident positron kinetic energies well below the electron work function.
This feature is explained by a single step process of the positron transition from an unbound scattering state to a bound surface state whereby the according transition energy is transferred to a valence electron which may leave the surface. 
This so-called Auger-mediated positron sticking process was shown to have an efficiency exceeding 10\% at positron energies of about 1\,eV.
At high temperature (700$^o$C) the low-energy electron peak was still present with about the same intensity as at room temperature whereas the PAES signal vanishes due to the formation and desorption of Ps. 
This in turn  indicates that the characteristic low energy electron peak is not associated with positron annihilation induced Auger emission.
The surface state binding energy was determined to 2.8(2)\,eV and 2.9(2)\,eV for Cu and Au, respectively \cite{Muk10}.

In a subsequent work, PAES spectra free of beam induced background have been recorded for Cu and Au by using positrons with an energy below the threshold for secondary-electron emission \cite{Muk11}. 
Thus, the spectra consist exclusively of electrons emitted as a direct result of the Auger transition and electrons emitted via intrinsic loss processes.
The spectra show an intense low-energy tail associated with the Auger peak extending to 0\,eV.
Due to its high intensity the low-energy tail comprises both, inelastic scattering of the Auger electrons from the surface and the core hole decay by emission of more than one Auger electron. 
It could be estimated that  about 47\% and 50\% of the core holes in Cu and Au, respectively, decay via multi-electron emission. 
Consequently, considering solely the Auger peak intensity underestimates the number of initial core holes, and hence better agreement with calculations is found by accounting for the multi-electron processes \cite{Muk11}. 

\subsubsection{PAES studies on semiconductor surfaces}
\label{sec:PAESStudiesOnSemiconductorSurfaces}
In order to compare first-principles calculations of positron surface states with experimental results PAES was performed on Si(100), Si(111) \cite{Faz03b, Faz06a} and  GaAs(100) \cite{Faz03a}.
Besides on clean semiconductor surfaces, PAES studies were performed on Si(100)  covered with Au \cite{Yan95a, Yan96}, terminated with  O and H \cite{Kim98} or passivated with Se \cite{Zhu05}.
The reactivity of as-deposited and thermally treated sub-monolayer of Au on the Si(100) surface was studied with PAES in order to reveal changes of the  surface morphology. 
Oxygen exposure of the surface resulted in a  strong decrease in the Si PAES intensity while leaving the Au intensity unchanged. 
This observation was attributed to O sticking preferentially at the Si atoms \cite{Yan95a}. 
In a subsequent study PAES was combined with sputter depth profiling of thin films in order to study the structure of the annealed Au/Si(100) system.
Benefiting from its topmost layer sensitivity, PAES revealed that the first ML consists of pure Au, and then the Au concentration  decreases continuously to 0\% at depths of 13{\AA} and 28{\AA} for initial Au depositions of 5{\AA} and 10\AA, respectively \cite{Yan96}. 

In a PAES study on the adsorption of H and O on Si(100) a significant decrease of the Auger intensity of Si was observed. 
This decrease of the PAES intensity after H (O) adsorption was almost three (more than six) times stronger than observed with EAES.
The attenuation of the Si Auger signal versus gas exposure were fitted by exponential functions and revealed sticking coefficients for H$_2$ and 0$_2$ which amount to 
4.4$\cdot10^{-5}$ and 2.6$\cdot10^{-4}$, respectively \cite{Kim98}.
The stability of the  Si(100) surface passivated with a Se layer during air exposure and at various temperatures was studied with PAES.
It could be shown that defects in the Se passivation layer are responsible for O chemisorbtion on the Si surface \cite{Zhu05}.

Temperature dependent PAES performed on Ge(100) showed a significant decrease of the PAES intensity above 200$^o$C. 
This observation was explained by the competing formation and emission of Ps  \cite{Soi92}.
Desorption of Ps, however, does not fully describe the PAES intensities above 500$^o$C which level off at about 5\% of the room temperature value.
The discrepancy is explained by positron trapping in surface defects which are described by different  models \cite{Soi92}.

On the  MoS${_2}$ (0001) surface exposed to O$_2$ the appearance of an according strong Auger peak at PAES is explained by positron trapping in surface defects where impurity atoms are adsorbed preferentially \cite{Ohd96}. 
Finally, it was demonstrated that PAES can be applied to study the surface contamination with superior sensitivity than EAES and on carbon nanotubes \cite{Ohd02}.

\section{Surface studies using spin-polarized positrons}
\label{sec:Polarized}


Spin phenomena such as charge to spin conversion in non-magnetic materials attracted much interest due to its implications for spintronics applications where the production, injection, transport, manipulation, and detection of  electron spins is of outmost importance.
At the surface several physical effects lead to an electronic structure which might widely differ from the bulk properties.
The reduced coordination number of the surface atoms leads to  band narrowing and strong electron localization. 
Lattice distortion and reconstruction of the surface also modulates the electronic states with the consequence that e.g. magnetic properties of the surface differ from those of the bulk.
In particular, spin polarized currents can occur on non-magnetic material surfaces or interfaces either caused by the spin Hall effect or due to the energy splitting of spin bands induced by spin-orbit coupling (SOC) and the broken spatial symmetry (Rashba-Edelstein effect).

Spin-polarized experiments using $\beta^+$ sources directly, i.e.\,without positron moderation, enable the investigation of the magnetic properties and the electronic structure of the bulk in a non-destructive way.
For the investigation of spin phenomena at surfaces, however, the applied technique has to be both spin sensitive and surface selective (Section\ref{sec:PolarizedBeams}). 
Therefore, various PAS techniques using low-energy spin-polarized positron beams provide powerful tools to study the effective electron polarization at the surface giving rise to e.g. surface magnetism or spin-polarized  surface currents. 

In general, positrons can detect spin-polarized electrons through either direct annihilation with electrons or  by picking up surface electrons near the Fermi level to form  Ps and subsequent annihilation.
Since the formation probability of o-Ps is correlated with the spin polarization of the outermost surface electrons, analysis of the positron annihilation $\gamma$ spectra provides unique information on the spin polarization on metal surfaces.
It was pointed out by Kawasuso et al. that the low detection limit of this technique in the order of 10$^{-3}$ is highly beneficial for the determination of electron spin polarization \cite{Kaw13}.

\subsection{Spin-polarized positron experiments}
\label{sec:Polarizedexp}
\begin{figure}[htb]
\centering
\includegraphics[width=0.5\textwidth]{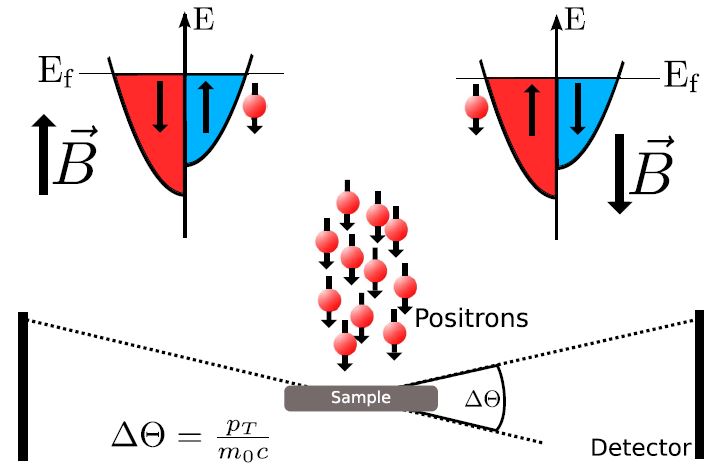} 
\caption{
Spin states in magnetic materials probed by spin-polarized positrons:
In electron positron annihilation the singlet configuration is preferred.
Hence, either the majority or minority spin electrons are probed by reversing the magnetization of the sample being parallel or anti-parallel to the polarization of the positrons.
In spin-polarized 2D-ACAR the angular correlation $\Delta\Theta$ of the two emitted 511\,keV annihilation quanta, and hence the transverse momentum of the annihilating pair p$_T$ is measured dependent on the magnetic field  
(figure from \cite{Cee16a}).
}
  \label{PolSource}
\end{figure}

The underlying  principle for the investigation of magnetism using spin-polarized positrons is the highly preferred annihilation  when the spins of positron and electron are aligned anti-parallel (see e.g. \cite{Wes73}).
Consequently, by applying an external magnetic field parallel or anti-parallel to the positron polarization the positrons will annihilate predominantly with electrons from the majority or the minority spin directions, respectively. 
The scheme for the case of spin-polarized ACAR experiments is depicted in Figure\,\ref{PolSource}.
For this reason, spin-polarized PAS can be applied to study the bulk magnetic properties of magnetic materials even at elevated temperatures.

In the bulk of ferromagnetic materials Doppler broadening spectroscopy (DBS) was applied to reveal the dependence on the relative spin orientations of positron and electron. 
The observed field-reversal asymmetry for Fe, Co, and Ni was found to correspond to the effective magnetization \cite{Kaw11}. 
As expected, the asymmetry disappears above the Curie  temperature as also measured for Gd, Tb, and Dy \cite{Kaw12}.
More recently, spin dependent PALS on Fe, Co, and  Ni  \cite{Mae15a} revealed similar asymmetries as predicted by theoretical considerations \cite{Lin14}.
Spin-polarized ACAR was demonstrated to be a powerful tool for the investigation of the bulk electronic structure of ferromagnetic materials.
In elemental crystals, the asymmetry of the annihilation rates dependent on the orientation of the magnetization using ACAR was firstly observed for Fe \cite{Han57, Ber64, Mij64},  and later for Ni \cite{Mih67, Jar86, Gen91}, Co \cite{Kon92} and Gd \cite{Hoh68}. 
In particular, by measuring the spin-difference of the Fermi surface topology  Ceeh et al.\, determined the absolute value of the electron-electron correlation strength in ferromagnetic Ni for the first time \cite{Cee16a}.
So far, only few (spin-polarized) ACAR studies have been performed on compounds since the synthesis of high-quality single crystals with low concentration of defects, which would act as positron trapping sites,  is often demanding.
Using spin-polarized ACAR Mijnarends et al. demonstrated that the ferromagnetic intermetallic compound NiMnSb fetures a band structure of a half-metal \cite{Han90}.
Applying  the same technique  Livesay et al.\,reported on excellent agreement between the experimental results and band structure calculations performed for the perovskite La$_{0.7}$Sr$_{0.3}$MnO$_{3}$ exhibiting colossal magnetoresistance
\cite{Liv01}.
More recently, Weber et al.\,were able to reconstruct the majority and minority Fermi surface sheets and succeeded in extracting the contribution of each Fermi surface sheet to the effective magnetization of the Heusler compound Cu$_2$MnAl \cite{Web15}.

\subsection{Applications of spin-polarized positron beams}
\label{sec:PolarizedBeams}
At the surface the formation and the subsequent annihilation of triplet Ps (o-Ps)  and singlet Ps (p-Ps) lead to distinct features in positron lifetime and energy  spectra of the annihilation photons.
For this reason, in the few surface studies with spin-polarized positron beams performed so far either the Ps lifetime is measured or the  annihilation $\gamma$'s are recorded with high purity Ge detectors.
In the following, experiments are discussed where the positron beam facilities presented in Section\,\ref{LowEBeam} have been used to reveal the electron spin polarization at surfaces. 

\subsubsection{Surface magnetism}
\label{sec:surfaceMagnetism}
The magnetic properties of the Ni(110) surface were studied by measuring the polarization P$^-$ of electrons captured at the surface \cite{Gid82}.
For this purpose, the formation probability of o-Ps at the surface of a  Ni(110) single crystal has been measured on reversing the magnetizing field or the polarization of the positron beam.
Due to the largely different lifetimes of (free) o-Ps (142\,ns) and p-Ps (125\,ps) asymmetries in o-Ps formation could be easily observed by recording annihilation lifetime spectra which were triggered by the detection of secondary electrons after positron impact.
Since the penetration depth of positrons with E$_+<$1.5\,keV is less than the thermal diffusion length virtually all positrons diffuse back to the surface where they may leave the surface as Ps by picking up an electron from the surface-trapped state.
A parallel alignment of the spin polarization of the positrons and of the surface-captured electrons leads to an increased probability for o-Ps formation.
The asymmetry in formation of o-Ps at the Ni(110) surface has been measured by reversing either the positron-beam polarization or the Ni magnetization.

\begin{figure}[htb]
\centering
\includegraphics[width=0.5\textwidth]{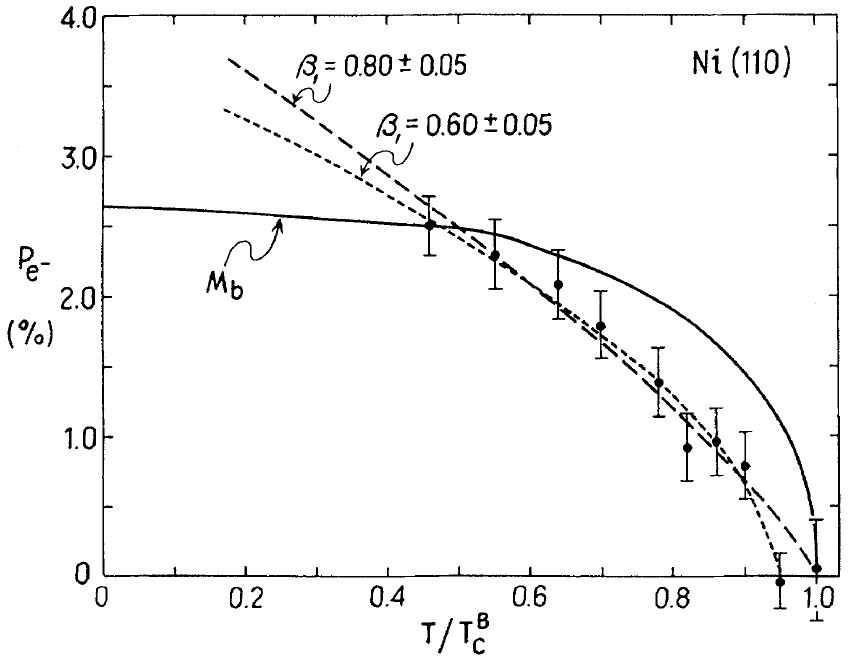} 
\caption{
Temperature dependence of the  electron spin polarization P$_e^-$ at the Ni(110) surface compared with the bulk magnetization M$_b$.
The fits (dashed lines) yield a mean critical exponent of $\beta=$0.7(1) 
(figure from \cite{Gid82}).}
  \label{NiPol_Gid82}
\end{figure} 

As shown in Figure\,\ref{NiPol_Gid82} the measured temperature dependence of P$^-$ differs significantly from that of the bulk magnetization.
A fit to the data yields a critical exponent of $\beta$=0.7(1) of the surface-layer magnetization.
This value is in agreement with both theoretical calculations and a polarized electron scattering experiment although discrepancies could have been anticipated since the latter probes several atomic layers of the sample.
As expected, P$^-$ disappears above the (bulk) Curie temperature $T_c$ ($T_c$(Ni)=633\,K).
In addition, rapid quenching of the ferromagnetic behavior is observed for sub-ML coverage of oxygen and hence successfully confirms the sensitivity of slow positrons probing spin-polarized electrons on the Ni surface \cite{Gid82}.

\subsubsection{Current-induced spin polarization on metal surfaces}
\label{sec:CurrentInducedSpinPolarizationOnMetalSurfaces}

\begin{figure}[htb]
\centering
\includegraphics[width=0.47\textwidth]{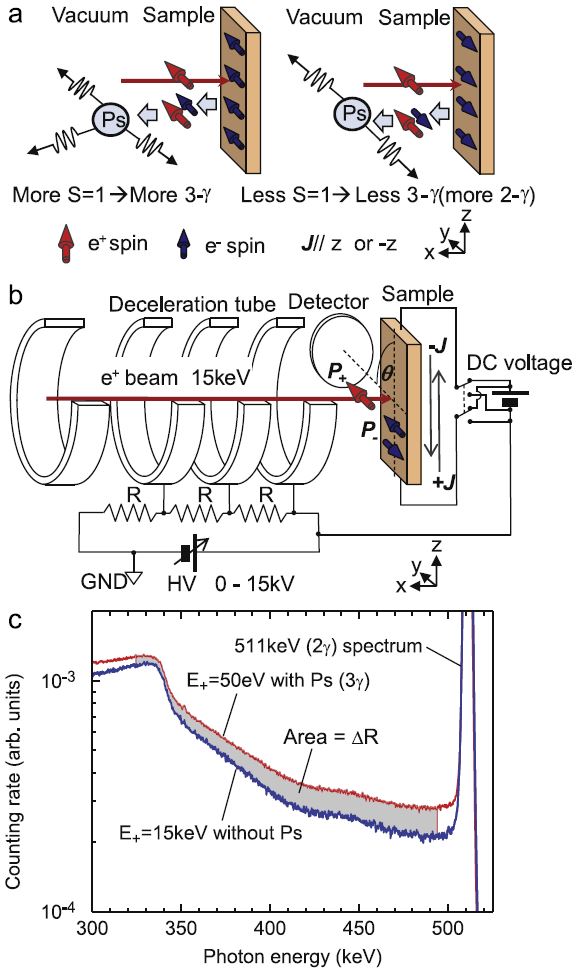} 
\caption{Principle of the electron polarization measurement using a positron beam with transverse spin polarization:
(a) After implantation into the sample positrons with a fixed polarization (along y-axis)  diffuse back to the surface and form Ps with surface conduction electrons. The direction of the electron polarization depends on the electric current which is applied (anti-)parallel to the z-axis. 
(b) Experimental setup shown for 15\,keV positrons which can be decelerated to about 50\,eV. 
(c) Energy spectra of annihilation photons recorded for positrons implanted in a Pt layer with E$_+$=15\,keV and 50\,eV normalized to the 511\,keV photo peak intensity. A measure for the triplet Ps annihilation rate is $\Delta$R (gray area) used to deduce the electron spin polarization (figure from \cite{Kaw13}).}
  \label{PolPrinciple_Kaw13}
\end{figure}

The current-induced spin polarization (CISP) on a Pt surface has been studied by recording annihilation spectra using  a spin-polarized  positron beam.
Kawasuso et al. implanted transversely polarized low-energy positrons (50\,eV)  in a 50\,nm Pt(001) single-crystalline film (grown epitaxially on 1\,nm thick Fe(001) seed layer on MgO(001)) and reversed repeatedly the direction of the current through the Pt film \cite{Kaw13}. 
The basic principle of the experiment is summarized in Figure\,\ref{PolPrinciple_Kaw13}.
After implantation of 50\,eV positrons, virtually all positrons diffuse back to the surface and form preferentially either o-Ps (S=1, triplet state) or p-Ps (S=0, singlet state) depending on the spin current and hence on the population of the individual spin channels at the surface.
The energy spectra of the annihilation photons are recorded with high-purity Ge detectors.
In order to obtain a reference spectrum with negligible contribution of the 3\,$\gamma$ emission positrons are implanted with E$_+$=15\,keV, since in the bulk no Ps is formed and, compared to 2\,$\gamma$ annihilation, the 3\,$\gamma$ annihilation is suppressed by a factor of 371.
Accordingly the 511\,keV photo peak (and Compton scattered photons at lower energy) can clearly be observed in the spectra.
In contrast, the continuous energy distribution between 0 to 511\,keV characteristic for the 3\,$\gamma$ annihilation leads to a much higher event rate below 511\,keV for E$_+$=50\,eV (see gray area quantified by $\Delta$R in Figure\,\ref{PolPrinciple_Kaw13}\,c).
Hence, the change of the triplet Ps annihilation rate is quantified by $\Delta$R from which changes in the  electron spin polarization due to a reversed current can be deduced.
In order to increase the sensitivity to the surface electrons the sample was heated to 250$^{o}$C leading to a cleaner surface and higher thermally activated Ps emission rate.

\begin{figure}[htb]
\centering
\includegraphics[width=0.4\textwidth]{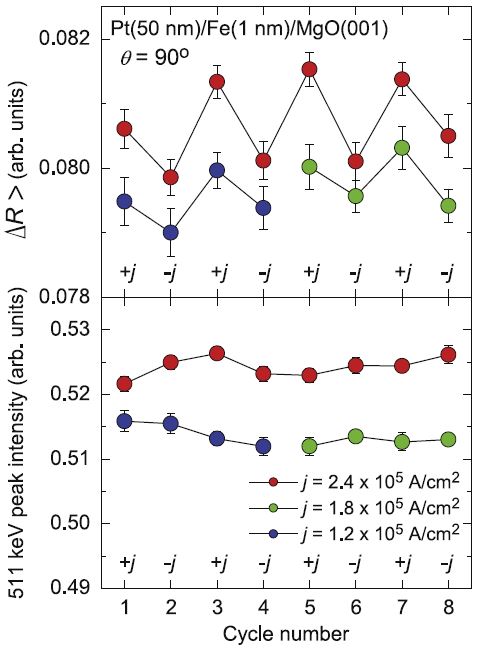} 
\caption{
$\Delta$R and the 511\,keV photo peak intensity as a function of the current reversal cycle number for different current densities through the Pt/Fe/MgO sample. $\Theta$=90$^{o}$ denotes the angle between the directions of the current  and the positron polarization
(figure from \cite{Kaw13}).
}
  \label{PolResult_Kaw13}
\end{figure}

As shown in Figure\,\ref{PolResult_Kaw13} the value of $\Delta$R  oscillates by switching the current direction. 
The higher (lower) $\Delta$R  corresponds to parallel (anit-parallel) spin alignment of positron and electron as depicted in Figure\,\ref{PolPrinciple_Kaw13}\,a.
This oscillation is attributed to the repetitive reversal of the in-plane spin-polarization of surface electrons of the Pt/Fe/MgO sample.
Additional measurements dependent on the angle $\Theta$ between the directions of the current  and the positron polarization resulted in a maximum asymmetry at $\Theta= $90$^{o}$.
The resulting spin polarization of the surface electrons was determined to P$^-\approx$0.04. 
With higher positron implantation energy the current dependent asymmetry decreases rapidly. 
In addition, the same measurement performed on a Au(001) thin film of a Au/Fe/MgO sample showed no oscillations. 
These observations underpin the conclusion that the effect is caused by CISP leading to a different o-Ps emission rate from the Pt surface.
The resulting maximum electron spin polarization was estimated to be $P^->$0.01.

Studies  on CISP on metal surfaces using spin-polarized positron beams have been extended to thin films of Cu, Pd, Ta, and W. 
A large CISP up to 15\,\% with an input charge current density of 10$^5$\,A/cm$^2$ was found for Pt, Pd, Ta, and W
\cite{Zha14}.
In addition, Ta and W showed opposing CISP compared to those of Pt and Pd. 
According to the spin Hall effect the sign of the CISP would suggest that SOC mainly causes the CISP.
 However, Zhang et al.\,explain the sign as well as the magnitude of the large CISP in the outermost layers of these materials in terms of the Rashba-Edelstein mechanism accounting for both, SOC and the broken spatial symmetry at the surface \cite{Zha14}.

\subsubsection{Charge-to-spin conversion and spin diffusion}
\label{sec:ChargeToSpinConversionAndSpinDiffusion}

It was experimentally demonstrated that both charge-to-spin conversion and spin diffusion can be observed using spin-polarized positrons for the determination of the surface polarization of electrons \cite{Zha15}.
Bilayer heterostructures of Bi/Ag/Al$_2$O$_3$ and Ag/Bi/Al$_2$O$_3$, where the Bi/Ag interface is well-known for a giant Rashba effect, were studied by spin-polarized Ps annihilation spectroscopy. 
By applying the analysis technique described above (see Figure\,\ref{PolPrinciple_Kaw13}) a variation of $\Delta$R, and hence of the surface spin polarization P$^-$ by reversing the direction of the charge current was observed for both systems.
In addition, direct evidence of spin diffusion was found by analyzing P$^-$ dependent on the thickness of the outermost surface layer.

\begin{figure}[htb]
\centering
\includegraphics[width=0.5\textwidth]{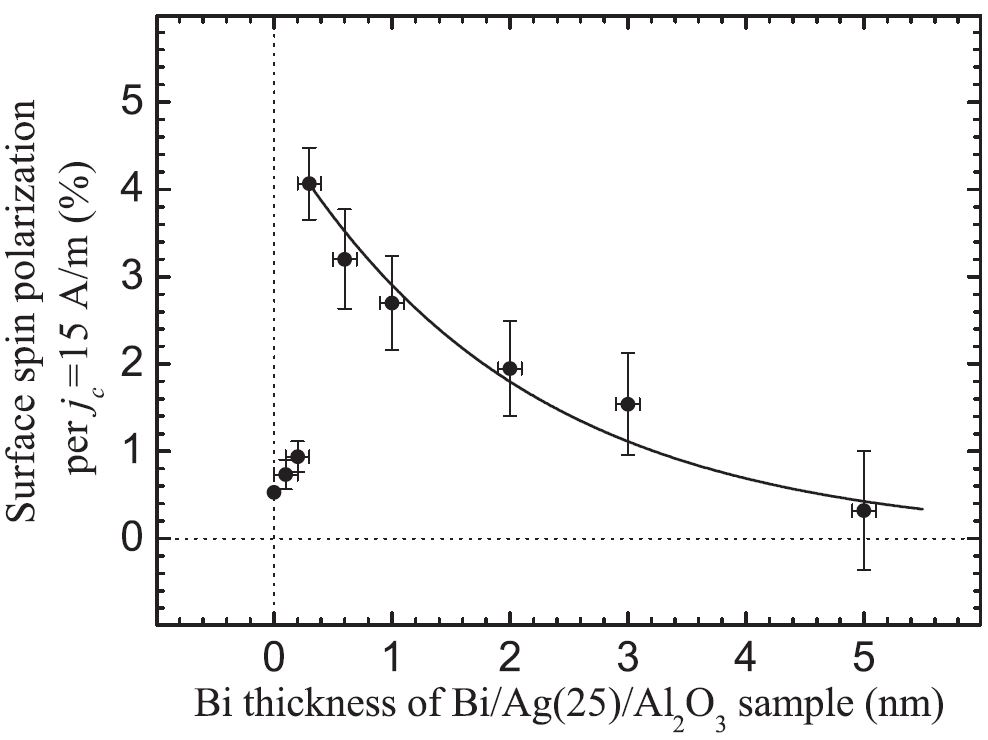} 
\caption{The surface spin polarization of Bi/Ag/Al$_2$O$_3$ samples as a function of Bi thickness; 0.3\,nm approximately corresponds to 1\,ML of Bi. The exponential fit (solid line) allows to extract the spin diffusion length (figure from \cite{Zha15}).
}
  \label{PolResult_Zha15}
\end{figure}

Figure\,\ref{PolResult_Zha15}  shows the electron spin polarization P$^-$ at the surface as function of the Bi layer thickness in the bilayer system Bi/Ag/Al$_2$O$_3$ with the thickness of the Ag layer fixed to 25\,nm.
P$^-$ increases up to a thickness of about 1\,ML of Bi ($\approx 0.3\,nm$) but continuously decreases with further layer thickness of Bi.
In the system Ag/Bi/Al$_2$O$_3$ an inverse behavior was observed.
With increasing layer thickness of Ag the surface spin polarization P$^-$ increases as well.
This suggests that excess electron spins generated at the Bi/Ag interface diffuse into both the Bi and  the Ag layer.
By fitting the experimental data with exponential functions the spin diffusion length could be determined to about 2.1\,nm and 357\,nm for the Bi layer and  for the Ag layer, respectively.
Within  this study, the measurement of the opposing electron spin polarizations at opposite surfaces of Bi/Ag bilayers enabled the direct observation of  the Rashba-Edelstein effect for the first time \cite{Zha15}.

\section{Conclusion and future prospects}
\label{} 
Within this review a comprehensive overview of surface studies using low-energy positrons has been given.
A great number of examples demonstrates that surface experiments clearly benefit from the aforementioned features of positron surface interaction giving rise to study the elemental composition, the atom positions and the electron polarization of the outermost atomic layer. 

In order to further increase the intensity and the brightness of positron beams both the positron moderation efficiency and the yield of the primary positron source have to be enhanced. 
Therefore, more efficient moderator materials such as SiC and new remoderation setups will be applied to provide more brilliant positron beams.
At present, large efforts are being made in constructing a new intense spin-polarized positron beam based on inverse Compton scattering at ELI-NP, Romania.

Several highly interesting applications would immediately profit from these new developments. 
Within the next years it is envisaged to install a 2D-ACAR spectrometer at the high intensity positron beam at NEPOMUC in order to enable depth dependent measurements of the electronic structure. 
Hence, the evolution of the Fermi surface from the surface to the bulk might be observed.
Using an intense spin-polarized positron beam would even allow for ACAR experiments on the spin channels of the electron density of states at the surface, at interfaces and in thin films.

The impressive results obtained by TRHEPD demonstrate that positron diffraction does not only complement established structure analysis e.g.\,with RHEED but leads to a higher level of accuracy in the determination of the atom position at reconstructed surfaces.
The obtained structure information is of particular importance as input for the calculation of the surface band structure which is needed to interpret electronic structure data obtained, e.g.\,by ARPES.
Due to the great potential of this technique funding was granted recently  to install a new TRHEPD apparatus at NEPOMUC.
In the future it is expected that TRHEPD will be expanded to all kinds of surfaces, 2D and 1D structural phase transitions of overlayers and self-assembled organic molecules at surfaces. 
In particular, buckling of 2D systems such as graphene and silicene on various substrates are expected to be investigated in order to deepen the understanding of their extraordinary electronic structure.
 
Time dependent PAES enables the investigation of dynamic processes at surfaces such as segregation and surface alloying.
This kind of effects are of great relevance for the stability of small (magnetic) structures on surfaces, in thin membranes, e.g.\,applied in fuel cells, or for thin film applications such as heterogeneous catalysis, ultrathin surface coatings, and surface corrosion.
Dependent on the surface topology one may profit from the high defect sensitivity of positrons.
Therefore, PAES is seen to be particularly suited  to probe the element distribution at surface defects such as steps or vacancies which play a major role in heterogeneous catalysis.
However, in order to become acquainted with this kind of studies it is recommended to accompany PAES by complementary techniques for the characterization of the surface topology  with STM and elemental analysis with XPS.
Another field of research might be the examination of organic molecules weakly bound to surfaces, since especially PAES is expected to be non-destructive. 

In general, spin-polarized positrons have been demonstrated to be particularly suitable to observe spin phenomena at surfaces. 
At present, however, for this kind of application only one experimental facility is in operation  worldwide.  
The currently long measurement times will be overcome by using new strong sources providing a low-energy spin-polarized positron beam. 

For a number of fundamental experiments the efficient formation of Ps and Ps$^-$ on well-prepared surfaces is crucial. 
In this research field, various projects are under way for the production of anti-hydrogen atoms, spectroscopy of bound leptonic systems, and the generation of a monoenergetic Ps beam by photo-detachment reactions of Ps$^-$.
Finally, it is also envisaged to create a Ps Bose-Einstein condensate  in the not too distant future. 

To sum up, surface investigations using low-energy positrons were shown to significantly contribute to the deeper understanding of all kinds of surface phenomena. 
Therefore, it is expected that positron beam studies will be further expanded to shed light on various open questions in surface science in the future. 
To name a few examples, (i) reconstructed surfaces and superstructures will be analyzed, 
(ii) the interaction of impurity atoms with surfaces, which plays an important role to understand catalytic processes, surface diffusion, surface alloying and segregation can be investigated, and 
(iii) spin-polarization experiments would be promising to study surfaces of topological insulators and functional surfaces for electronic and spintronics applications.

\section*{Acknowledgment}
\label{sec:Acknowledgment}
With regard to the present review the author would like to express his gratitude for fruitful and helpful discussions with various colleagues in particular with Prof. Dr. Toshio Hyodo (High Energy Accelerator Research Organization, KEK , Japan) and Prof. Dr. Alex Weiss ( University of Texas Arlington, USA).
I am much obliged to Dr. Hubert Ceeh, member of my research group, for proofreading of the completed manuscript. 
Furthermore, I owe particular thanks to JSPS for awarding a research fellowship in Japan (2014).
Finally, funding of several projects is greatfully acknowledged: DFG-projects ``Determination of the electronic structure using positrons'' within the Transregional Collaborative Research Center TRR\,80 ``From electronic correlations to functionality'' (2010-2017), and BMBF-projects 05K10WOB (2010-2013) and 05K13WO1 (2013-2016) for the development of various positron beam techniques in surface science and solid state physics.

\clearpage
\section*{Acronyms}
\label{Acronyms}
\begin{tabular}{ll}
ACAR &  Angular Correlation of Annihilation Radiation \\ 
AES	& Auger Electron Spectroscopy \\
AMOC & Age-MOmentum Correlation\\
APECS & Auger Photo Electron Coincidence Spectroscopy\\
ARPES & Angle-Resolved Photo Emission Spectroscopy\\
BEC & Bose-Einstein Condensate \\
CDBS & Coincident Doppler-Broadening Spectroscopy \\
CISP & Current-Induced Spin Polarization \\
CMM & Corrugated Mirror Model \\ 
cps & counts per second \\
DBS & Doppler-Broadening Spectroscopy\\
DFT & Density Functional Theory\\ 
EAES	& Electron induced Auger Electron Spectroscopy \\
FWHM & Full Width at Half Maximum\\
IMFP & Inelastic Mean Free Path\\
LEED & Low-Energy Electron Diffraction \\
LEPD & Low-Energy Positron Diffraction \\
MCP & Micro Channel Plate \\
ML &  MonoLayer \\ 
NEPOMUC & NEutron induced POsitron source MUniCh\\
PALS & Positron Annihilation Lifetime Spectroscopy\\
PAS & Positron Annihilation Spectroscopy\\
PAES	& Positron annihilation induced Auger Electron Spectroscopy \\
Ps & Positronium (o-Ps and p-Ps: ortho-Ps and para-Ps)\\
RHEED & Reflection High-Energy Electron Diffraction \\
RHEPD & Reflection High-Energy Positron Diffraction \\
SNR & Signal-to-Noise Ratio\\
SOC & Spin-Orbit Coupling \\ 
SPM & Scanning Positron Microscope\\
STM &Scanning Tunneling Microscope\\
STS & Scanning Tunneling Spectroscopy \\
SXRD & Surface X-Ray Diffraction \\
TEM &Transmission Electron Microscope\\
TOF &	Time Of Flight\\
TRHEPD & Total Reflection High-Energy Positron Diffraction \\
UHV & Ultra High Vacuum\\
UPS & Ultraviolet induced Photo electron Spectroscopy\\
XAES	& X-ray induced Auger Electron Spectroscopy \\
XPS & X-ray induced Photo electron Spectroscopy\\
XRD & X-Ray Diffraction \\
\end{tabular}  

\clearpage
\section{Appendix}
\label{App}

\subsection{Selected positron emitters}
\label{sec:OverviewOfPositronEmitters}
\begin{table}[htb]
\begin{center}
\begin{small}
\begin{tabular}{ccclllcc}

\textbf{Nuclide }& \textbf{Half-life}  & \textbf{\textit{f}$_{e^+}$} & \textbf{\textit{E}$_{max}$} & \textbf{\textit{E}$_{av}$} &  \textbf{\textit{v$_{av}/$c} }& \textbf{\textit{E}$_{\gamma}$} & \textbf{ \textit{I}$_{\gamma}$} \\

       &    &       &      [keV] &     [keV] &       &     [keV]       &            \\[6pt]

\hline \\[3pt]

       $^{11}$C &   20.4 min &      0.998 &      960.0 &      385.6 &      0.822 &            &            \\[3pt]
       $^{13}$N &   9.97 min &      0.998 &     1198.3 &      491.8 &      0.860 &            &            \\[3pt]
      $^{15}$O &   2.04 min &      0.999 &     1731.7 &      735.3 &      0.912 &            &            \\[3pt]
      $^{18}$F  &   110 min &       0.967  &   633.2  &    249.8  &    0.741  &          &          \\[3pt]
    $^{22}$Na  &  2.60 y  &    0.898  &    545.4  &    215.5  &    0.711  &       1275 &      0.999 \\[3pt]
           &            &      0.001 &     1819.7 &      835.0 &      0.925 &            &            \\[3pt]
      $^{27}$Si &     4.16 s &      0.997 &     3788.8 &     1719.8 &      0.973 &       2211 &      0.002 \\[3pt]
 $^{58}$Co  &    70.8 d  &    0.150  &    475.2  &    201.3  &    0.697  &     811  &    0.994  \\[3pt]
 $^{64}$Cu  &    12.7 h  &    0.179  &    652.5  &    278.1  &    0.762  &    1346  &    0.005 \\[3pt]
 $^{68}$Ge/$^{68}$Ga  &    271 d  &    0.880  &    1899.0  &    836.0  &    0.925  &    1077  &    0.030  \\[3pt]
           &            &      0.011 &     821.7 &      352.6 &      0.806 &            &            \\[3pt]
      $^{89}$Zr &     3.27 d &      0.228 &      902.0 &      395.8 &      0.826 &        909 &      0.999 \\[3pt]
           &            &            &            &            &            &            &            \\
\end{tabular}  
\end{small}
	\caption{Selected $\beta^+$ emitting nuclides with positron branching ratio $f_{e^+}$, end-point energy $E_{\mathrm{max}}$, average energy $E_{\mathrm{av}}$, helicity $v_{av}/c$, energy $E_{\gamma}$ and intensity $I_{\gamma}$ of the most dominant $\gamma$ transition. 
The lifetime of the $^{68}$Ge/$^{68}$Ga  generator system is dominated by the lifetime of the mother nucleus 
(data from NuDat database \cite{URL-NuDat}).}
	\label{tabBetas}
	\end{center}
\end{table}			  

\newpage
\subsection{Overview of positron sources at large-scale facilities}
\label{sec:large-scale}

\begin{table}[h!tb]
\begin{center}
\begin{small}
\begin{tabular}{lcccc}
{\bf Location } & {\bf Facility}  & {\bf Intensity} & {\bf Status} & {\bf References} \\
 &  & [e$^+_{\mathrm{slow}}$/s] & &  \\[6pt]
\hline \\[6pt]
KEK IMSS  &      linac & 5$\cdot10^7$ & in operation & \cite{Hyo11, Wad12, Wad13}  \\ 
 \quad Tsukuba, Japan & 55\,MeV, 11\,$\mu$A, 25\,Hz        &  &  \\[6pt]  
AIST &      linac, &    $10^7$  & in operation & \cite{Suz97a,Aka90}\\  
 \quad Tsukuba, Japan & 75\,MeV, 4\,$\mu$A, 50-100\,Hz        &  &  \\[6pt]
BEPC  &      linac & 2.5$\cdot10^5$  & to be upgraded  & \cite{Wan04, Bao08} \\ 
 \quad Beijing, China  &   1.3\,GeV, 12.5\,Hz      &  &  \\[6pt]
EPOS at ELBE &      linac & $10^7$-$10^8$  & in operation & \cite{Kra08} \\ 
 \quad HZDR Dresden, Germany &  40\,MeV, 1\,mA, 13\,MHz       &  &  \\[6pt]
POSH at RID &    reactor & 8$\cdot10^7$  & in operation & \cite{Sch04b} \\
 \quad TU Delft, Netherlands &    2\,MW     &  &  \\[6pt]
NEPOMUC at FRM II  &    reactor  & $10^9$  & in operation & \cite{Hug08b, Hug13b} \\
\quad TU M\"unchen, Germany &   20\,MW      &  &  \\[6pt]
PULSTAR at NCSU &    reactor  & 5$\cdot10^8$ & in operation & \cite{Haw11}  \\
\quad Raleigh, USA  &   1\,MW     &  &  \\[6pt]
KUR Kyoto Univ.&    reactor  & 1.4$\cdot10^6$  & to be upgraded & \cite{Sat15} \\
\quad Osaka, Japan  &  5\,MW        &  &  \\[6pt]
MNR McMaster Univ. &    reactor     & ($>10^8$)  &   under construction & \cite{Mas15}  \\ 
\quad Hamilton, Canada &     5\,MW     &  &  \\[6pt]
ELI-NP  &    $\gamma$-beam  &   (1-2$\cdot 10^6$)  &    under construction & \cite{Hug12a, Djo16} \\
\quad Bucarest, Romania & $2.4\cdot 10^{10}\,s^{-1}$, E$_{\gamma}<3.5\,$MeV        &  &  \\
\end{tabular}
\end{small}
    \caption{Positron sources at large-scale facilities with main parameters energy, current and repetition rate at electron linacs; thermal power at reactors; photon intensity with maximum energy of the $\gamma$-beam. }
   \label{tabSources}
   \end{center}
\end{table}

\newpage

\subsection{Calculated core-annihilation probabilities}
\label{tab:Core-annihilation}
		
\begin{table}[htb]
\centering
\begin{tabular}{|l|c|c|c|c|c|}
\hline
   Element &    Level &            &            &            &            \\
\hline
           &         1s &         2s &         2p &         3s &         3p \\
\hline\hline
   $_3$Li &       5.42 &            &            &            &            \\
\hline
 $_{4}$Be &       4.37 &            &            &            &            \\
\hline
$_{11}$Na &      0.049 &       2.11 &       6.86 &            &            \\
\hline
$_{12}$Mg &      0.032 &       1.56 &       4.87 &            &            \\
\hline
$_{13}$Al &      0.021 &       1.18 &       3.53 &            &            \\
\hline
$_{14}$Si (100) &       &        &       2.01 \cite{Faz04}&            &            \\
\hline
$_{19}$K &            &      0.028 &      0.069 &       1.27 &       5.25 \\
\hline
$_{22}$Ti &            &      0.045 &       0.11 &       2.31 &       8.70 \\
\hline
$_{23}$V &            &      0.053 &       0.12 &       2.76 &      10.21 \\
\hline
$_{24}$Cr &            &      0.059 &       0.14 &       3.07 &      11.28 \\
\hline
$_{26}$Fe &            &      0.034 &      0.076 &       1.98 &       7.20 \\
\hline
$_{28}$Ni &            &      0.034 &      0.075 &       2.07 &       7.41 \\
\hline
$_{29}$Cu &            &      0.027 &      0.058 &       1.66 &       5.93 \\
\hline
$_{30}$Zn &            &      0.018 &      0.038 &       1.18 &       4.23 \\
\hline
\end{tabular}  
\caption{Core-annihilation probabilities in $\%$  calculated by Jensen et al.\,\cite{Jen90}}
\end{table}
\vspace{1.5cm}
\begin{table}[htb]
\centering
\begin{tabular}{|l|c|c|c|c|c|}
\hline
   Element &    Level &            &            &            &            \\
\hline
           &         3s &         3p &         3d&         4s&         4p \\
\hline\hline
   $_{37}$Rb &       0.021 &0.072            &  0.17          & 1.05 &5.31\\
\hline
 $_{40}$Zr &      0.039 &0.13 & 0.26&2.05&8.71\\
\hline
$_{41}$Nb &0.042& 0.14 &0.26 &2.18& 9.07  \\
\hline
$_{42}$Mo &    0.040 &0.13 &0.24 &2.12&8.74\\
\hline
$_{46}$Pd &0.026 &0.080 &0.13 &1.51&6.07 \\
\hline
$_{47}$Ag & 0.022&0.069 &0.11 &1.38 &5.53 \\
\hline
$_{48}$Cd & 0.015 &0.047 &0.073 &1.01 &4.06\\
\hline
\end{tabular}
\caption{Core-annihilation probabilities in $\%$  calculated by Jensen et al.\,\cite{Jen90}}  
\end{table}
\vspace{1.5cm}

\begin{table}[t!]
\centering
\begin{tabular}{|l|c|c|c|c|c|c|}
\hline
   Element &    Level &            &     & &&  \\
\hline
           &         4s &4p &4d&4f& 5s& 5d\\
\hline\hline
   $_{55}$Cs & 0.020&0.078 & 0.26&  &0.93& 4.41\\
\hline
 $_{58}$Ce &0.052 &0.19 & 0.18& &1.48&5.60\\
\hline
$_{64}$Gd &0.039& 0.14 &0.39 & & 2.03 &9.01 \\
\hline
$_{73}$Ta & 0.034 &0.12 &0.30 &1.82&1.94&8.43\\
\hline
$_{74}$W &0.036 &0.13 &0.31 &1.64&2.05&8.80 \\
\hline
$_{78}$Pt & 0.026&0.089 &0.21 &0.75 &1.56 &6.63\\
\hline
$_{79}$Au & 0.021&0.071 &0.16 &0.56 &1.32 &5.58\\
\hline
$_{81}$Tl & 0.009&0.031 &0.071 &0.22 &0.68 &2.89\\
\hline
$_{82}$Pb & 0.005&0.018 &0.039 &0.12 &0.40 &1.72\\
\hline
\end{tabular}
\caption{Core-annihilation probabilities in $\%$  calculated by Jensen et al.\,\cite{Jen90}} 
\end{table}
\vspace*{13cm}

\newpage

\end{document}